\documentclass[useAMS,usenatbib]{mn2e}
\makeatletter                                   %
\def\endtable{\end@float}                       %
\def\endfigure{\end@float}                      %
\def\@makecaption{\SFB@makefigurecaptio}        %
\makeatother                                    %
\usepackage{fixltx2e}
\usepackage{natbib}
\usepackage{graphicx}
\usepackage{threeparttable}
\usepackage{longtable,lscape}
\usepackage{subfig}
\usepackage[below]{placeins}
\usepackage[bookmarks,breaklinks]{hyperref}
   \hypersetup{
     colorlinks,
     citecolor=blue,
     linkcolor=blue,
    }
\newcommand{\apj}{ApJ}
\newcommand{\aaps}{A{\&}AS}
\newcommand{\cjaa}{ChJAA}
\newcommand{\mnras}{MNRAS}

\newcommand{\aap}{A{\&}A}
\newcommand{\pasp}{PASP}

\newcommand{\apjl}{ApJL}
\newcommand{\apjs}{ApJS}
\newcommand{\aj}{AJ}

\newcommand{\nat}{Nature}
\title[High-frequency very long baseline interferometry studies of NRAO\,530]{High-frequency very long baseline interferometry studies of NRAO\,530}
\author[R.-S. Lu et al ]{R.-S. Lu$^{1,2}$\thanks{E-mail:
rslu@mpifr-bonn.mpg.de}, T. P. Krichbaum$^{1}$
\& J. A. Zensus$^{1}$ \\
$^{1}$ Max-Planck-Institut f\"ur Radioastronomie, Auf dem H\"ugel 69, D-53121 Bonn, Germany\\
$^{2}$ Shanghai Astronomical Observatory, Chinese Academy of Sciences, 80 Nandan Road, 200030 Shanghai, China}

\begin{document}

\date{Accepted 2011 July 29. Received 2011 July 14; in original form 2010 December 24}

\pagerange{\pageref{firstpage}--\pageref{lastpage}} \pubyear{2011}

\maketitle

\label{firstpage}

\begin{abstract}
NRAO\,530 is an optically violent variable source and has been studied with multi-epoch multi-frequency high-resolution very long baseline interferometry (VLBI) observations. NRAO\,530 was monitored with the Very Long Baseline Array (VLBA) at three frequencies (22, 43 and 86\,GHz) on 10 consecutive days in 2007 May during observations of the Galactic center (Sgr\,A*). Furthermore, analysis of archival data of NRAO\,530 at 15\,GHz over the last 10 years allows us to study its detailed jet kinematics. We identified the compact component located at the southern-end of the jet as the VLBI core, consistent with previous studies.  The 10-d monitoring data at the three high frequencies were shown to produce high-quality and self-consistent measurements of the component positions, from which we detected for the first time a two-dimensional frequency-dependent position shift. In addition, the repeated measurements also permit us to investigate the interday flux density and structure variability of NRAO\,530. We find that it is more variable for the inner jet components than those further out. We obtained apparent velocities for eight jet components with $\beta_{\rm app}$ ranging from 2 to $26c$. Accordingly, we estimated physical jet parameters with the minimum Lorentz factor of 14 and Doppler factors in the range of 14--28 (component {\sl f}). The changes in the morphology of NRAO\,530 were related to the motion of separate jet components with the most pronounced changes occurring in the regions close to the core. For NRAO\,530, we estimated a position angle swing of 3\fdg4 per year for the entire inner jet (components {\sl d} and {\sl e}). The non-ballistic motion and change of jet orientation make this source another prominent example of a helical and possibly ``swinging" jet.

\end{abstract}


\section{INTRODUCTION}
At a redshift of 0.902 \citep{1984PASP...96..539J}, NRAO\,530 (also known as B1730-130) is a well-known blazar, which is classified as an optically violent variable object. Strong and erratic broad-band variability has been observed in the radio \citep{1997ApJ...484..118B}, optical~\citep{1988AJ.....95..374W}, and the $\gamma$-ray regime ~\citep{1997ApJ...490..116M}. In $\gamma$-rays, it appeared as a relatively quiescent source after the launch of $Fermi$~\citep{2009ApJ...700..597A}, but started to flare from the end of 2010 October~\citep{Ammando}. Interestingly, \citet{2006A&A...450...77F} serendipitously detected a short hard X-ray flare on 2004 February 17 with the $INTEGRAL$ satellite.

On kpc scales, NRAO\,530 exhibits two-sided lobes in the east-west direction. The western lobe is stronger than the eastern one and is connected to the core with weak arch-structured emission~\citep{2008ChJAA...8..179H}. Centimetre very long baseline interferometry (VLBI) images showed a core-jet structure on pc scales with an oscillating trajectory extending to the north from the assumed core \citep[e.g.][]{1997AJ....114.1999S}. Space VLBI observations revealed brightness temperatures of NRAO\,530 significantly in excess of both, the inverse Compton and the equipartition limit \citep{1998ApJ...507L.117B}. Superluminal motion of several jet components is detected with apparent velocities in the range of 10--40$c$ \citep{1997ApJ...484..118B,2001ApJS..134..181J,2006A&A...456...97F,2008ChJAA...8..179H}.

Images of NRAO\,530 obtained at 5\,GHz in the early 1990s indicated that the position angle (PA) of the jet components changed significantly \citep{1999A&AS..134..201H}. Our new images in 2007 reveal a jet structure for the inner region very different from those previously reported \citep[e.g. in ][]{2006A&A...456...97F}. NRAO\,530 has undergone two moderate flares \citep*[][]{2009arXiv0912.3176A,2010MNRAS.401.1240J} since the dramatic flare around 1997 \citep{1997ApJ...484..118B,2006A&A...456...97F}. This stimulates us to also investigate the pc-scale jet kinematics and its possible relation to the long-term radio flux variations in NRAO\,530.

Following a brief explanation of the observations and data reduction in Section~\ref{sec:2}, we present in Section~\ref{sec:3} the results from Very Long Baseline Array (VLBA) observations at three frequencies (22, 43 and 86\,GHz). The structure of the inner jet was found to have changed significantly when compared with previous VLBI studies. In order to better understand the jet kinematics, we therefore have made use of VLBA data at\,2 cm (15\,GHz), from the Monitoring Of Jets in Active galactic nuclei with VLBA Experiments (MOJAVE) programme, which allows us to conduct a detailed study of structural variations with good time coverage. We briefly discuss in Section~\ref{sec:4} the implications for our understanding of jet morphology and kinematics. A summary of the results is presented in Section~\ref{sec:5}.

Throughout this paper, we adopt the luminosity distance to NRAO\,530 $D_{\mathrm L}$ = 5.8\,Gpc, 1\,mas of angular separation corresponding to 7.8\,pc, and a proper motion of 1\,mas yr$^{-1}$ corresponding to a speed of $\beta_{\mathrm app}$ = 48.5\,c \citep[$H_{0}$ = 71\,km\,s$^{-1}$ Mpc$^{-1}$, $\Omega_{\mathrm M} = 0.27$, $\Omega_{\mathrm \Lambda} = 0.73$,][]{2009ApJS..180..330K}.
\section{OBSERVATIONS AND DATA ANALYSIS}
\label{sec:2}

\begin{figure}
\includegraphics[width=0.45\textwidth,clip]{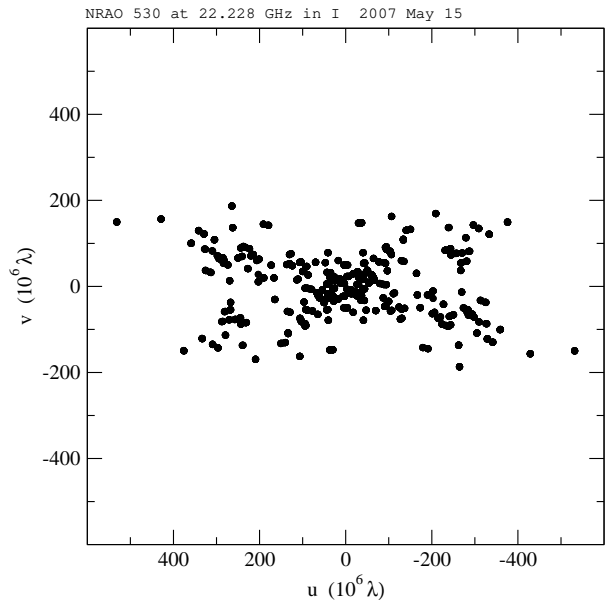}
\caption{An example of uv coverage at 22\,GHz on 2007 May 15.}
\label{fig:uv_plt}
\end{figure}

During the global campaign on Sgr\,A* in 2007 May, NRAO\,530 was monitored as a calibrator with the VLBA from 2007 May 15 to 24 (2007.370--2007.395) at 22, 43 and 86\,GHz every day. NRAO\,530 was observed as a fringe finder and was used to check the amplitude calibration of Sgr\,A*~\citep{2011A&A...525A..76L}. NRAO\,530 was observed shortly before, during, and after the time of visibility of the target source, Sgr\,A*. Dual circular polarization at a recording rate of 512 Mbps [8 intermediate frequency (IF) channels, 16 MHz per IF, and 2 bits per sample] was recorded at each station. The individual VLBI scans on NRAO\,530 had a duration of 3\,min at 22\,GHz and 4\,min at 43 and 86\,GHz, respectively. Five scans at each frequency were observed, resulting in a total time on source of $\sim$ 15\,min at 22\,GHz and $\sim$ 20\,min at 43 and 86\,GHz, respectively. An example of the \textit{uv} coverage at 22\,GHz is shown in Fig.~\ref{fig:uv_plt}.
The data analysis was performed within the \textsc{aips} software in the usual manner. Opacity corrections were made at all three frequencies by solving for receiver temperature and zenith opacity for each antenna. In addition, we have also used the VLBA observations at 15\,GHz from the ``2\,cm Survey'' and the follow up MOJAVE\footnote{http://www.physics.purdue.edu/MOJAVE/}program between 1999 and 2009, spanning 10 years. These data were provided as automatically self-calibrated \textit{uv} FITS file, which we re-do the mapping and self-calibration in \textsc{difmap} before model-fitting.

In all cases of the model fitting, circular Gaussian components have been chosen in order to simplify and unify the comparison. The final fitting of the jet components was stopped when no significant improvement of the reduced $\chi_\nu^2$ values was obtained. The formal errors of the fit parameters were estimated by using the formalisms described in \citet{1999ASPC..180..301F}. The position errors were estimated by $\sigma_{\mathrm r} = \frac{\sigma_{\rm{rms}}\ d}{2\ I_{\rm{peak}}}$, where $\sigma_{\rm{rms}}$ is the post-fit rms, and $d$ and $I_{\rm{peak}}$ are the size and the peak intensity of the component. In case of very compact components, this tends to underestimate the measurement error. We therefore included an additional minimum error criterion according to the map grid size, which roughly matches one-fifth of the beam size at each frequency and corresponds to 0.06\,mas at 15 and 22\,GHz, 0.03\,mas at 43\,GHz, and 0.02\,mas at 86\,GHz. For the weighted mean of parameters from the 10 consecutive epochs, we have chosen the max $(\sqrt{1/\Sigma{\frac{1}{\sigma_{\rm i}^2}}},\frac{\sigma}{\sqrt{N}})$ as the standard error for the few cases when the former is smaller than the latter, with $\sigma$ being the standard deviation. By doing so, we have assumed stationarity of the corresponding parameter on a daily time-scale. For the spectral analysis, we estimate absolute flux-density calibration errors of 3-5 per cent at 22 and 43\,GHz and 20 per cent at 86\,GHz. Our 10-d observations have revealed a consistent jet structure at all three frequencies. Cross-identification of jet components is straightforward and unambiguous. The components seen at 15\,GHz were cross-identified based on their flux density evolution, core separation, PA, and full width at half-maximum (FWHM). We noted that although the archival data at 15\,GHz have a good time resolution, large time gaps between 2000 and 2002 pose some difficulties in tracing the weak jet components (see Section~\ref{sec:3.4}).

\section{RESULTS}
\label{sec:3}

\begin{figure*}%
\centering
\hspace{25pt}
\subfloat[][15\,GHz, 2007/06/10 (2007.411)]{%
\label{fig:maps2007a}%
\begin{minipage}[t]{0.34\textwidth}
{\raisebox{0.1in}{
\includegraphics[width=1.0\textwidth,clip]{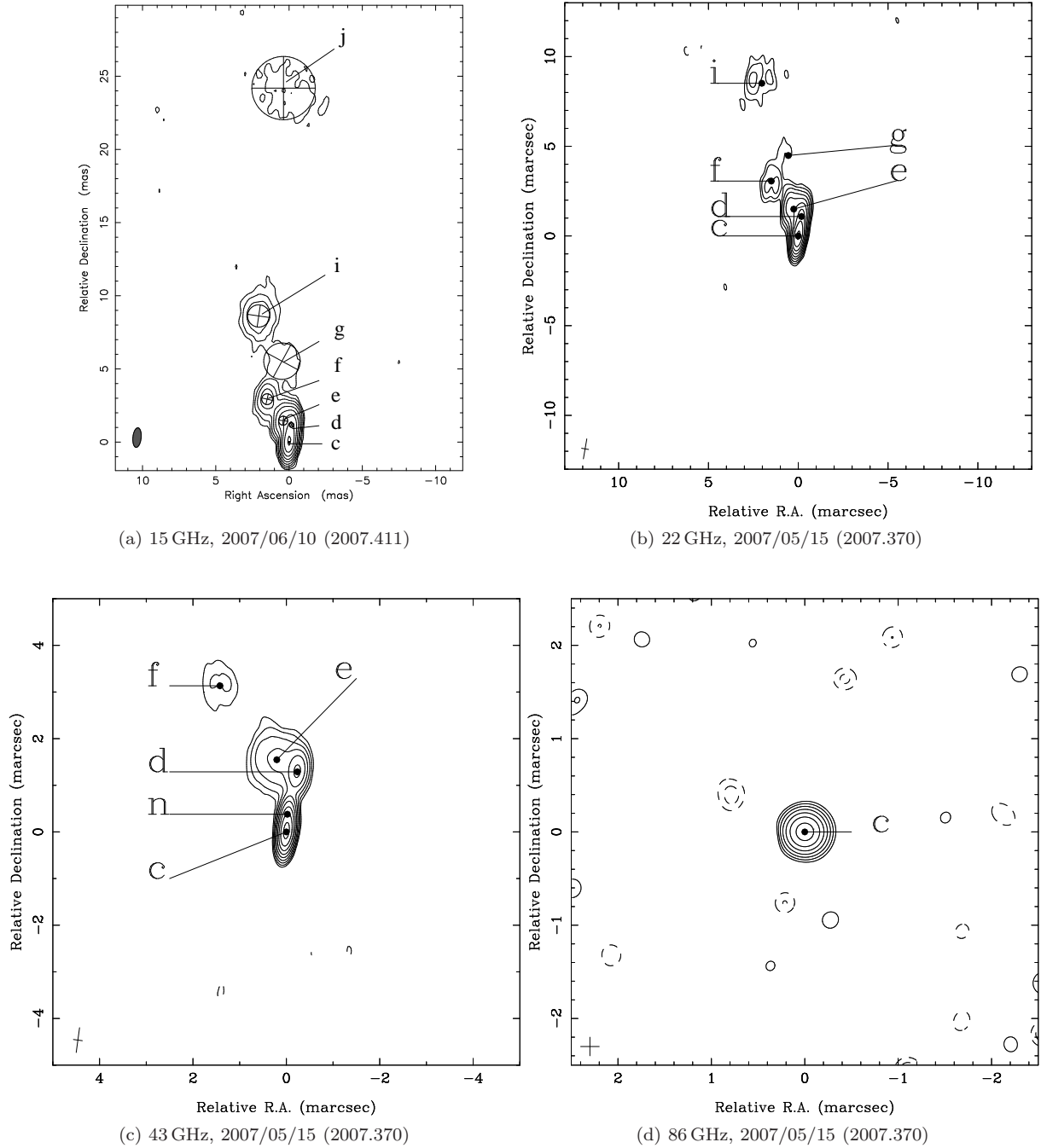}}}
\end{minipage}}
\hspace{25pt}
\subfloat[][22\,GHz, 2007/05/15 (2007.370)]{%
\label{fig:maps2007b}%
\includegraphics[width=0.45\textwidth,clip]{fg2b.ps}}\\%
\vspace{10pt}
\hspace{3pt}%
\subfloat[][43\,GHz, 2007/05/15 (2007.370)]{%
\label{fig:maps2007c}%
\includegraphics[width=0.45\textwidth,clip]{fg2c.ps}}
\hspace{3pt}%
\subfloat[][86\,GHz, 2007/05/15 (2007.370)]{%
\label{fig:maps2007d}%
\includegraphics[width=0.45\textwidth,clip]{fg2d.ps}}
\caption{Example of clean maps of NRAO\,530 at (a) 15\,GHz, (b) 22\,GHz, (c) 43\,GHz, and (d) 86\,GHz. See Table~\ref{tab:para_n} for the map parameters, and the Supporting Information section for all images.}
\label{fig:maps2007}%
\end{figure*}

In Fig.~\ref{fig:maps2007}, we show as an example one of the resulting clean maps of NRAO\,530 at 15\,GHz at epoch 2007 June 10 (2007.441) and at 22, 43, and 86\,GHz at epoch 2007 May 15 (2007.370). Maps for all epochs at all frequencies are included as Supporting Information in the online version of this paper (Figs~\ref{fig:maps2007_online} and \ref{fig:u_online}). Table~\ref{tab:para_n} lists the map parameters and Table~\ref{tab:nrao530_model} summarizes the epoch of observation, the identification of components, and the model fit parameters and their uncertainties (all errors are 1 $\sigma$ throughout) of the maps shown in Fig.~\ref{fig:maps2007} (only a small portion of each table is shown here; the full tables are available online in Tables~\ref{tab:para_n_online} and \ref{tab:nrao530_model_online} -- see Supporting Information). On 2007 May 15 (2007.370), the sparse \textit{uv} coverage at 86\,GHz did not permit us to model the source structure by more than a single Gaussian component, and we will not consider this data set for further analysis.

\begin{table*}
\caption[Description of VLBA images of NRAO\,530]{Description of VLBA images of NRAO\,530 shown in Fig.~\ref{fig:maps2007}. The columns list the epoch of observation, the observing frequency, the peak flux density, the parameters of the restoring elliptical Gaussian beam (the FWHM of the major and minor axes and the PA of the major axis), and the contour levels of the image, expressed in percentage of the peak intensity. See the Supporting Information section for the full table.}
\label{tab:para_n}
\begin{tabular}{lllllrr}
\hline
&&&\multicolumn{3}{c}{Restoring Beam}&\\
Epoch&$\nu$&$S_{\rm peak}$&Major&Minor&P.A.&Contours\\
&(GHz)&(Jy beam$^{-1}$)&(mas)&(mas)&($^\circ$)&\\
(1)&(2)&(3)&(4)&(5)&(6)&(7)\\
\hline
2007.441&15&1.20&1.34&0.568&$-5.7$&$-0.35$, 0.35, 0.7, ..., 89.6\\
2007.370&22&1.32&1.19&0.344&$-12.8$&$-0.3$, 0.3, 0.6, ..., 76.8\\
2007.370&43&1.33&0.537&0.193&$-7.7$&$-0.3$, 0.3, 0.6, ..., 76.8\\
2007.370&86&0.74&0.2&0.2&0&$-1$, 1, 2, ..., 64\\
\hline
\end{tabular}
\end{table*}

\begin{table*}
\centering
\caption[Parameters of the Gaussian model fit components for NRAO\,530]{Parameters of the Gaussian model fit components of the images shown in Fig.~\ref{fig:maps2007}. See the Supporting Information section for the full table.}
\label{tab:nrao530_model}
\begin{tabular}{lcrcrc}
\hline
Epoch & Id. & Flux& Core Separation& P.A.& Size\\
&&[mJy]&[mas]&[degree]&[mas]\\
\hline
&&\multicolumn{2}{|c|}{$\nu$ = 15\,GHz}&&\\
2007.441&c &1138.4$\pm$60.1& 0.00$\pm$0.00&  0.0$\pm$0.0&0.07$\pm$0.01\\
       &d & 545.9$\pm$20.6& 1.17$\pm$0.06& $-8.6\pm$2.9&0.31$\pm$0.01\\
       &e & 310.9$\pm$47.4& 1.52$\pm$0.06& 15.4$\pm$2.3&0.62$\pm$0.09\\
       &f & 109.6$\pm$19.5& 3.29$\pm$0.06& 27.0$\pm$1.0&0.74$\pm$0.11\\
       &g &  67.1$\pm$19.4& 5.55$\pm$0.35&  5.0$\pm$3.6&2.49$\pm$0.70\\
       &i & 139.5$\pm$22.7& 8.87$\pm$0.12& 13.5$\pm$0.8&1.52$\pm$0.23\\
       &j & 121.9$\pm$37.0&24.19$\pm$0.65&  0.9$\pm$1.5&4.33$\pm$1.30\\

&&\multicolumn{2}{|c|}{ $\nu$ = 22\,GHz}&&\\
2007.370&c&1332.9$\pm$23.6& 0.00$\pm$0.00&  0.0$\pm$0.0&0.08$\pm$0.01\\
          &d& 385.9$\pm$40.5& 1.10$\pm$0.06& $-9.5\pm$3.1&0.28$\pm$0.02\\
          &e& 377.8$\pm$72.1& 1.52$\pm$0.08&  9.6$\pm$2.9&0.87$\pm$0.15\\
          &f&  99.1$\pm$26.5& 3.41$\pm$0.10& 26.1$\pm$1.6&0.81$\pm$0.20\\
          &g&  29.5$\pm$11.2& 4.52$\pm$0.22&  7.0$\pm$2.8&0.32$\pm$0.09\\
          &i& 138.3$\pm$44.2& 8.75$\pm$0.30& 13.4$\pm$2.0&1.99$\pm$0.61\\
&&\multicolumn{2}{|c|}{ $\nu$ = 43\,GHz}&&\\
2007.370&c&1258.0$\pm$60.7& 0.00$\pm$0.00&  0.0$\pm$0.0&0.03$\pm$0.01\\
          &n& 272.9$\pm$22.4& 0.38$\pm$0.03& $-3.2\pm$4.5&0.08$\pm$0.01\\
          &d& 221.2$\pm$27.7& 1.31$\pm$0.03&$-10.4\pm$1.3&0.21$\pm$0.02\\
          &e& 250.3$\pm$40.2& 1.56$\pm$0.05&  7.5$\pm$2.0&0.70$\pm$0.11\\
          &f&  56.1$\pm$20.2& 3.44$\pm$0.12& 24.4$\pm$2.0&0.68$\pm$0.24\\
&&\multicolumn{2}{|c|}{ $\nu$ = 86\,GHz}&&\\
2007.370&c&1214.4$\pm$186.3&0.00$\pm$0.00&0.0  $\pm$0.0&0.16$\pm$0.02\\
\hline
\end{tabular}
\end{table*}

On VLBI scales, NRAO\,530 is characterized by a one-sided jet following a curved trajectory of $\sim$25\,mas to the north  \citep[e.g.][]{1984A&A...135..289R,1996A&A...308..415B}.
We find no naming convention for the VLBI components in the literature that we could follow for NRAO\,530. We therefore labelled the modelled components in an alphabetical order with increasing core separation, starting from the core ({\sl c}) to the outermost jet component ({\sl j}). A very faint jet feature, which faded away after 2002 is labeled as {\sl x}. An inner jet component, which was not resolved at frequencies $\leq$ 22\,GHz in 2007 May, is named as {\sl n} and the newly ejected components after 2008 is named as {\sl n$^\prime$}.
In our analysis, we identify the southernmost component (component {\sl c}) to be the stationary core of the radio emission, because of its flat radio spectrum (see Section~\ref{sec:3.1}) and high compactness (see Table~\ref{tab:nrao530_model}). This is in agreement with the identification scheme proposed by \citet{2001ApJS..134..181J}, \citet{2006A&A...456...97F}, and \citet*{2010MNRAS.408..841C}.

As can be seen in Fig.~\ref{fig:maps2007}, our three-frequency data sets of 2007 revealed a very consistent jet morphology within the central $\sim$ 10\,mas, with richer structure seen at lower frequencies. The innermost jet component ({\sl n}) detected at 43 and 86\,GHz was not seen at 22\,GHz, which may be due to the insufficient angular resolution at this frequency and the relative weakness, when compared with adjacent components. The inner jet extends slightly to the north-west direction (PA $\sim -10^{\circ}$) out to a core separation of $\sim$ 1.5\,mas and then bends sharply by $\sim 90^{\circ}$ towards a PA of $25^{\circ}$ at a core separation of 3.5\,mas. From here, the jet bends gently northward following a curved path with underlying diffuse emission (see Fig.~\ref{fig:maps2007}). Finally, it fades away at a core separation of $\sim$ 9\,mas to a PA $\sim$ $10^\circ$. The jet component ({\sl i}) located at $\sim$25\,mas north of the core is not seen at the three highest frequencies, but is clearly visible at 15\,GHz across all the epochs (Table~\ref{tab:nrao530_model}). This is probably due to the lower resolution at the 15 GHz and the steep spectrum of this jet component.

\subsection[Component spectrum]{\label{sec:3.1}Component spectrum}
\begin{figure*}
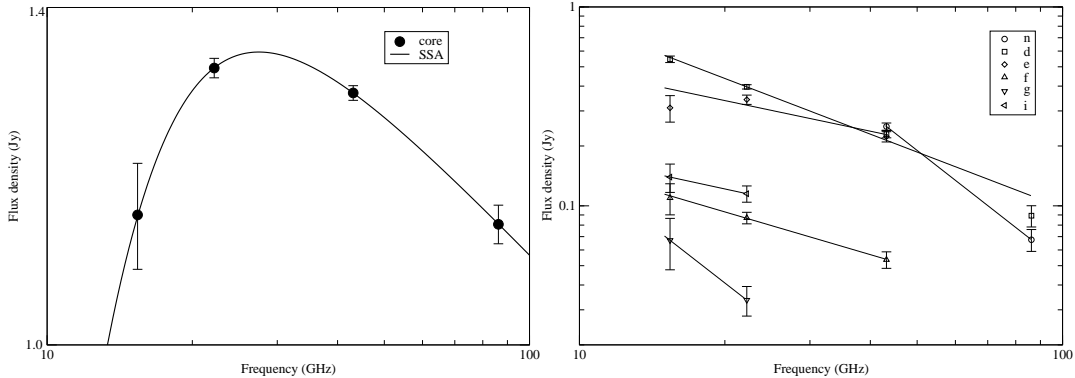

\centering
\includegraphics[width=0.4\textwidth]{fg3a.eps}%
\includegraphics[width=0.4\textwidth]{fg3b.eps}
\caption{Component spectra of NRAO\,530 in 2007.4. Left: Spectrum of the core. The SSA fit is shown as a solid curve. Right: Spectra of the jet components.}
\label{fig:spectra}%
\end{figure*}

In order to study the spectra of the VLBI components, we also used the 15-GHz data of 2007.441, taken close in time to our multifrequency observations in 2007 May. The spectra of the components are shown in Fig.~\ref{fig:spectra} based on the flux densities obtained by model-fitting of Gaussian components to the visibility data.
The spectrum of the core component {\sl c}, which is located at the south end of the VLBI jet, is shown in Fig.~\ref{fig:spectra} (left).
The presence of an unambiguous turnover and the flattest optically thin spectral index ($-0.19\pm0.03$ between 43 and 86\,GHz) among all the components allow us to identify this component as the compact VLBI core.

By assuming that the spectral turnover is due to synchrotron self-absorption~\citep*[SSA;][]{1974ApJ...192..261J}, the convex spectrum of the core can be fitted by the following equation:
\begin{equation}
S_{\nu}=S_{0}\nu^{2.5}[1-{\rm exp}(-\tau_{\rm s}\nu^{\alpha-2.5})],
\end{equation}
where $\nu$ is the observing frequency in GHz, $S_{0}$ is the intrinsic flux density in Jy at 1\,GHz, $\tau_{\rm s}$ is the SSA opacities at 1\,GHz, and $\alpha$ is the optically thin spectral index.
The best-fit parameters are $\alpha=-0.22\pm0.01$, $S_{0}=2.2\pm0.1$\,mJy and $\tau_{\rm s}=1342.7\pm14.4$. The fitted spectra are shown in Fig.~\ref{fig:spectra} (left).

The magnetic field $B$ of a homogeneous synchrotron self-absorbed source component can be calculated via
\begin{equation}
B^{\mathrm syn}=10^{-5}b(\alpha)\nu_{\mathrm max}^5\theta^{4}S_{\mathrm max}^{-2}\delta/(1+z),
\end{equation}
where $\nu_{\rm max}$ is the peak frequency in GHz, $\theta$ the source angular size in mas, $S_{\rm max}$ the peak flux density in Jy and $b(\alpha)$ a tabulated function of the spectral index $\alpha$~\citep{1983ApJ...264..296M}. Assuming similar Doppler factors $\delta$ for the core and jet component {\sl f} ($\delta=14.1$, see Section~\ref{sec:3.4}), we obtain $B^{\rm syn}=76.1$\,mG with the following parameters: $b(\alpha) = 1.8$, $S_{\rm max}$ = 1.339\,Jy, $\nu_{\rm max}=27.5$\,GHz, $\theta=0.09$\,mas. This is not in agreement with the magnetic field estimation in \citet{2006A&A...456...97F}, who obtained 8\,mG for the core component by assuming $\delta=1$.

The spectra of the outer jet components are shown in Fig.~\ref{fig:spectra} (right), along with their spectral fitting. Component {\sl n} is most likely self-absorbed at frequencies below 43\,GHz. From 43 and 86\,GHz, the extrapolated flux density at 22\,GHz is $\sim$ 0.9\,Jy. Without self-absorption, such a bright feature should have been seen at 22\,GHz at a core distance of $\sim$ 0.4\,mas (see also Fig.~\ref{fig:slice}). Other jet components ({\sl d}, {\sl e}, {\sl f}, {\sl g} and {\sl i}) displayed steep spectra, except for some indication of absorption for component {\sl e} at 15\,GHz. Further observations with wider frequency coverage for component {\sl e} would be needed to clarify this. The corresponding optically thin spectral indices $\alpha$ ($S_\nu \propto \nu^\alpha$) for these jet components are $\alpha_{\rm n}$ = $-1.90\pm0.19$, $\alpha_{\rm d}$ = $-0.93\pm0.09$, $\alpha_{\rm e}$ = $-0.51\pm0.21$, $\alpha_{\rm f}$ = $-0.71\pm0.03$, $\alpha_{\rm g}$ = $-1.89\pm0.92$, and $\alpha_{\rm i}$ = $-0.53\pm0.51$, where the index labels each component.
\subsection[Frequency-dependence of component positions]{\label{3.2}Frequency-dependence of component positions}
\begin{figure*}
\centering
\includegraphics[width=0.7\textwidth,clip]{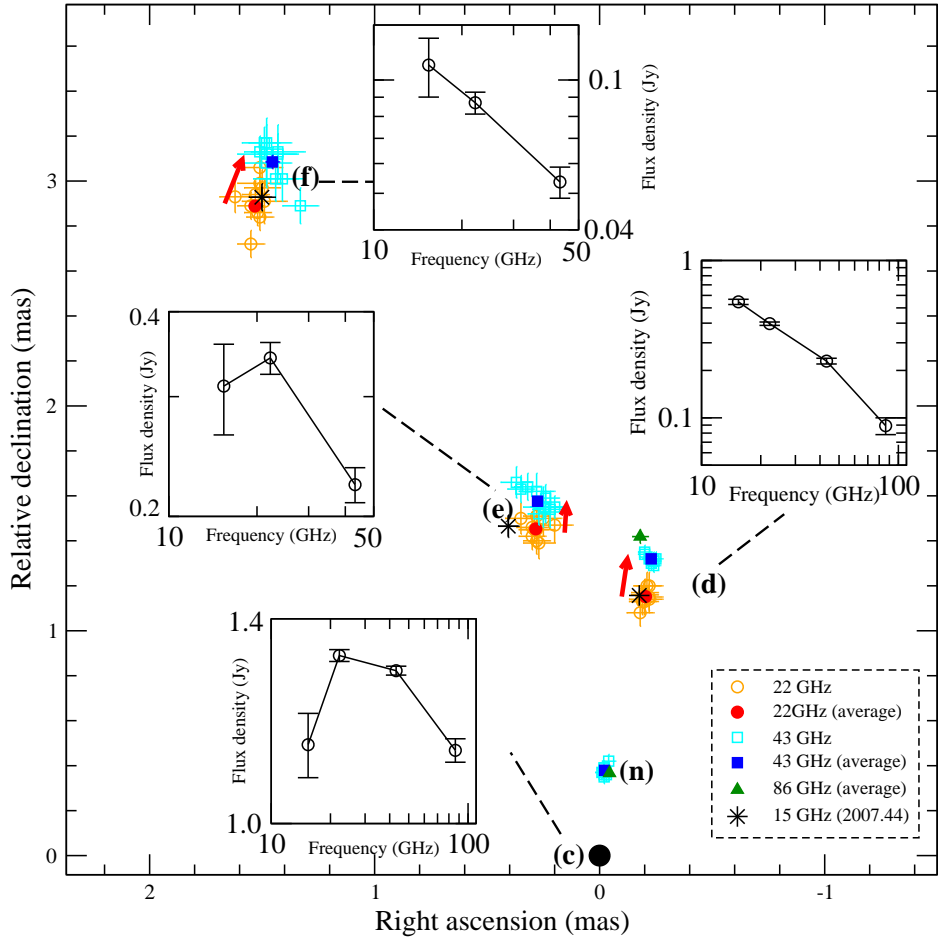}
\caption[Frequency dependence of jet components]{Plot of the positions for the inner jet components. Different symbols denote different frequencies with the filled symbols representing the mean positions at 22 (filled circles), 43 (filled squares) and 86\,GHz (filled triangles). To prevent confusion, only the averaged mean positions are shown at 86\,GHz. A clear frequency-dependent displacement of the jet component positions can be seen and is indicated by red arrows. Spectra are shown for those components, which are visible at more than 2 frequencies (insets).}
\label{fig:shift}
\end{figure*}

The 10-d repeated measurements at 22, 43 and 86\,GHz allow us to investigate the frequency-dependent position shift of the jet components. In Fig.~\ref{fig:shift}, we show the positions of the components detected at any two of these three frequencies. By using the additional data at 15\,GHz (epoch 2007.441), the spectra of the core and components {\sl d}, {\sl e} and {\sl f} are shown as insets in Fig.~\ref{fig:shift}. The time averaged component positions are summarized in Table~\ref{tab:shift}.

\begin{table*}
\caption[Position shift of jet components]{\label{tab:shift}Position shift of jet components. Listed are the component Id., the time-averaged value of 10 d of components' core separation (r) and PA ($\theta$) at 22, 43 and 86\,GHz. The last four columns summarize the magnitude and direction (in degrees east of north) of the position shift between the two frequency pairs (22--43\,GHz and 43--86\,GHz) for each component.}
\newsavebox{\tablebox}
\begin{lrbox}{\tablebox}
\begin{threeparttable}[b]
\begin{tabular}{lccccccrrrr}
\hline
Id.&\multicolumn{2}{c}{22\,GHz}&\multicolumn{2}{c}{43\,GHz}&\multicolumn{2}{c}{86\,GHz}&&\cr
&r&$\theta$&r&$\theta$&r&$\theta$&$\rmn{\triangle r_{22-43}}$&$\rmn{\theta_{22-43}}$&$\rmn{\triangle r_{43-86}}$&$\rmn{\theta_{43-86}}$\cr
&(mas)&($^\circ$)&(mas)&($^\circ$)&(mas)&($^\circ$)&(mas)&($^\circ$)&(mas)&($^\circ$)\cr
\hline
n&...&...&0.38$\pm$0.01&$-3.3\pm1.4$&0.37$\pm$0.01&$-6.8\pm1.1$&...&...&0.03$\pm$0.01&$-118.6\pm30.0$\cr
d&1.17$\pm$0.02&$-10.0\pm0.9$&1.34$\pm$0.01&$-9.8\pm0.4$&1.43$\pm$0.03&$-7.2\pm1.4$&$0.17\pm0.02$&$-8.5\pm7.1$&$0.11\pm0.03$&$26.3\pm18.3$\cr
e&1.48$\pm$0.02&11.0$\pm$0.8&1.60$\pm$0.02&9.9$\pm$0.7&...&...&$0.12\pm0.03$&$-3.7\pm13.5$&...&...\cr
f&3.27$\pm$0.02&27.8$\pm$0.4&3.41$\pm$0.03&25.1$\pm$0.5&...&...&$0.21\pm0.04$&$-21.7\pm10.0$&...&...\cr
\hline
\end{tabular}
\end{threeparttable}
\end{lrbox}
\resizebox{1.0\textwidth}{!}{\usebox{\tablebox}}
\end{table*}

\begin{figure}
\centering
\includegraphics[width=0.45\textwidth,clip]{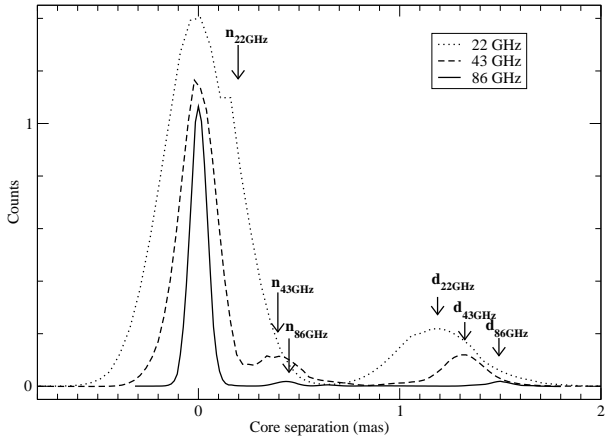}
\caption[Slice for the inner jet of NRAO\,530 along PA = $-10^{\circ}$]{Slice
along PA = $-10^{\circ}$ through the total intensity maps of the inner 2\,mas of NRAO\,530 at 22 and 43\,GHz on  2007 May 15 and at 86\,GHz on 2007 May 16. An increase of the core separation of components {\sl n} and {\sl d} with frequency can be interpreted as opacity shift of the core position (see text).}
\label{fig:slice}
\end{figure}

As it can be clearly seen in Fig.~\ref{fig:shift}, the relative positions of three components ( {\sl d}, {\sl e}, and {\sl f}) are systematically displaced between 22 and 43\,GHz. These are components whose positions are most reliably measured in our experiments. For all three components, the core separation at 43\,GHz is larger than that at 22\,GHz, at a $> 3\,\sigma$ level. For the components {\sl d}, {\sl e} and {\sl f}, the position shift between 22 and 43\,GHz is $0.17$, $0.12$ and $0.21$\,mas, respectively (see Table~\ref{tab:shift}).
The measured position of component {\sl d} at 86\,GHz is consistent with the lower frequency data in the sense that its core distance at 86\,GHz is larger than that at lower frequencies. For component {\sl n}, we are unable to compare the position shift between 43 and 86\,GHz, limited mainly by the large position scatter at 86\,GHz. At the same time, the observations at 15\,GHz also provide consistency checks for the components {\sl d}, {\sl e} and {\sl f}. To further verify this position shift effect, we show in Fig.~\ref{fig:slice} radial slices made along PA of $-10^{\circ}$ for the inner $\sim$ 2\,mas of the jet. One can clearly see that the peak of the component {\sl d} is displaced between 22, 43 and 86\,GHz. For the innermost jet component {\sl n}, however, it is impossible to reliably model it at frequencies $\leq$ 22\,GHz, and its position measured at 86\,GHz suffers from large scatter.

\begin{figure}
\centering
\includegraphics[width=0.35\textwidth,clip]{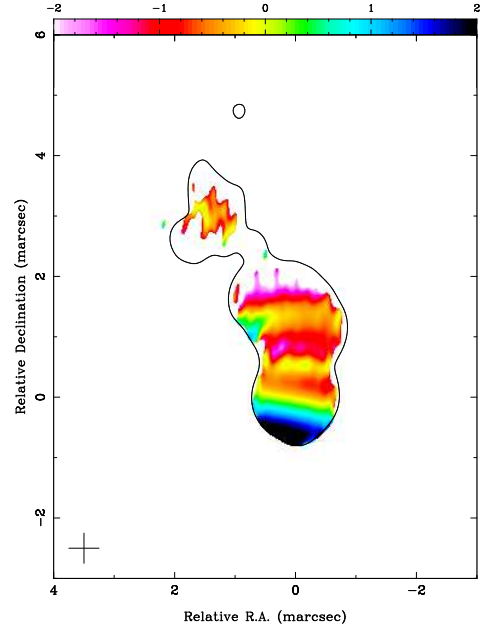}
\caption[Spectral index map of NRAO\,530]{Spectral index map of NRAO\,530 with a single contour (5\,mJy beam$^{-1}$) of the 22\,GHz Stokes I image superimposed. The map was made with 22 and 43\,GHz data on 2007 May 15 after applying for the 43\,GHz map a relative shift of 0.02\,mas to the east and -0.16\,mas to the north (see text for details). Note that the oblique boundary between the regions of optically thick and thin emission near the core is just perpendicular to the inner jet direction along a PA = $-10^{\circ}$.}
\label{fig:spectral_index}
\end{figure}

We also used spectral index mapping to pin down this effect. In Fig.~\ref{fig:spectral_index}, we show a spectral index map constructed between 22 and 43\,GHz. Prior to making the spectral index map, we have corrected the misalignments between the two maps by shifting one map relative to the other. Such an alignment is necessary as suggested by the appearance of obvious false features in spectral index maps made without any corrections. The amount of the shift was determined by aligning the optically thin part of the jet based on a two-dimensional cross-correlation technique, a method very similar to that described by \citet{2008MNRAS.386..619C}. In our case, we restored both images with a circular beam of 0.5\,mas. The derived shift of the 43\,GHz-image relative to the 22\,GHz-image was 0.02\,mas to the east and $-0.16$\,mas to the north ($\rmn{\triangle r_{22-43}=0.16\,mas}, \theta_{\rm{22-43}}=-7.1^{\circ}$), in a direction very close to the inner jet (PA $\sim -10^{\circ}$). These shifts are consistent with the results obtained from model fitting (average shift of $\rmn{\triangle r_{22-43}=0.16\,mas}, \theta_{\rm{22-43}}=-11.5^{\circ}$ for components {\sl d}, {\sl e} and {\sl f}), indicating an overall relative shift effect between the two images. The derived shifts for other selected epochs also agree well with these values.

\begin{figure}
\centering
\includegraphics[width=0.45\textwidth,clip]{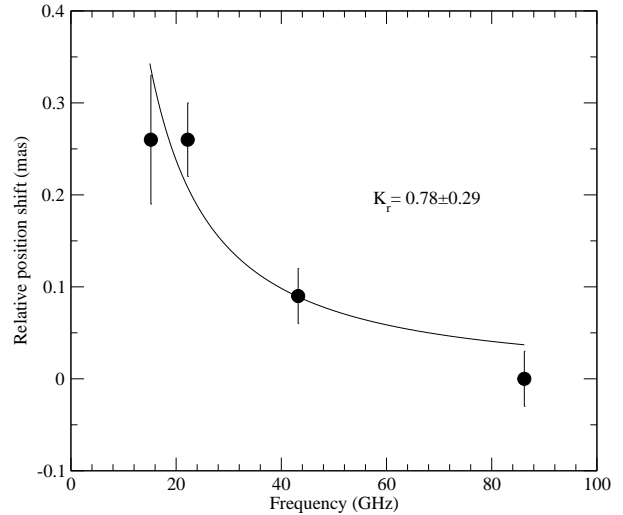}
\caption[Position shift plot]{Position shift plot for component {\sl d} with 86\,GHz as reference frequency.}
\label{fig:Kr_plt}
\end{figure}

A well known effect which could cause the frequency dependence of jet component position is the opacity-induced position shift of the VLBI core~\citep[`core shift',][]{1984ApJ...276...56M, 1998A&A...330...79L} in a synchrotron-self absorbed jet \citep{1979ApJ...232...34B}.
The VLBI core position, which is defined by the $\tau=1$ surface at the jet base, depends on the frequency, with the distance to the central engine, $r_{\rm core}$ $\propto \nu^{-1/k_{\rm r}}$, where $k_{\rm r}$ is related to the electron energy distribution, the magnetic field, and the particle number density \citep{1981ApJ...243..700K}. For a self-absorbed core which is in energy equipartition, one can show that $k_{\rm r} = 1$~\citep{1998A&A...330...79L}. Recent measurements indeed show a good agreement with $k_{\rm r} = 1$ \citep[e.g.][]{2009MNRAS.400...26O}. Between two frequencies ($\nu_2 > \nu_1$), the core shift can be expressed as
\begin{equation}
\label{eq:core_shift}
\triangle r_{\rm core} =(r_{\rm core}^{\nu_1}-r_{\rm core}^{\nu_2}) \propto (\nu_1^{-1/k_{\rm r}}-\nu_2^{-1/k_{\rm r}})
\mid_{k_{\rm r}=1} \propto \frac{\nu_2-\nu_1}{\nu_2\nu_1}.
\end{equation}
Since the absolute position information for VLBI observations is lost due to the phase self-calibration performed in the imaging, core shifts are normally measured by referencing the core to the optically thin jet components, whose positions are expected to be frequency-independent.
We used the optically thin component {\sl d} (which is visible at all frequencies between 15 and 86\,GHz, see Table~\ref{tab:nrao530_model_online}) to measure the VLBI core positions shift between 86\,GHz and the lower frequencies. We estimated $k_{\rm r}=0.78\pm0.29$, consistent within error with $k_{\rm r} = 1$ as expected if the region in question is in equipartition (Fig.~\ref{fig:Kr_plt}).

The number of extragalactic radio sources with reported core shift has been increasing~\citep[][and references therein]{2007A&A...468..963C,2008A&A...483..759K,2009MNRAS.400...26O}.
\citet{2008A&A...483..759K} studied a core shift sample of 29 AGN between 2.3 and 8.4\,GHz. They found the magnitude of measured core shifts between 2.3 and 8.4\,GHz to range from $-0.1$ to 1.4\,mas with a median value of 0.44\,mas\footnote{Positive values mean that the separation of the jet component measured at higher frequency is larger than the one measured at
lower frequency in one frequency pair.}. For higher frequency pairs, we would expect that the core shift decreases since opacity effects become less important (equation~\ref{eq:core_shift}). For a few sources where core shifts between high frequency pairs are available \citep{2009MNRAS.400...26O}, the reported core shifts between 22 and 43\,GHz (0.02--0.04\,mas) are significantly smaller than those measured here for components {\sl d}, {\sl e}, and {\sl f} (0.12--0.17\,mas). In addition, the displacement of the jet components' position are not all in the same direction. In other words, we observe non-radial shifts, which may indicate two dimensional gradients of the jet parameters (magnetic field, particle number density), an effect which is expected to occur if a bent jet is observed at a small viewing angle.

\subsection[Flux density and structure variability on daily timescales]{\label{sec:3.3}Flux density and structure variability on daily time-scales}
\begin{figure*}
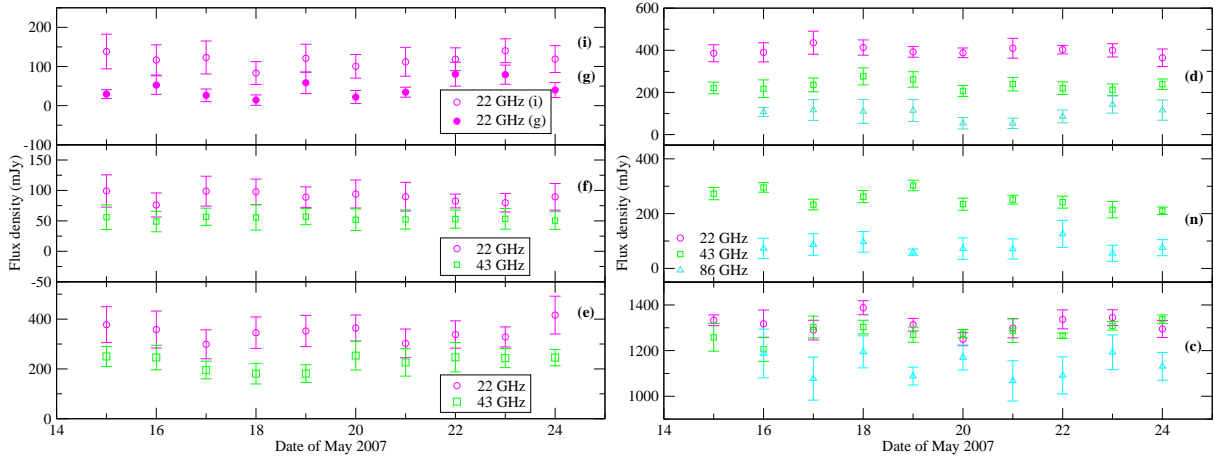

\centering
\includegraphics[width=0.45\textwidth,clip]{fg8a.eps}
\includegraphics[width=0.45\textwidth,clip]{fg8b.eps}
\caption{Variations of the flux density of model fit components on daily time-scales. Left: outer jet components ({\sl i}, {\sl g}, {\sl f} and {\sl e}), Right: inner components ({\sl d}, {\sl n} and {\sl c}). Note that components {\sl i} and {\sl g} are only visible at 22\,GHz.}
\label{fig:flux_var}
\end{figure*}

Our data also permit the study of flux density and structure variability in the VLBI jet of NRAO\,530 on a daily time-scale. We show in Fig.~\ref{fig:flux_var} the variations of the flux density of the model fit components with their modulation indices summarized in Table~\ref{tab:flux_var}. On daily time-scales, the flux density of most model fit components shows no significant variations. Component {\sl g} seems to be more variable (m = 53 per cent). However, it is the weakest component representing a diffuse region with flux density of only $\sim$ 30\,mJy. This may in turn reflect our modelling uncertainties. Through a $\chi^2$-test (see Table~\ref{tab:flux_var}) we found the probability for variability for most of the components to be low. For component {\sl c} and component {\sl n}, the $\chi^2$-test indicates the presence of interday variability with probability of 85 per cent at 22\,GHz and 92 per cent at 43\,GHz for component {\sl c} and 99.9 per cent at 43\,GHz for component {\sl n}. We therefore conclude that the inner jet components are more variable than those further out.
However, since component {\sl n} is located very close to component {\sl c}, the apparent variability may be caused by variations in how the flux of component {\sl c} and {\sl n} is divided up due for example to different \textit{uv} coverage. An anti-correlation between the fluxes of component {\sl c} and {\sl n} at 43\,GHz (correlation coefficient: $-0.75$) seems to support this scenario, while it is less clear at 86\,GHz (correlation coefficient: $-0.21$). Future VLBI studies are needed to confirm this effect.

\begin{table*}
\caption{\label{tab:flux_var}Interday flux variability characteristics of
model fit components of NRAO\,530. The columns list the component Id., the
modulation index, reduced $\chi_\nu^2$ and probability for the observable being variable.}
\begin{center}
\begin{tabular}{*{11}{r}}
\hline
&\multicolumn{3}{c}{22\,GHz}&\multicolumn{3}{c}{43\,GHz}&\multicolumn{3}{c}{86\,GHz}\cr
Id.&$m$&$\chi_\nu^2$&$p$&$m$&$\chi_\nu^2$&$p$&$m$&$\chi_\nu^2$&$p$\cr
    &(per cent)&&(per cent)&(per cent)&&(per cent)&(per cent)&&(per cent)\cr
\hline
c&3.1&1.5&84.6&2.3&1.7&92.3&4.4&0.6&19.4\cr
n&-&-&-&13.3&3.0&99.9&23.3&0.4&6.8\cr
d&3.5&0.2&0.5&8.7&0.4&7.7&34.9&0.9&49.3\cr
e&9.1&0.3&2.2&13.8&0.5&15.7 &-&-&-\cr
f&8.7&0.2&0.3&15.2&$<$0.1&$<0.1$&-&-&-\cr
g&53.2&1.1&67.6&-&-&-&-&-&-\cr
i&15.5&0.3&1.9&-&-&-&-&-&-\cr
\hline
\end{tabular}
\end{center}
\end{table*}

\begin{figure*}
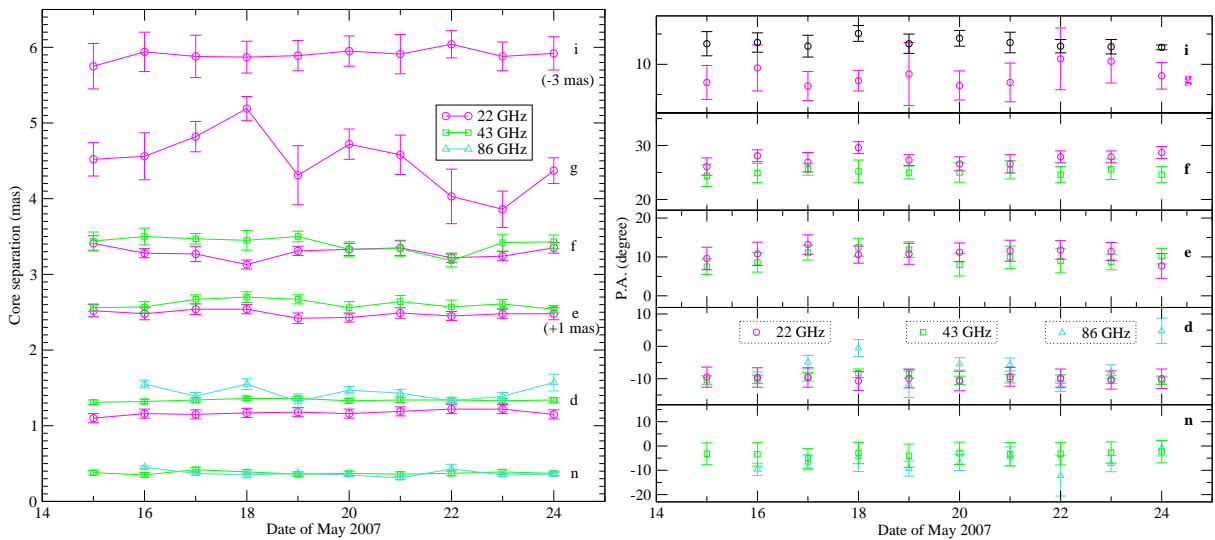

\centering
\includegraphics[width=0.45\textwidth,clip]{fg9a.eps}
\includegraphics[width=0.45\textwidth,clip]{fg9b.eps}%
\caption{Variations of the core separation (left) and PA (right) of model fit components on daily time-scales. Different symbols denote different frequencies (circles: 22\,GHz; squares: 43\,GHz; triangles: 86\,GHz). The core separation of component {\sl i} and {\sl e} has been shifted for display purposes.}
\label{fig:structure_var}
\end{figure*}

\begin{table*}
\caption[Interday variability characteristics of core separation for the jet components]{\label{tab:r_var}Interday variability characteristics of core separation for the jet components. The columns list the component Id., the modulation index, reduced $\chi_\nu^2$ and probability for the observable being variable.}
\begin{center}
\begin{tabular}{*{11}{r}}
\hline
&\multicolumn{3}{c}{22\,GHz}&\multicolumn{3}{c}{43\,GHz}&\multicolumn{3}{c}{86\,GHz}\cr
Id.&$m$&$\chi_\nu^2$&$p$&$m$&$\chi_\nu^2$&p&$m$&$\chi_\nu^2$&$p$\cr
    &(per cent)&&(per cent)&(per cent)&&(per cent)&(per cent)&&(per cent)\cr
\hline
n&-&-&-&5.3&0.4&9.2&11.3&3.5&$>99.9$\cr
d&3.0&0.4&4.3&1.2&0.3&1.8&6.0&2.5&98.8\cr
e&2.9&0.4&6.9&3.6&0.8&42.5 &-&-&-\cr
f&2.4&1.3&74.6&3.2&1.4&82.0 &-&-&-\cr
g&8.8&3.3&$>99.9$&-&-&-&-&-&-\cr
i&0.8&0.1&$<0.1$&-&-&-&-&-&-\cr
\hline
\end{tabular}
\end{center}
\end{table*}

\begin{table*}
\caption[PA variability characteristics of
model fit components for NRAO\,530]{\label{tab:pa_var}Interday PA variability characteristics of model fit components for NRAO\,530. The columns list the component Id., the modulation index, reduced $\chi_\nu^2$ and probability for the observable being variable.}
\begin{center}
\begin{tabular}{*{11}{r}}
\hline
&\multicolumn{3}{c}{22\,GHz}&\multicolumn{3}{c}{43\,GHz}&\multicolumn{3}{c}{86\,GHz}\cr
Id.&$m$&$\chi_\nu^2$&$p$&$m$&$\chi_\nu^2$&$p$&$m$&$\chi_\nu^2$&$p$\cr
    &(per cent)&&(per cent)&(per cent)&&(per cent)&(per cent)&&(per cent)\cr
\hline
n&-&-&-&24.5&$<0.1$&$<0.1$&44.9&0.8&38.6\cr
d&4.8&$<0.1$&$<0.1$&8.2&0.4&5.8&60.0&4.0&$>99.9$\cr
e&11.8&0.3&1.4&17.5&0.6&19.1&-&-&-\cr
f&3.7&0.7&30.2&1.7&$<0.1$&$<0.1$&-&-&-\cr
g&16.7&0.2&0.7&-&-&-&-&-&-\cr
i&5.0&0.5&9.2&-&-&-&-&-&-\cr
\hline
\end{tabular}
\end{center}
\end{table*}

Similarly, the variations of the core separation and PA of the jet components are displayed in Fig.~\ref{fig:structure_var}. The modulation indices $m$, $\chi_{\nu}^2$, and the probability for the studied source parameters to be variable are tabulated in Tables~\ref{tab:r_var} and~\ref{tab:pa_var}. For the core separation, the scatter in component {\sl g} is again much larger than what we expected from statistical errors. In addition, the large scatter for the core separation of component {\sl n} and {\sl d} at 86\,GHz also indicates the presence of additional errors that are not accounted for by the general accepted `rule of thumb', i.e. one-fifth beam size. This may be due to the sparse \textit{uv} coverage at 86\,GHz and the weakness of the components. For the PA, all the components show stationarity except component {\sl d} at 86\,GHz, which shows similar large scatter to the core separation. The limited \textit{uv} coverage and calibration accuracy at 86\,GHz, however, may cause artificial variability, making it necessary to confirm this effect in future VLBI studies.

\subsection[Jet kinematics at 15 GHz]{\label{sec:3.4}Jet kinematics at 15 GHz}
\begin{figure*}
\centering
\includegraphics[width=0.75\textwidth,clip]{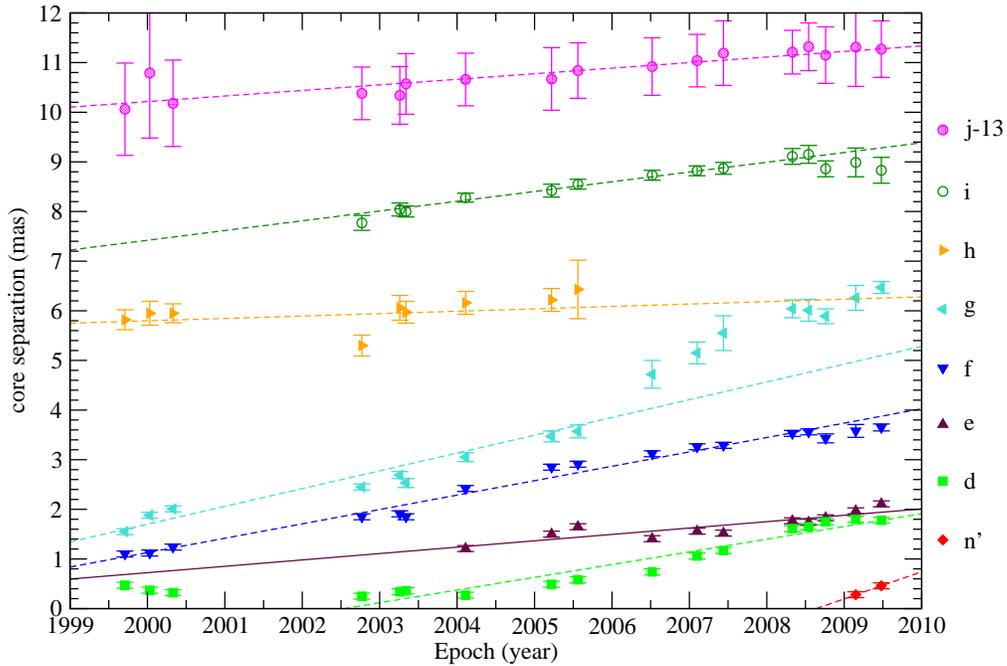}
\caption[Core separation vs. time for jet components in NRAO\,530]{Separation from the core as a function of time for the VLBI jet components seen in NRAO\,530. The core separations of component {\sl j} have been shifted 13\,mas downwards for display purposes, and the component {\sl x} is not shown. The dashed lines are linear regression fits to the data.}
\label{fig:r_t}
\end{figure*}

In this section, we investigate the kinematics of the jet components at 15\,GHz. The plot of the core separation of the jet components as a function of time, as shown in Fig.~\ref{fig:r_t}, served as a tool for the cross-identification across epochs.
We, however, note that the identification of the innermost jet component {\sl d} at epochs before 2001 may be subject to ambiguity due to the gaps during 2000 and 2002. The systematic change of flux density and PA (see later in this section), however, supports our identification scheme also for component {\sl d}. Linear regression fits to the data after 2002 yielded for the epoch of ejection a date, coinciding with the time of a radio flare \citep[$\sim$ 2002.6,][]{2008ChJAA...8..179H,2009arXiv0912.3176A}. This may indicate that component {\sl d} was newly ejected during that flare. Component {\sl e} appears to separate from component {\sl f} after 2003. This scenario is supported by a sudden change in the flux density of component {\sl f} (whose flux density would be maintained after the separation by adding the flux density of component {\sl e}), the similar PAs of {\sl f} and {\sl e} in 2004, and the lack of components between {\sl d} and {\sl f} in earlier epochs before 2004. The most reliable identification corresponds to outer components {\sl f}, {\sl g}, {\sl h}, {\sl i}, {\sl j}, and the newly ejected component {\sl n$^\prime$}.

In order to derive estimates for the separation speeds of the individual components, we applied linear regression fits [$f(t)=a\cdot(t-t_{0}$), where $t_0$ is the component ejection time] to the separation of the components from the core versus the epoch of observations. The fitting results are shown in Fig.~\ref{fig:r_t} as dashed lines. The angular separation rates, the time of ejection, and the apparent speeds are summarized in Table~\ref{tab:linear_fit}.

\begin{table}
\begin{minipage}[t]{\hsize}
\caption[Results of the linear fits for the component motion at 15\,GHz]{\label{tab:linear_fit}Results of the linear fits for the component motion at 15\,GHz. Listed are the component Id., number of epochs for the fit, proper motion $\mu$, apparent velocity $\beta_{\rm app}$ and ejection time of each component $t_0$.}
\centering
\begin{tabular}{rrrrrr}
\hline
Id. &\#  & $\mu$& $\beta_{\rm app}$ & $t_{0}$\\

&& (mas yr$^{-1}$] &($c$) & (yr) \\
\hline
n$^\prime$&2&0.55$\pm$0.26&26.5$\pm$12.5&2008.6$\pm$0.4\\
d&14&0.26$\pm$0.02&12.4$\pm$1.0&2002.5$\pm$0.3\\
e&11&0.13$\pm$0.02&6.2$\pm$1.1&1994.4$\pm$2.4\\
f&17&0.29$\pm$0.01&14.1$\pm$0.5&1996.0$\pm$0.4\\
g&11&0.44$\pm$0.03&21.3$\pm$1.5&1996.4$\pm$0.5\\
h&9&0.05$\pm$0.05&$2.3\pm$2.5&1876.6$\pm$133.6\footnote{From formal back-extrapolation}\\
i&14&0.19$\pm$0.01&9.2$\pm$0.5&1960.8$\pm$3.6\\
j&17&0.13$\pm$0.01&6.3$\pm$0.5&1823.1$\pm$14.7\\
\hline
\end{tabular}
\end{minipage}
\end{table}

Component {\sl n$^\prime$} seems to show the maximum apparent speed ($\beta_{\rm app}= 26.5\pm12.5$). However, the speed had to be calculated on the basis of only two epochs and is therefore subject to large uncertainties. We note that component {\sl n$^\prime$} apparently was ejected around 2008.6, which coincides with the second moderate flare seen in late 2008 at millimeter wavelengths\footnote{cf. http://sma1.sma.hawaii.edu/callist/callist.html?data=1733-130} and in early 2009 at centimeter wavelengths \citep{2009arXiv0912.3176A}. Component {\sl f} and {\sl g} can be identified with components B and E seen by \citet{2006A&A...456...97F}. The derived apparent speeds (14\,$c$ and 21\,$c$) are slightly faster than their findings (10\,$c$ and 14\,$c$), based on their shorter time coverage ($\sim$ 4 yr). The ejection time of both components lies in the range of the dramatic outburst phase between 1994 and 1998. However, due to the expansion effects, the uncertainty of core separation for component {\sl g} gets systematically larger at later epochs and, at the same time, it shows an acceleration indicating that a linear fit to its motion is only a rough approximation. Component {\sl e} and the three outer components ({\sl h}, {\sl i} and {\sl j}) reveal relatively slow motions with $\beta_{\rm app}$ in the range of 2--9\,$c$.

The measured apparent velocities can be used to estimate the parameters of the jet components according to the superluminal motion equation
\begin{equation}
\label{eq:superluminal}
\beta_{\rm app}=\frac{\beta\sin\theta}{1-\beta\cos\theta},\\
\end{equation}
where $\beta_{\rm app}$ and $\beta$ are the apparent and the true velocity in units of speed of light $c$ and $\theta$ is the angle between the direction of motion and the line of sight.
Making use of the apparent speed of component {\sl f} ($\beta_{\rm app}=14.1$\,$c$), which is the most reliably determined fastest component, we can set a lower limit on the Lorentz factor $\Gamma$, $\Gamma_{\rm min}=\sqrt{1+\beta_{\rm app}^2}=14.1$, and derive maximum allowable viewing angle $\theta_{\rm max}$ of 8\fdg1. The critical viewing angle, which maximizes the $\beta_{\rm app}$ for a given $\beta$, is $\theta_{\rm cri}=\arcsin(\frac{1}{\Gamma})$. For component {\sl f}, $\theta_{\rm cri}$ is 4\fdg1.
Correspondingly, we can also derive a Doppler factor $\delta_{\rm min}$ using the minimum Lorentz factor $\Gamma_{\rm min}$ and $\theta_{\rm cri}$. For component {\sl f}, $\delta_{\rm min}$ is 14.1. For even smaller angles ($\theta\rightarrow 0$), $\delta$ tends to the limit of $\sim$ 2$\Gamma=28.2$.

\begin{figure}
\centering
\includegraphics[width=0.35\textwidth,clip]{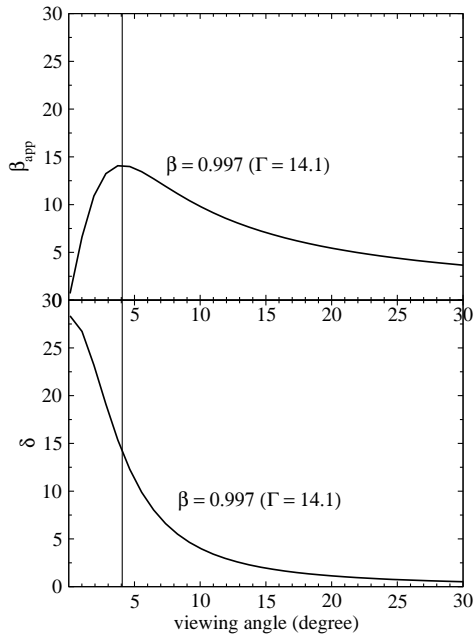}
\caption[Relativistic effects in NRAO\,530]{Apparent speed (upper panel) and Doppler factor (lower panel) as a function of viewing angle for component {\sl f} with a Lorentz factor of 14.1. The vertical line indicates the critical angle (4\fdg1) at which the apparent speed is maximized.}
\label{fig:beta}
\end{figure}

\begin{figure*}
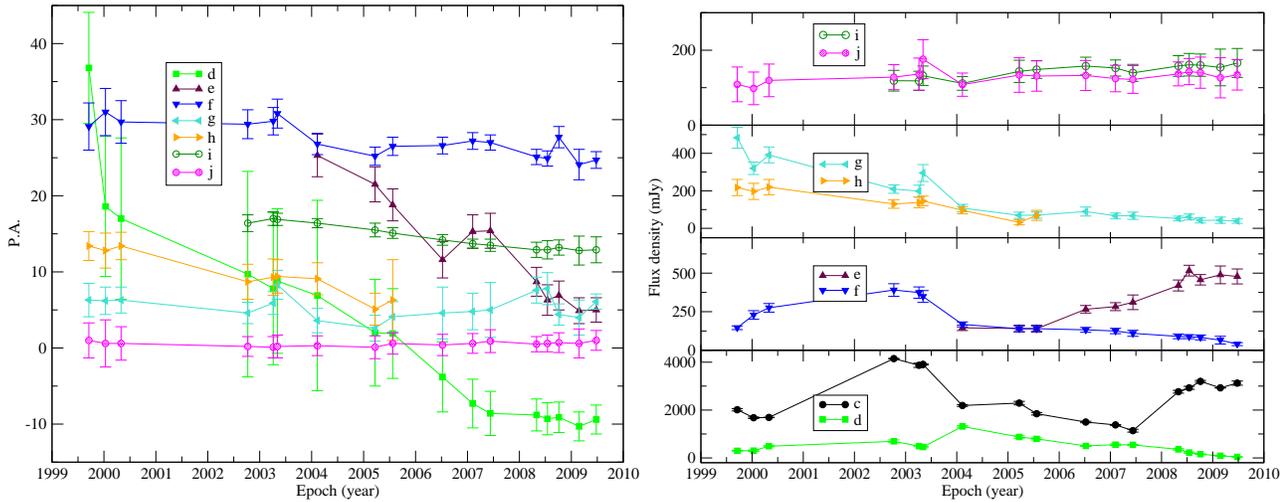

\centering
\includegraphics[width=0.475\textwidth,clip]{fg12a.eps}
\includegraphics[width=0.475\textwidth,clip]{fg12b.eps}
\caption{Time evolution of PA for components {\sl d}--{\sl j} (left) and 15\,GHz flux density of each model fit component (right).}
\label{fig:pa_flux_t}
\end{figure*}

Fig.~\ref{fig:beta} shows the relation between the apparent speed, the Doppler factor and the viewing angle. For a given Lorentz factor, the apparent speed first shows a shallow increase with decreasing viewing angle and then decreases sharply towards very small angles. Therefore, equation~(\ref{eq:superluminal}) has, for a given apparent speed, two solutions for the viewing angle. However, the Doppler factor, which has a strong influence on the observed flux density due to the beaming effects, shows a different orientation dependence and increases continuously with decreasing viewing angle. Based on these signatures and how the apparent speed and flux density of the jet component evolves, we can estimate the jet geometry.
For component {\sl g}, we found that the apparent jet speed increases and the flux density decreases at later epochs [see Figs~\ref{fig:r_t} and \ref{fig:pa_flux_t} (right)], indicating that the viewing angle takes the small viewing angle solution and is approaching the critical value.
These signatures are reminiscent of a three-dimensional (spatially) bent jet.

In Fig.~\ref{fig:pa_flux_t} (left), we present the evolution of the PA of jet components with time. One can clearly see a systematic variation of the PAs for components {\sl d}, {\sl e}, {\sl h} and {\sl i} is in contrast to the stationarity of the components {\sl f}, {\sl g} and {\sl j}. The newly ejected component {\sl n$^\prime$} also shows a similar position angle rotation (not shown). The changes of PA with time clearly indicate a non-radial, non-ballistic motion of the jet components.

In Fig.~\ref{fig:pa_flux_t} (right), the flux density evolution of the modelled components at 15\,GHz is displayed. Interestingly, the flux density evolution of the modelled components is characterized by several distinct behaviors. For components {\sl i} and {\sl j}, the flux density remains nearly constant, while for components {\sl g} and {\sl h}, the flux density continuously decreases with time. It seems that the drop in the light curve of component {\sl f} around 2004 was due to the separation of {\sl e} from it. Component {\sl e}, however, shows a clearly increasing trend. Component {\sl d} shows convolved behavior before the peak around 2004 and fades afterwards as it travels down the jet. The core component {\sl c} shows two intermediate flares, which coincide with the ejection of component {\sl n$^\prime$} and probably also with the ejection of component {\sl d}, if it was ejected around 2002 according to the straightforward back-extrapolation.
Component {\sl c} and {\sl f} show correlated variations before 2005 (significant at the 90 per cent confidence level), and the flux changes in components {\sl c} and {\sl e} are also correlated after $\sim 2004$, significant at the level of 95 per cent, for which the physical reason is unclear, but could be due, for example, to differential Doppler boosting.

\subsection[Morphology and its evolution]{\label{sec:3.5}Morphology and its evolution}
\begin{figure*}
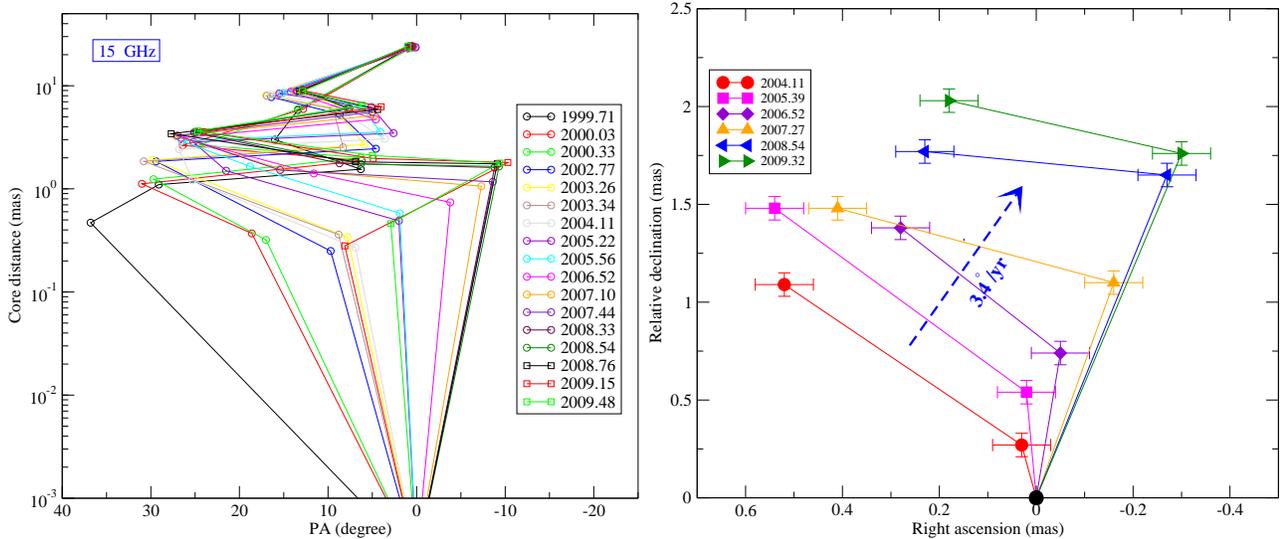

\centering
\includegraphics[width=0.475\textwidth,clip]{fg13a.eps}
\includegraphics[width=0.475\textwidth,clip]{fg13b.eps}
\caption{Evolution of the projected jet axis. Left: the jet ridge line between 1999 and 2009. Right: motion of inner jet component {\sl d} and {\sl e} in rectangular coordinates between 2004 and 2009. Close epochs are not shown in order to avoid cluttering the plot.}
\label{fig:ridge_line}
\end{figure*}

A comparison of the projected trajectories of jet components in 2007 and 1997 \citep{2006A&A...456...97F} indicated that the morphology of the jet changed significantly. It is obvious that the major differences are in the regions close to the core, in contrast to the relative `position' stationarity of the outermost component. This can be attributed to the relatively fast motion of the inner jet components. Fig.~\ref{fig:ridge_line} (left) shows the evolution of the jet ridge line over the last 10 years at 15\,GHz. This plot illustrates that the morphology of the jet changes from epoch to epoch due to the motion of the jet components. Shown in Fig.~\ref{fig:ridge_line} (right) is the evolution of the inner jet axis using component {\sl d} and {\sl e} as tracers.  A jet PA swing of $\sim$ 3\fdg4 per year can be estimated for both components between 2004 and 2009. For component {\sl d}, the average swing speed is also $\sim$ 3\fdg4 per year between 1999 and 2009. This indicates that the entire inner jet is swing together.

\section{DISCUSSION}
\label{sec:4}
In this section, we discuss the implications of the jet bending and swing which we observed for the pc-scale jet in NRAO\,530. Recently, an increasing number of blazars show clear evidence of jet PA swing, either periodic or erratic, e.g. 3C 120~\citep{1999ApJ...521L..29G}, BL Lac~\citep{2003MNRAS.341..405S}, OJ 287~\citep{2004ApJ...608..149T}, 3C 273~\citep{2006A&A...446...71S} and NRAO 150~\citep{2007A&A...476L..17A}. The change of the direction of the inner jet axis may be also responsible for the jet curvature observed at pc and kpc scales~\citep[e.g.][]{1999ApJ...521L..29G}. Since the jet axis is close to the observer's line of sight (as indicated by the observed apparent superluminal motion), it is certain that the apparently curved morphology results from amplification effects by projection. However, the physical reason for the intrinsic jet bending and swing is still unclear.

Changes in the PA of the jet can be attributed to precession effects. Jet precession could arise from scenarios relying on a single rotating black hole~\citep[e.g.][]{2004ApJ...616L..99C} or a binary black hole system~\citep[e.g.][]{1980Natur.287..307B,2005A&A...431..831L, 2008A&A...483..125R,2010A&A...511A..57B}. (Quasi)-periodicities in the radio and optical light curves and helical motions of jet components argue in favor of precession models~\citep[e.g.][]{2011A&A...526A..51K,2010A&A...522A...5L}. Furthermore, regular variations of the component ejection PA with a likely period of a few years were observed in a number of sources, e.g. 3C 273~\citep{1999A&A...344...61A}.  \citet{1999A&A...344...61A} proposed a simple ballistic precessing model of the jet `nozzle' to explain the kinematics in the quasar 3C 273. Similar models have been proposed and applied to other objects, e.g. 3C 279~\citep{1998ApJ...496..172A}, 3C 345~\citep{2004ApJ...602..625C}, and OJ 287~\citep{2004ApJ...608..149T}. In the case of NRAO 530, however, the non-ballistic motion of the components suggests that a ballistic treatment of the jet is probably not appropriate.

Another alternative explanation are plasma instabilities in the context of general relativistic magnetohydrodynamical (GRMHD) jet formation and acceleration \citep{1977MNRAS.179..433B,1982MNRAS.199..883B}, most importantly, Kelvin-Helmholtz (KH) instability~\citep[e.g.][]{1988ApJ...334...70H, 2006A&A...456..493P}. KH instabilities can be driven by velocity shear at the boundary layers between the jet and the ambient medium. Ideas of current-driven (CD) kink instability are also considered \citep[e.g.][]{2009ApJ...700..684M}. It is shown that the strength of CD kink instability strongly depends on the magnetic pitch profile. However, it is still not clear whether velocity shear or current driven instability are responsible for the observed bending structure and jet swing, or whether they act at different spatial scales. Furthermore, instabilities may only determine the morphology and kinematics for jet regions at large scales (beyond $\sim$100 pc), while the inner part of the flow is strongly shock-dominated ~\citep{1999ApJ...521..509L,2010arXiv1010.2856L}. For the jet features in NRAO 530, the non-ballistic motion is observed in the inner few pc, a region where the growth of instabilities may be weakened by the magnetic field, similar to the case of NRAO 150 \citep{2007A&A...476L..17A}.

\section{CONCLUSIONS}
\label{sec:5}
We studied the pc-scale jet properties of NRAO\,530 by using high-resolution multi-epoch VLBI observations at 15, 22, 43 and 86\,GHz. We identified component {\sl c} as the core based on its unambiguous spectral turnover, compactness and variability using the quasi-simultaneous VLBI data obtained in 2007. The frequency-dependent position shifts are found to be significant for three components (component {\sl d}, {\sl e} and {\sl f}) with mean radial shift of $0.15\pm0.02$\,mas between 22 and 43\,GHz. The position shift can be interpreted as being due to a core shift effect in a synchrotron self-absorbed jet. Interestingly, we find, for the first time, that jet components show a two-dimensional position shift. This indicates that the variations of the optical depth, which induces the core-shift, have also a non-radial dependence. Our 10-d monitoring observations allow us to investigate the jet variability characteristics on daily time-scales for NRAO\,530. We find the inner jet to more variable than the outer jet.

We studied the jet kinematics with individual components using the archival data at 15\,GHz. We observed superluminal motion for eight components with apparent speeds $\beta_{\rm app}$ in the range of 2--26$c$. From back-extrapolation, the ejection of component {\sl n$^\prime$} can be associated with the moderate flare in the radio light curve around 2009. However, the identification of the ejection time for another component {\sl d} is subject to uncertainties, and so is the link between its ejection and the moderate flare around 2002. From the measured apparent jet speed, we set limits to the jet parameters. For the reliably measured fastest component, we obtained a minimum Lorentz factor of 14 and Doppler factors in the range of 14--28. Different jet components show different kinematical behavior and flux density evolution, which can be understood as a consequence of a rotating jet. Using the variations of the apparent speed and flux density, we can estimate the jet geometry. The morphology of NRAO\,530 was found to have changed significantly over the past 10 years between 1997 and 2007. The variations are more pronounced near the core in contrast to the apparent `stationarity' of components further out. Moreover, we found significant variations of the jet PA and no common path existed for all the jet components. For components {\sl d} and {\sl e} , we estimated a mean swing of $\sim$3\fdg4 per year, which is partly responsible for the variations of the inner jet structure and indicates the swing of the entire inner jet. We discuss our results in the light of jet precession and models of jet instabilities.

To summarize, we conclude that the jet of NRAO\,530 is far more complex than previously known. The observed non-ballistic motion and a pronounced change of the jet orientation with a rate of 3\fdg4 per year suggest that NRAO\,530 is another example of a ``swinging'' jet, similar to e.g. BL Lac \citep{2003MNRAS.341..405S} and NRAO\,150 \citep{2007A&A...476L..17A}.

\section*{Acknowledgments}

We thank the referee, Dr D.~Gabuzda, for a careful reading of the manuscript and her insightful comments that improved this paper. The authors are grateful to Dr T.~Savolainen for providing software
to align images. This research has made use of data from the MOJAVE database that is maintained by the MOJAVE team (Lister et al., 2009, AJ, 137, 3718). The National Radio Astronomy Observatory is a facility of the National Science Foundation operated under cooperative agreement by Associated Universities, Inc. This work was supported in part by the National Natural Science Foundation of China (grants 10803017 and 10803015). The authors thank Dr I. Mart\'i-Vidal for his useful and helpful comments.

\clearpage
\section*{SUPPORTING INFORMATION}
Additional Supporting Information may be found in the online version of this article:

Fig.~\ref{fig:maps2007_online}. Clean maps of NRAO\,530 at 22, 43 and 86\,GHz obtained during our 10-d observations in 2007 May.

Fig.~\ref{fig:u_online}. Clean maps of NRAO\,530 at 15\,GHz from the archival data between 1999--2009.

Table~\ref{tab:para_n_online}. Description of VLBA images of NRAO 530 shown in Figs~\ref{fig:maps2007_online} and \ref{fig:u_online}.

Table~\ref{tab:nrao530_model_online}. Model-fitting results including the epoch of observation, the identification of components and the fitted parameters and their uncertainty.

\renewcommand{\thesubfigure}{\thefigure\,(\arabic{subfigure})}
\captionsetup[subfigure]{labelformat=simple,labelsep=colon,
listofformat=subsimple}
\makeatletter
\renewcommand{\p@subfigure}{}
\makeatother

\begin{figure*}
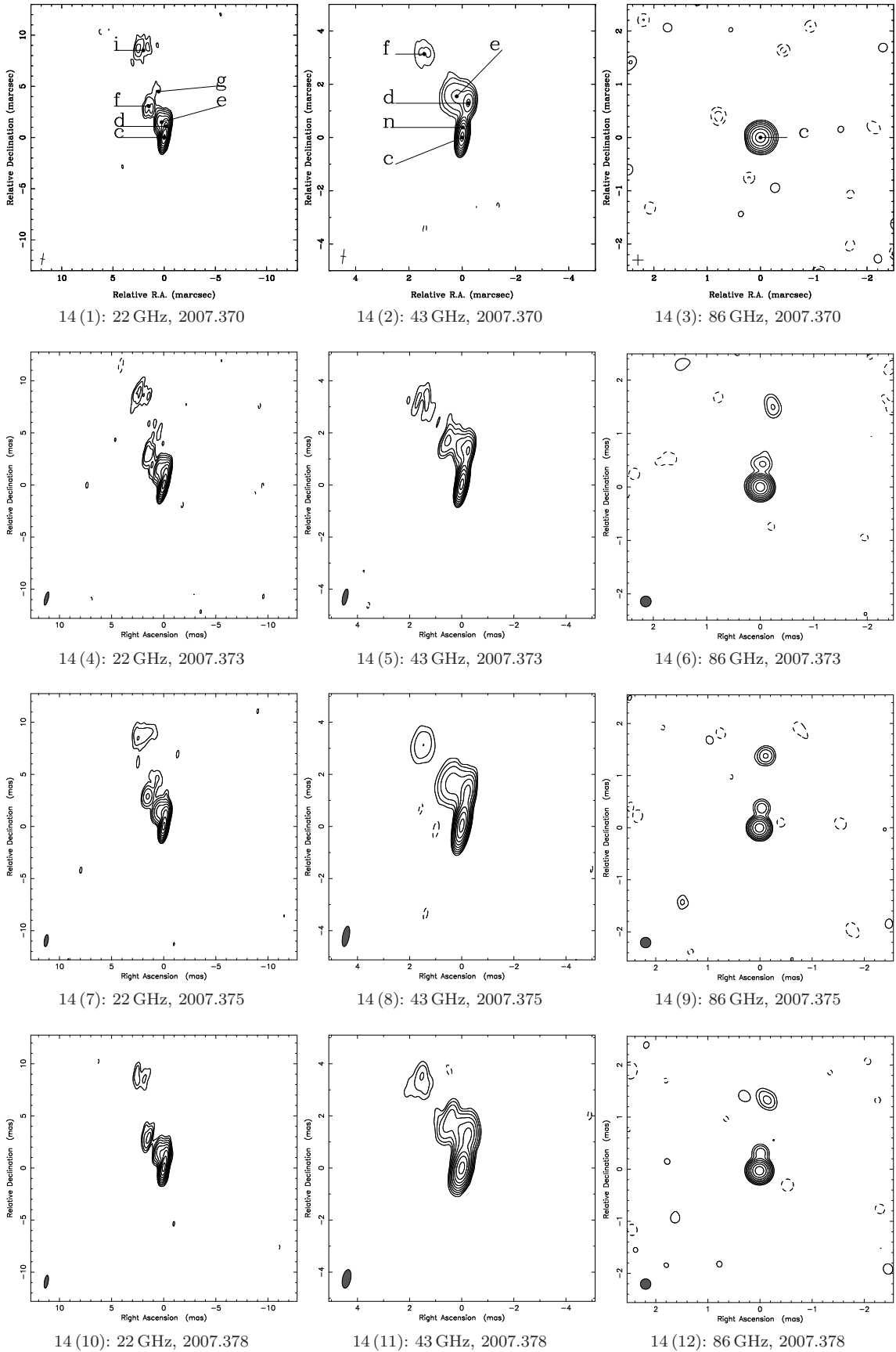
%
\centering
\subfloat[][22\,GHz, 2007.370]{%
\label{fig:maps2007-1}%
\includegraphics[width=0.28\textwidth,clip]{13mmA.ps}}
\hspace{3pt}%
\subfloat[][43\,GHz, 2007.370]{%
\label{fig:maps2007-2}%
\includegraphics[width=0.28\textwidth,clip]{7mmA.ps}}
\hspace{3pt}%
\subfloat[][86\,GHz, 2007.370]{%
\label{fig:maps2007-3}%
\includegraphics[width=0.28\textwidth,clip]{3mmA.ps}}\\
\subfloat[][22\,GHz, 2007.373]{%
\label{fig:maps2007-4}%
\includegraphics[width=0.28\textwidth,clip]{13mmB.ps}}
\hspace{3pt}%
\subfloat[][43\,GHz, 2007.373]{%
\label{fig:maps2007-5}%
\includegraphics[width=0.28\textwidth,clip]{7mmB.ps}}%
\hspace{3pt}%
\subfloat[][86\,GHz, 2007.373]{%
\label{fig:maps2007-6}%
\includegraphics[width=0.28\textwidth,clip]{3mmB.ps}}\\%
\subfloat[][22\,GHz, 2007.375]{%
\label{fig:maps2007-7}%
\includegraphics[width=0.28\textwidth,clip]{13mmC.ps}}
\hspace{3pt}%
\subfloat[][43\,GHz, 2007.375]{%
\label{fig:maps2007-8}%
\includegraphics[width=0.28\textwidth,clip]{7mmC.ps}}%
\hspace{3pt}%
\subfloat[][86\,GHz, 2007.375]{%
\label{fig:maps2007-9}%
\includegraphics[width=0.28\textwidth,clip]{3mmC.ps}}\\
\subfloat[][22\,GHz, 2007.378]{%
\label{fig:maps2007-10}%
\includegraphics[width=0.28\textwidth,clip]{13mmD.ps}}%
\hspace{3pt}%
\subfloat[][43\,GHz, 2007.378]{%
\label{fig:maps2007-11}%
\includegraphics[width=0.28\textwidth,clip]{7mmD.ps}}
\hspace{3pt}%
\subfloat[][86\,GHz, 2007.378]{%
\label{fig:maps2007-12}%
\includegraphics[width=0.28\textwidth,clip]{3mmD.ps}}
\caption{Clean maps of NRAO\,530 at 22, 43, and 86\,GHz in May 2007.}%
\label{fig:maps2007_online}%
\end{figure*}
\begin{figure*}%
\ContinuedFloat
\subfloat[][22\,GHz, 2007.381]{%
\label{fig:maps2007-13}%
\includegraphics[width=0.28\textwidth,clip]{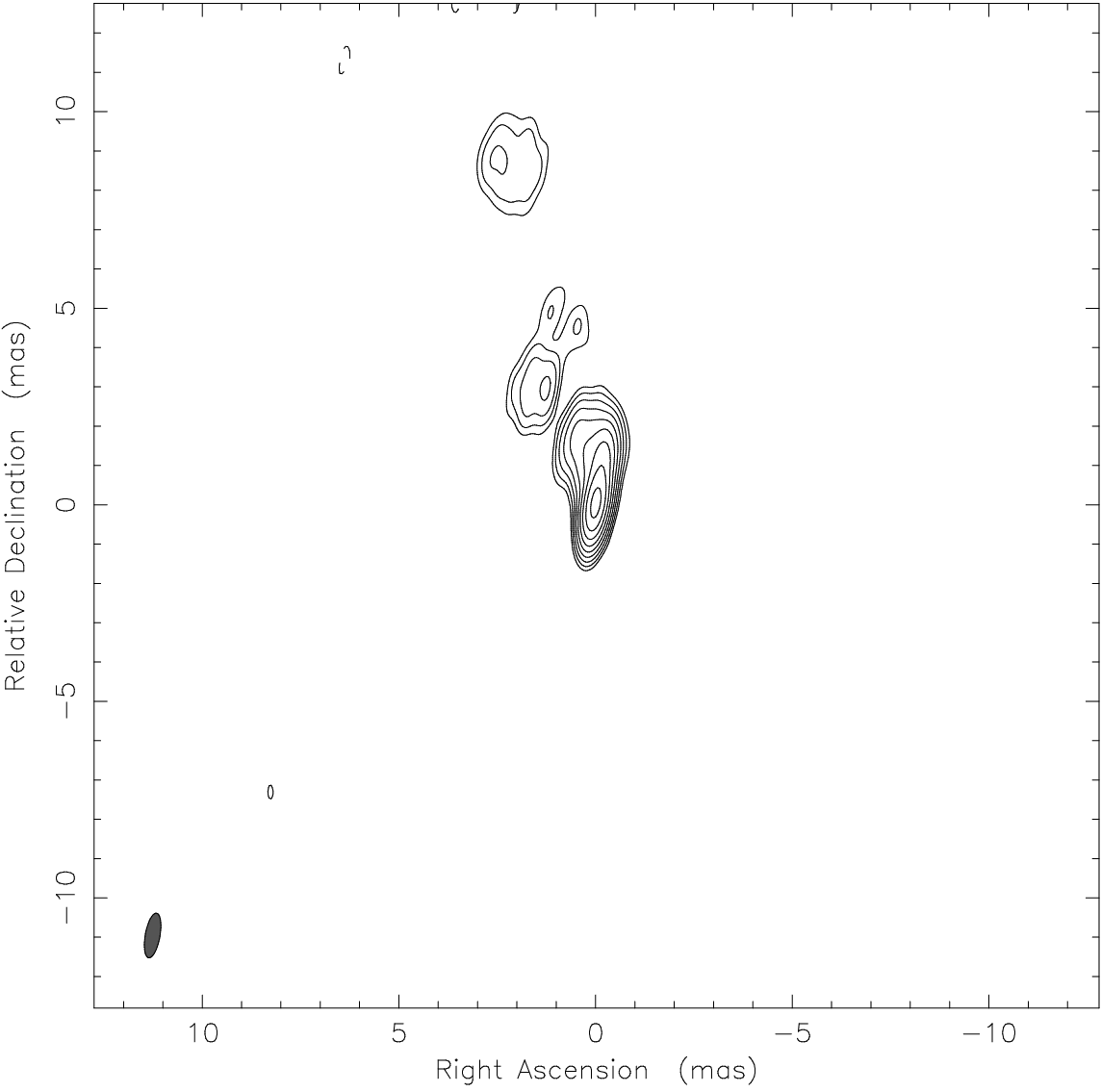}}%
\hspace{3pt}%
\subfloat[][43\,GHz, 2007.381]{%
\label{fig:maps2007-14}%
\includegraphics[width=0.28\textwidth,clip]{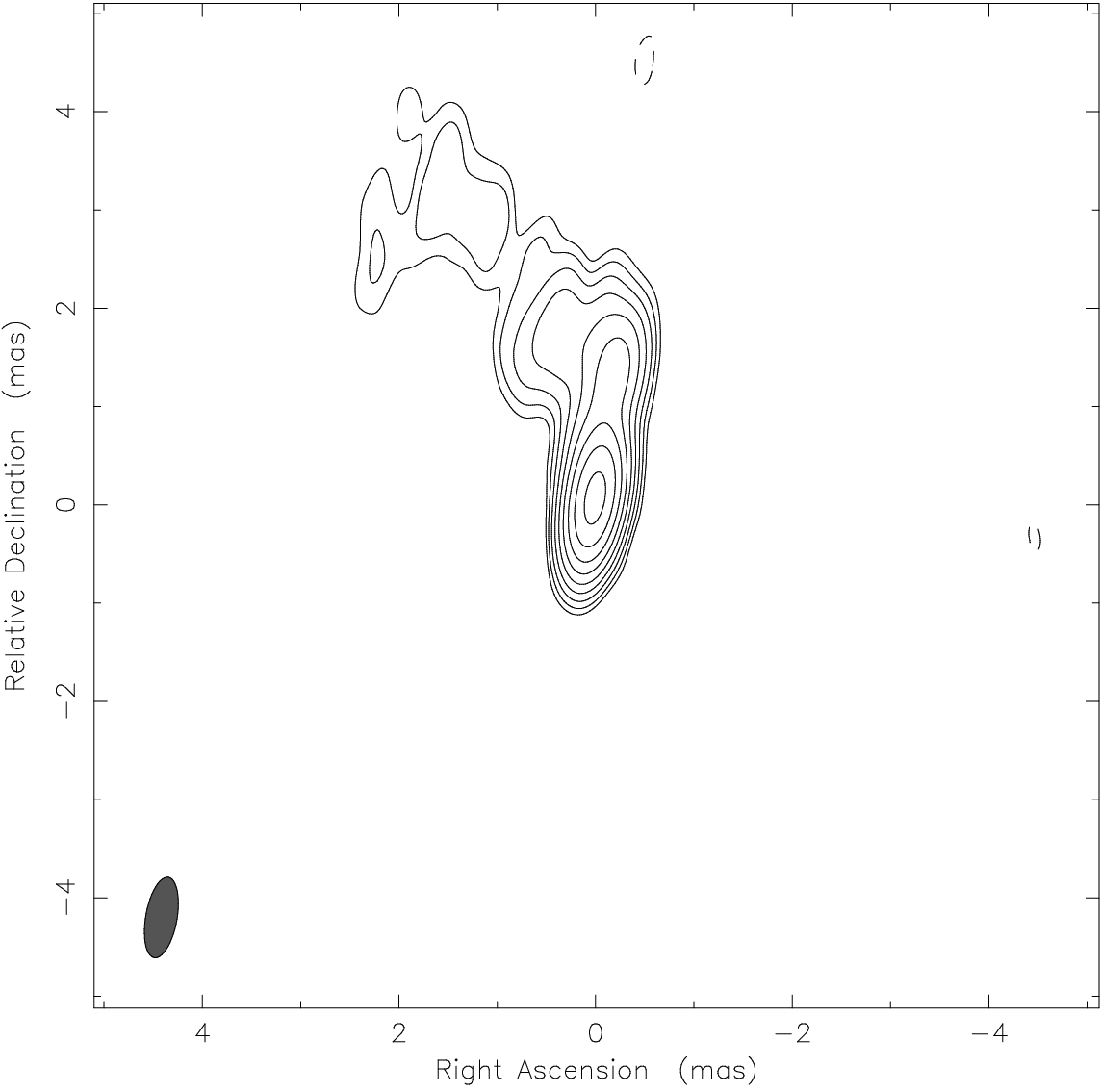}}
\hspace{3pt}%
\subfloat[][86\,GHz, 2007.381]{%
\label{fig:maps2007-15}%
\includegraphics[width=0.28\textwidth,clip]{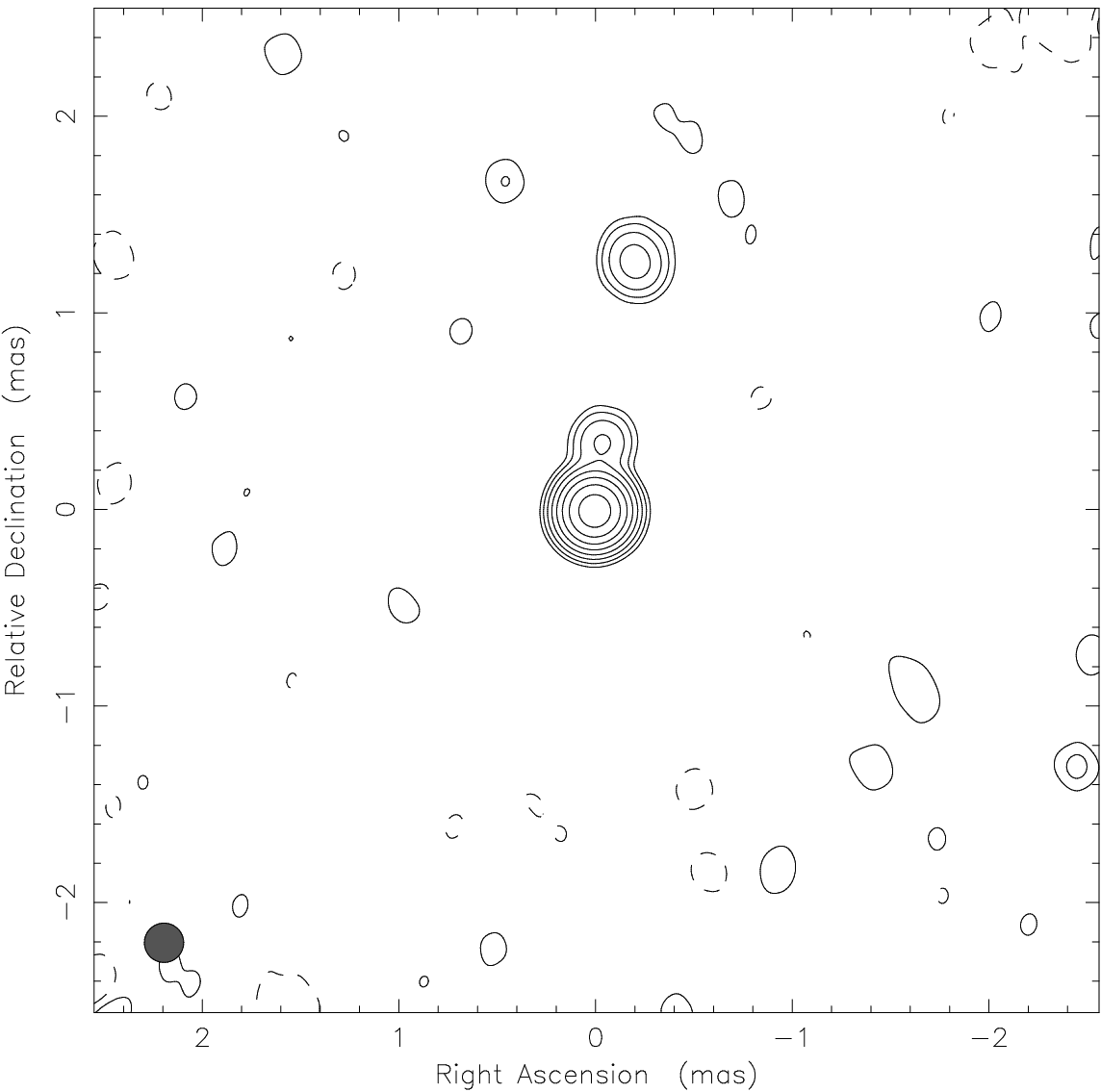}}\\
\subfloat[][22\,GHz, 2007.384]{%
\label{fig:maps2007-16}%
\includegraphics[width=0.28\textwidth,clip]{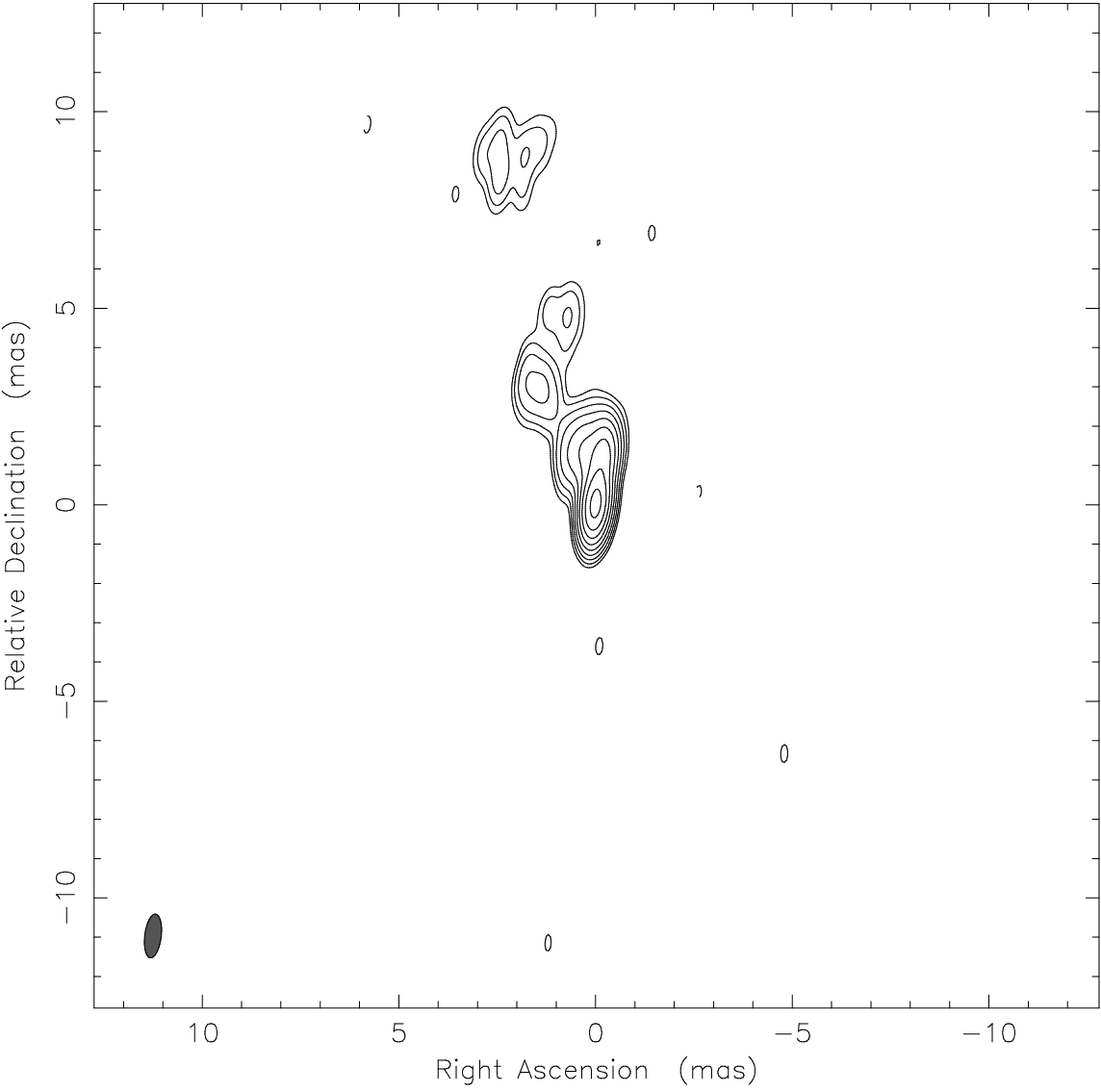}}%
\hspace{3pt}%
\subfloat[][43\,GHz, 2007.384]{%
\label{fig:maps2007-17}%
\includegraphics[width=0.28\textwidth,clip]{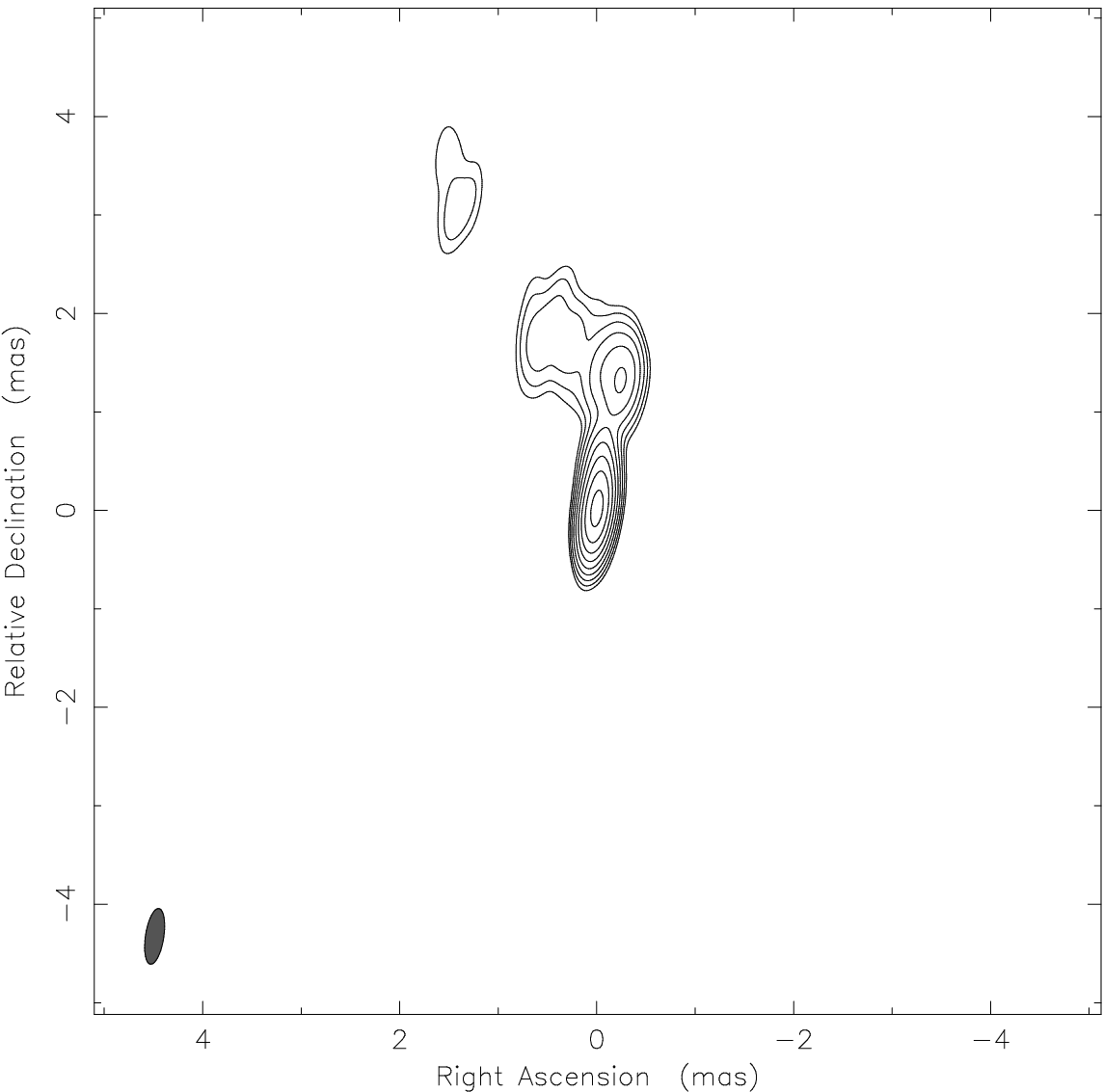}}
\hspace{3pt}%
\subfloat[][86\,GHz, 2007.384]{%
\label{fig:maps2007-18}%
\includegraphics[width=0.28\textwidth,clip]{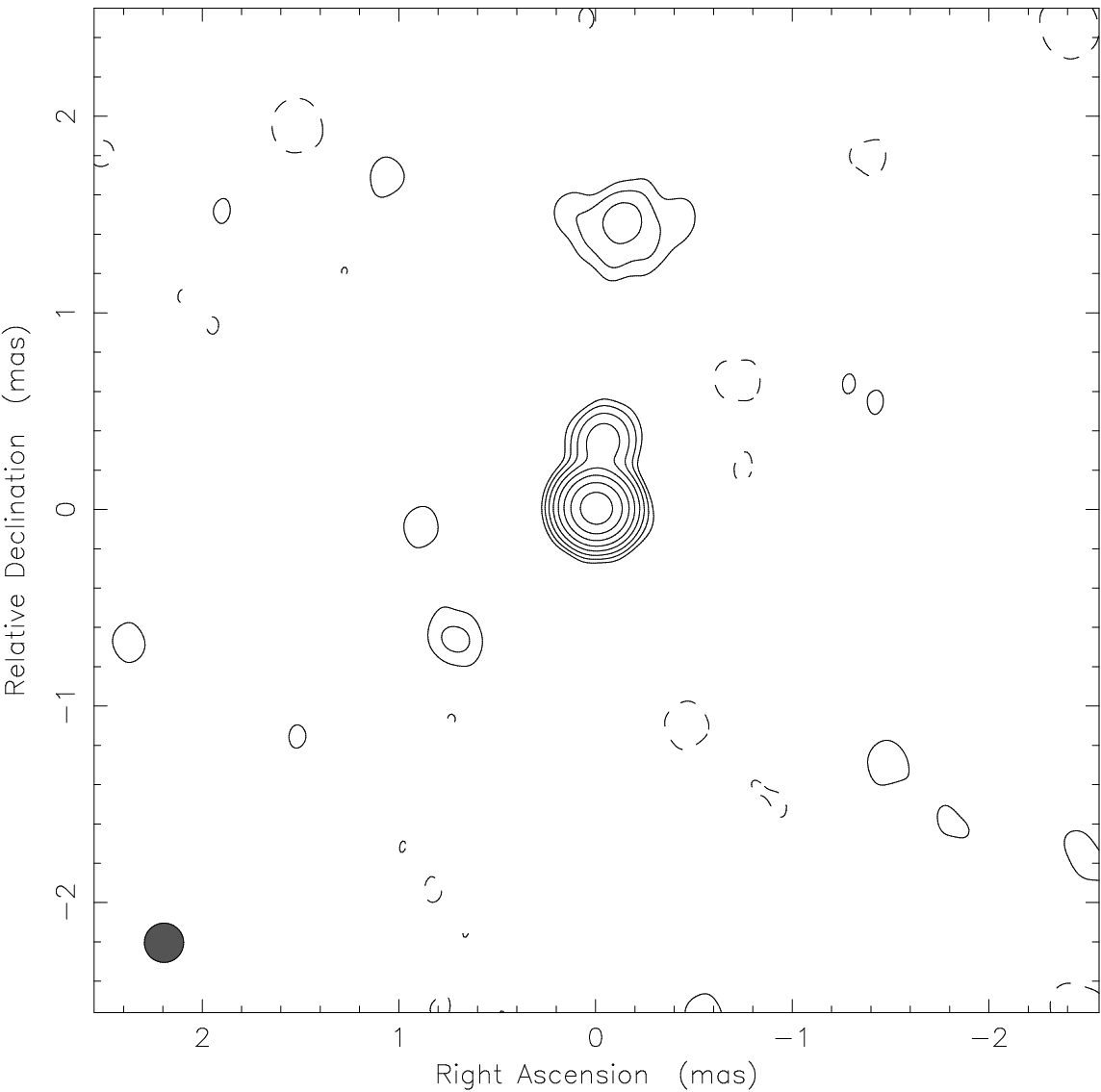}}\\
\subfloat[][22\,GHz, 2007.386]{%
\label{fig:maps2007-19}%
\includegraphics[width=0.28\textwidth,clip]{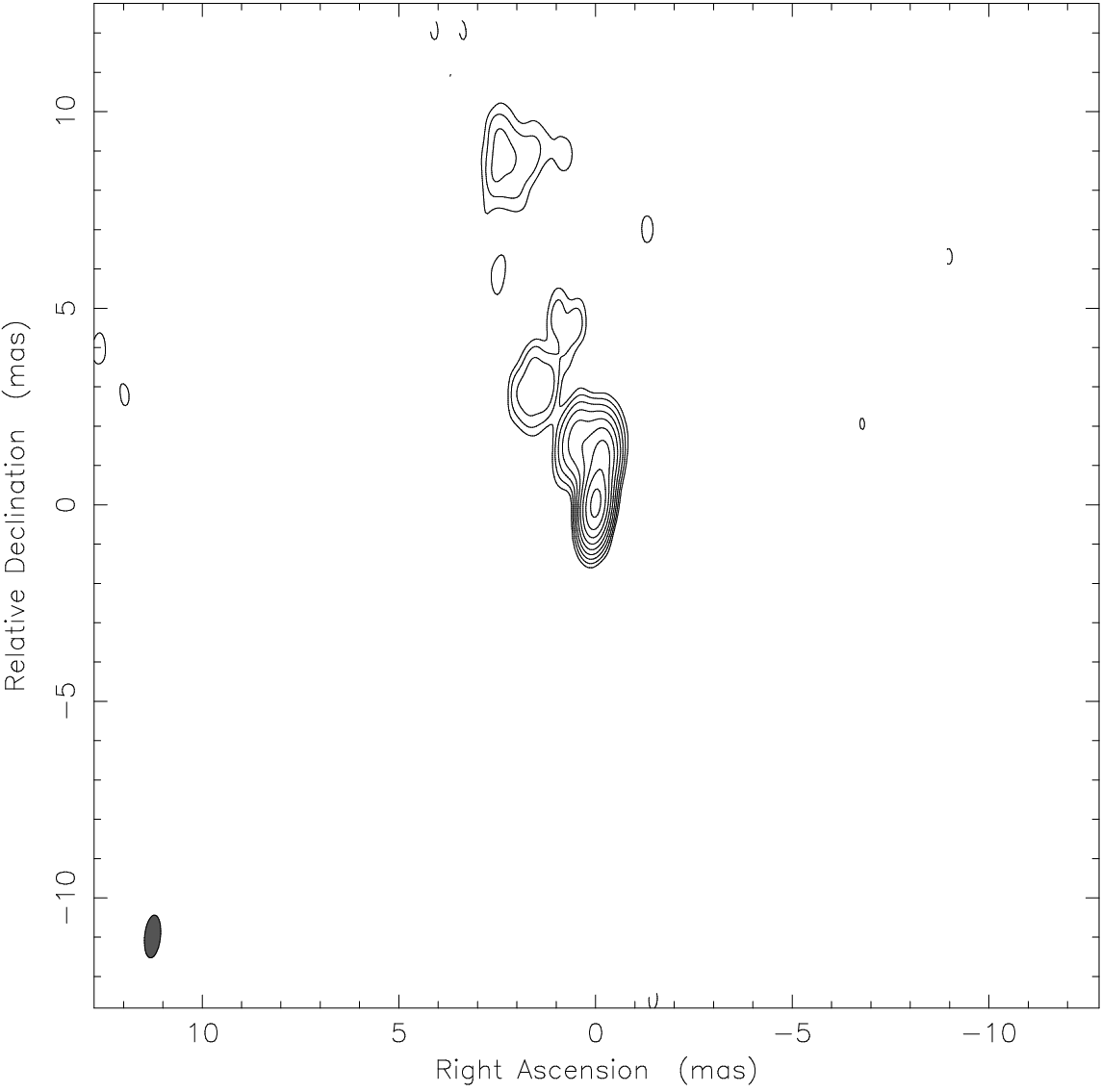}}%
\hspace{3pt}%
\subfloat[][43\,GHz, 2007.386]{%
\label{fig:maps2007-20}%
\includegraphics[width=0.28\textwidth,clip]{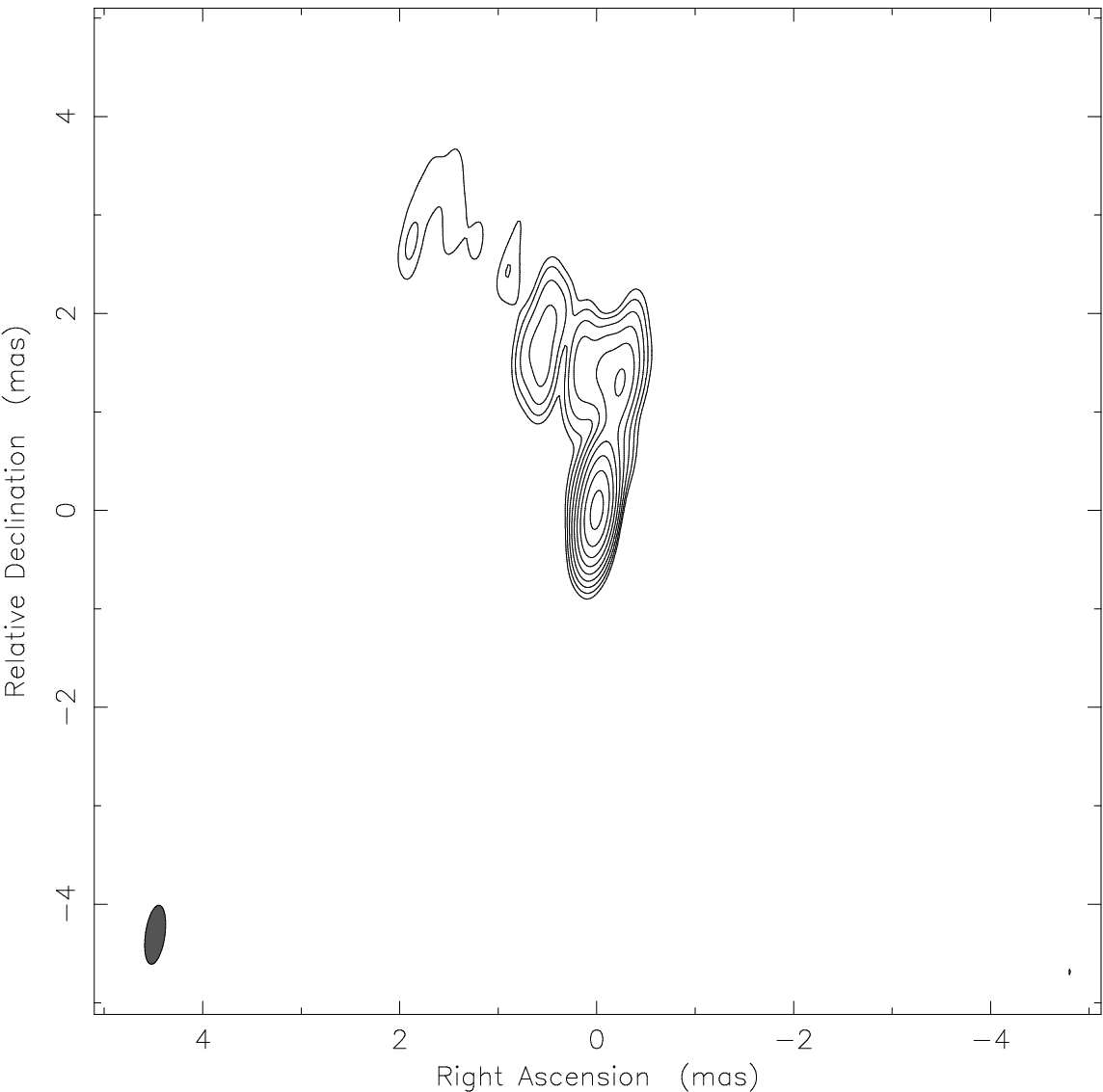}}
\hspace{3pt}%
\subfloat[][86\,GHz, 2007.386]{%
\label{fig:maps2007-21}%
\includegraphics[width=0.28\textwidth,clip]{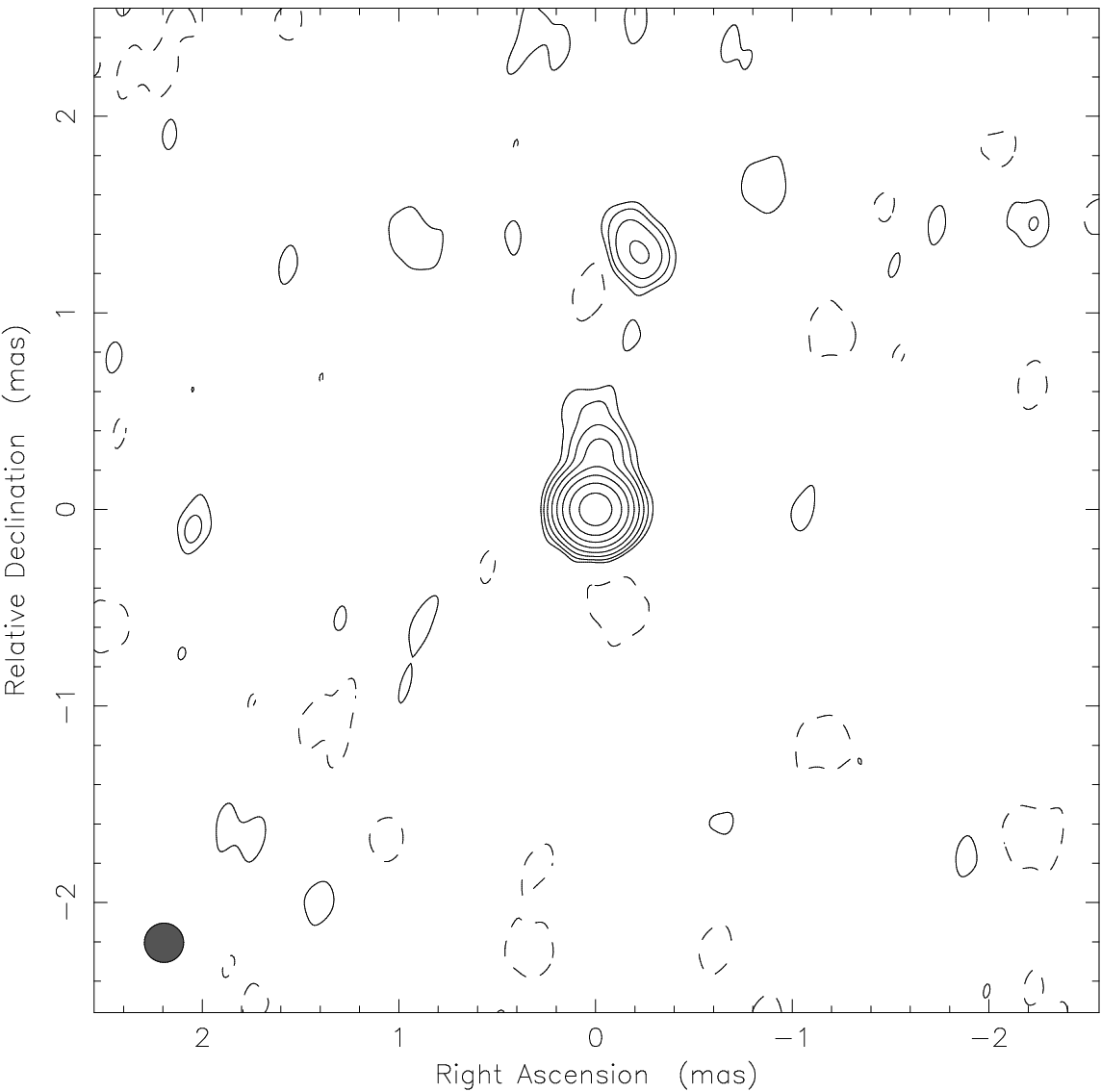}}\\
\subfloat[][22\,GHz, 2007.389]{%
\label{fig:maps2007-22}%
\includegraphics[width=0.28\textwidth,clip]{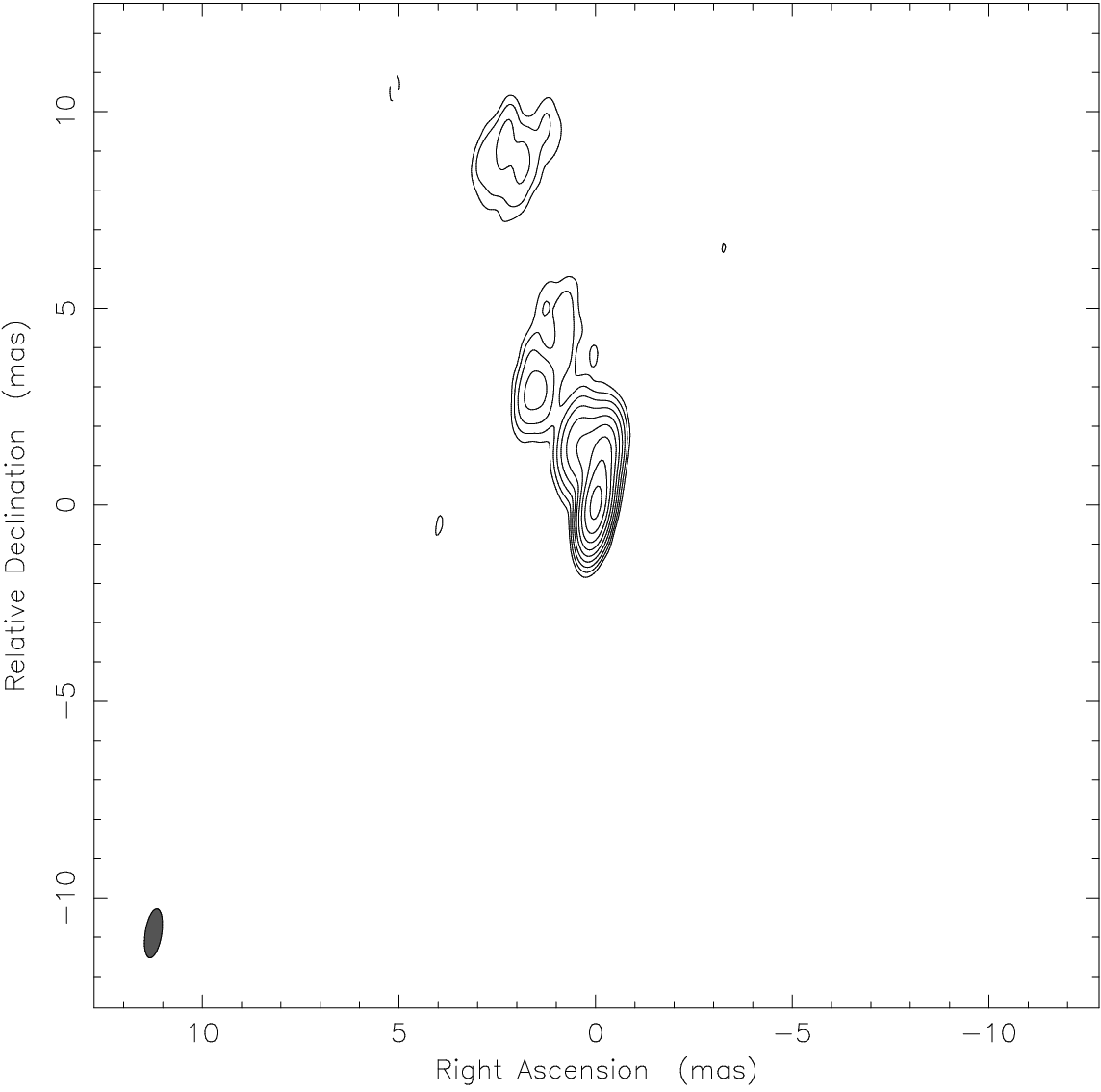}}%
\hspace{3pt}%
\subfloat[][43\,GHz, 2007.389]{%
\label{fig:maps2007-23}%
\includegraphics[width=0.28\textwidth,clip]{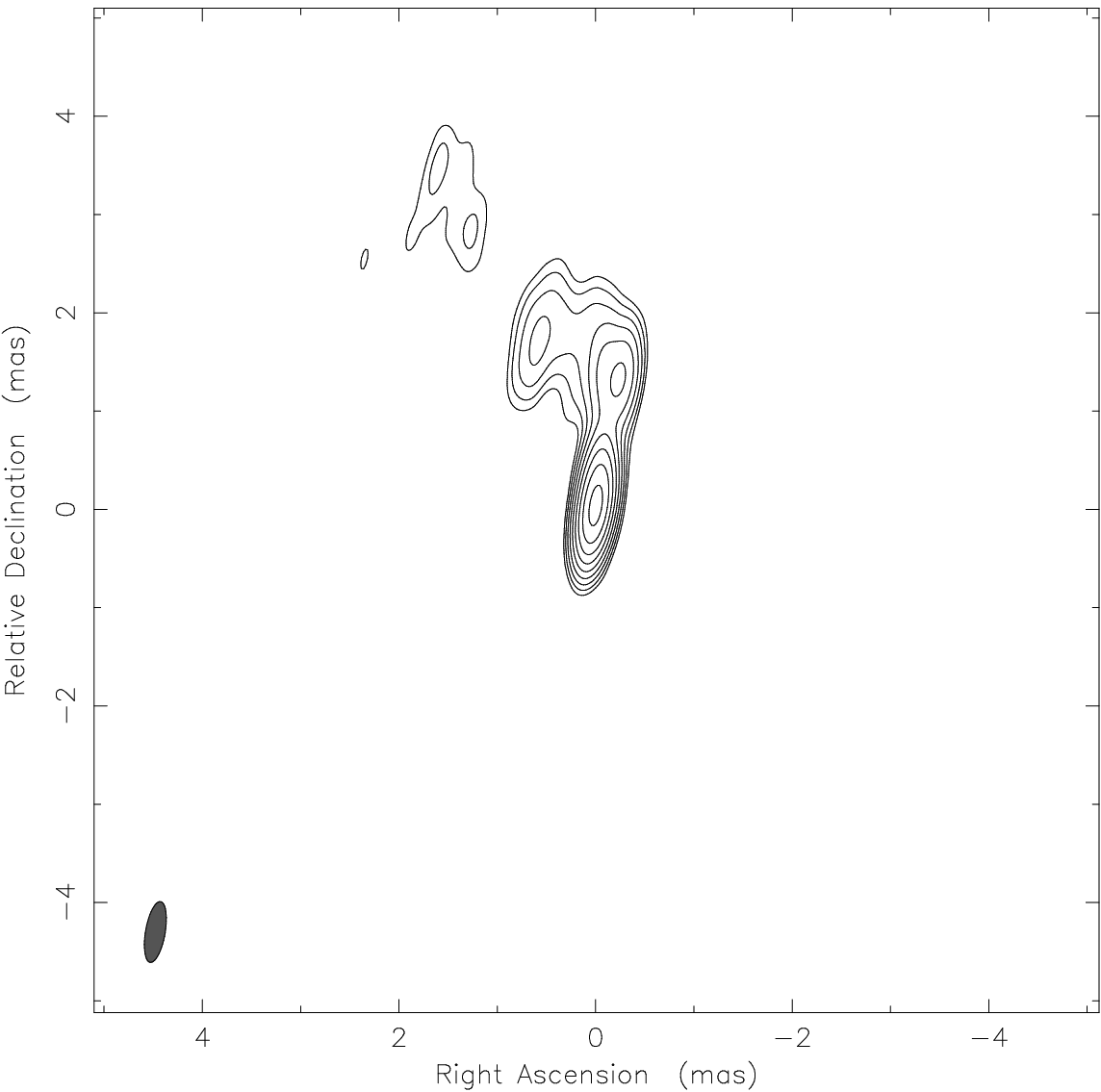}}
\hspace{3pt}%
\subfloat[][86\,GHz, 2007.389]{%
\label{fig:maps2007-24}%
\includegraphics[width=0.28\textwidth,clip]{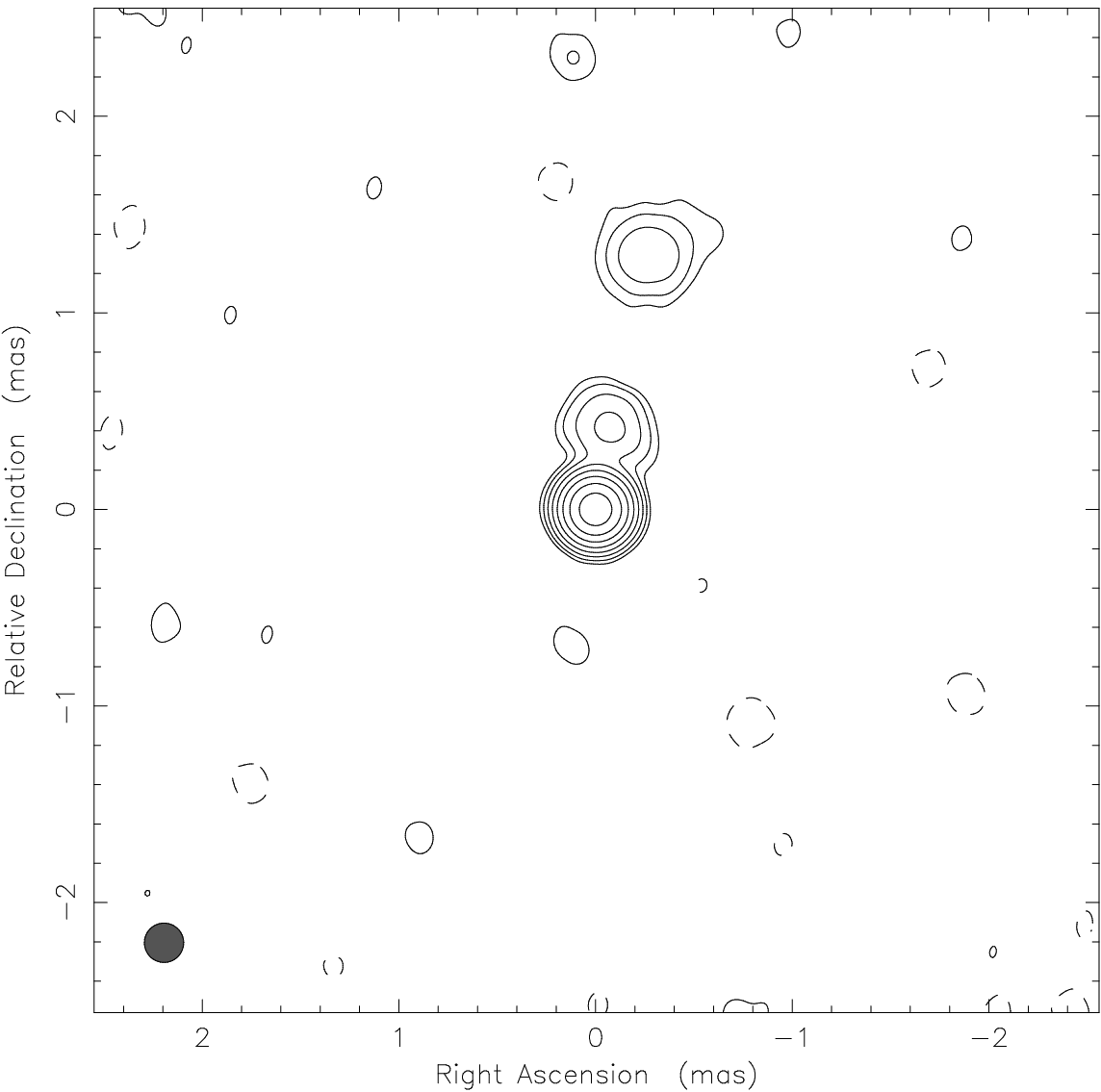}}
\caption[\it{-continued}]{\it{-continued}.}%
\end{figure*}
\begin{figure*}%
\ContinuedFloat
\subfloat[][22\,GHz, 2007.392]{%
\label{fig:maps2007-25}%
\includegraphics[width=0.28\textwidth,clip]{13mmI.ps}}%
\hspace{3pt}%
\subfloat[][43\,GHz, 2007.392]{%
\label{fig:maps2007-26}%
\includegraphics[width=0.28\textwidth,clip]{7mmI.ps}}
\hspace{3pt}%
\subfloat[][86\,GHz, 2007.392]{%
\label{fig:maps2007-27}%
\includegraphics[width=0.28\textwidth,clip]{3mmI.ps}}\\
\subfloat[][22\,GHz, 2007.395]{%
\label{fig:maps2007-28}%
\includegraphics[width=0.28\textwidth,clip]{13mmJ.ps}}%
\hspace{0pt}%
\subfloat[][43\,GHz, 2007.395]{%
\label{fig:maps2007-29}%
\includegraphics[width=0.28\textwidth,clip]{7mmJ.ps}}
\hspace{0pt}%
\subfloat[][86\,GHz, 2007.395]{%
\label{fig:maps2007-30}%
\includegraphics[width=0.28\textwidth,clip]{3mmJ.ps}}
\caption[\it{-continued}]{\it{-continued}.}%
\end{figure*}

\begin{figure*}%
\centering
\subfloat[][15\,GHz, 1999.712]{%
\label{fig:u-1}%
\includegraphics[width=0.30\textwidth,clip]{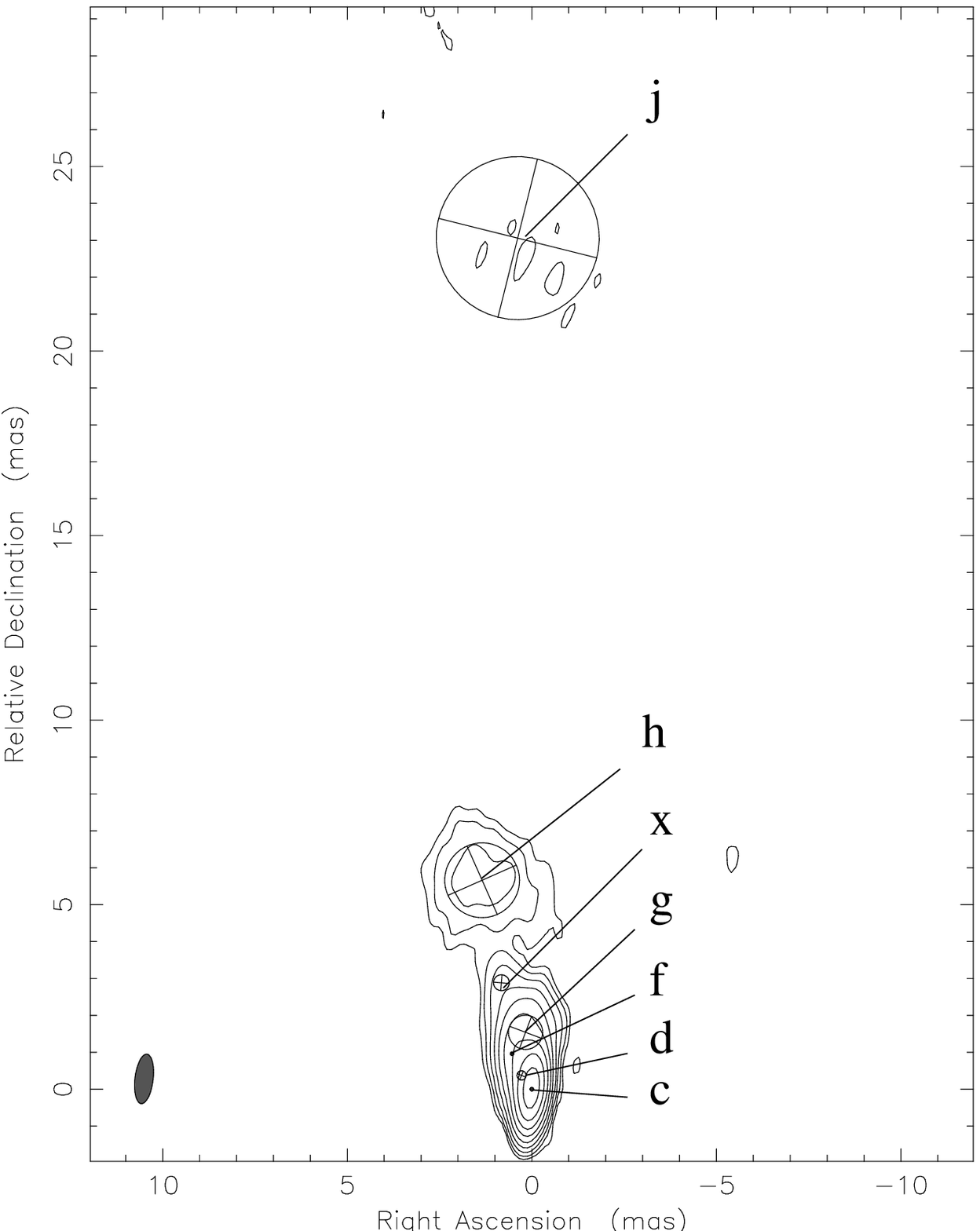}}
\hspace{3pt}%
\subfloat[][15\,GHz, 2000.030]{%
\label{fig:u-2}%
\includegraphics[width=0.30\textwidth,clip]{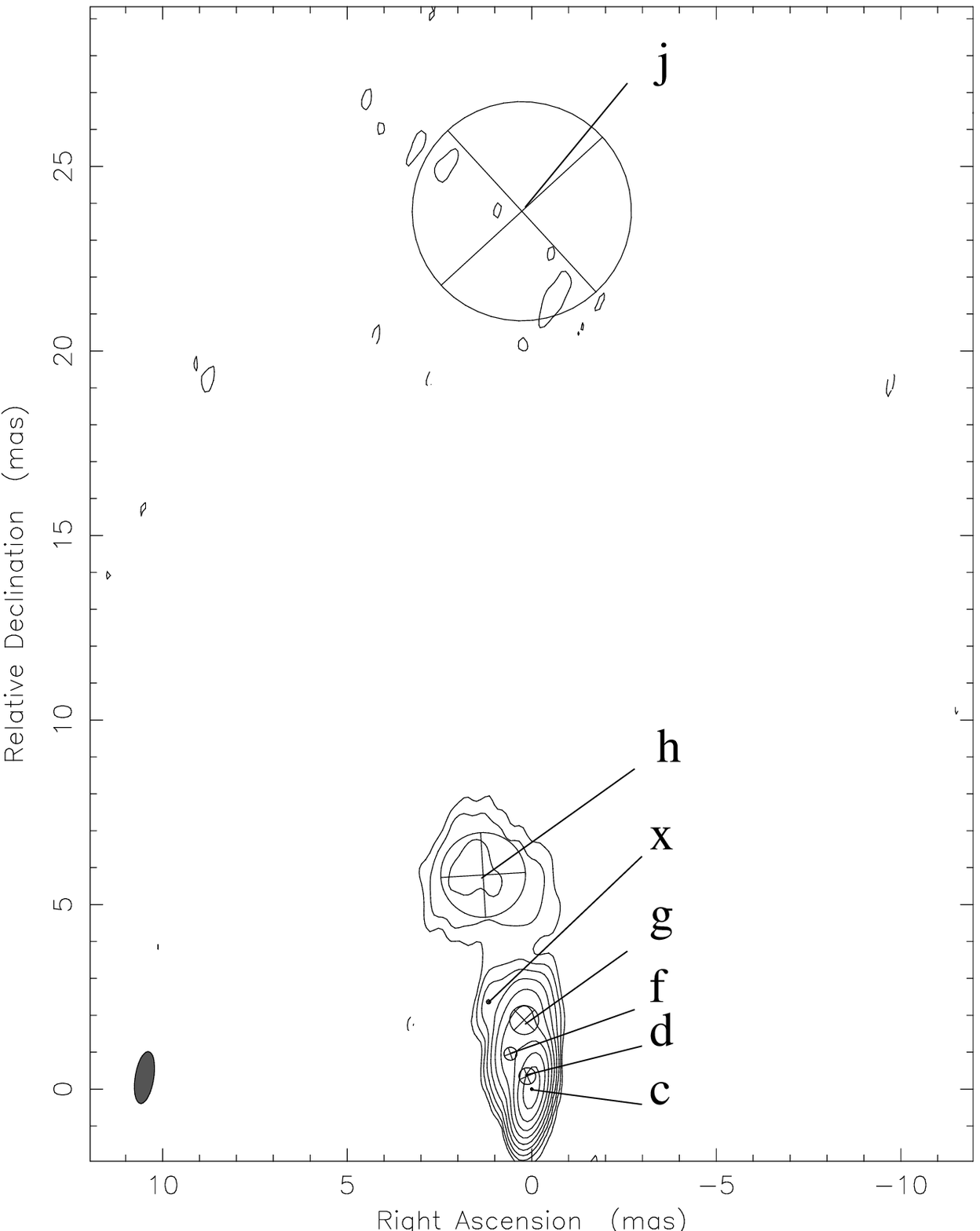}}
\hspace{3pt}%
\subfloat[][15\,GHz, 2000.333]{%
\label{fig:u-3}%
\includegraphics[width=0.30\textwidth,clip]{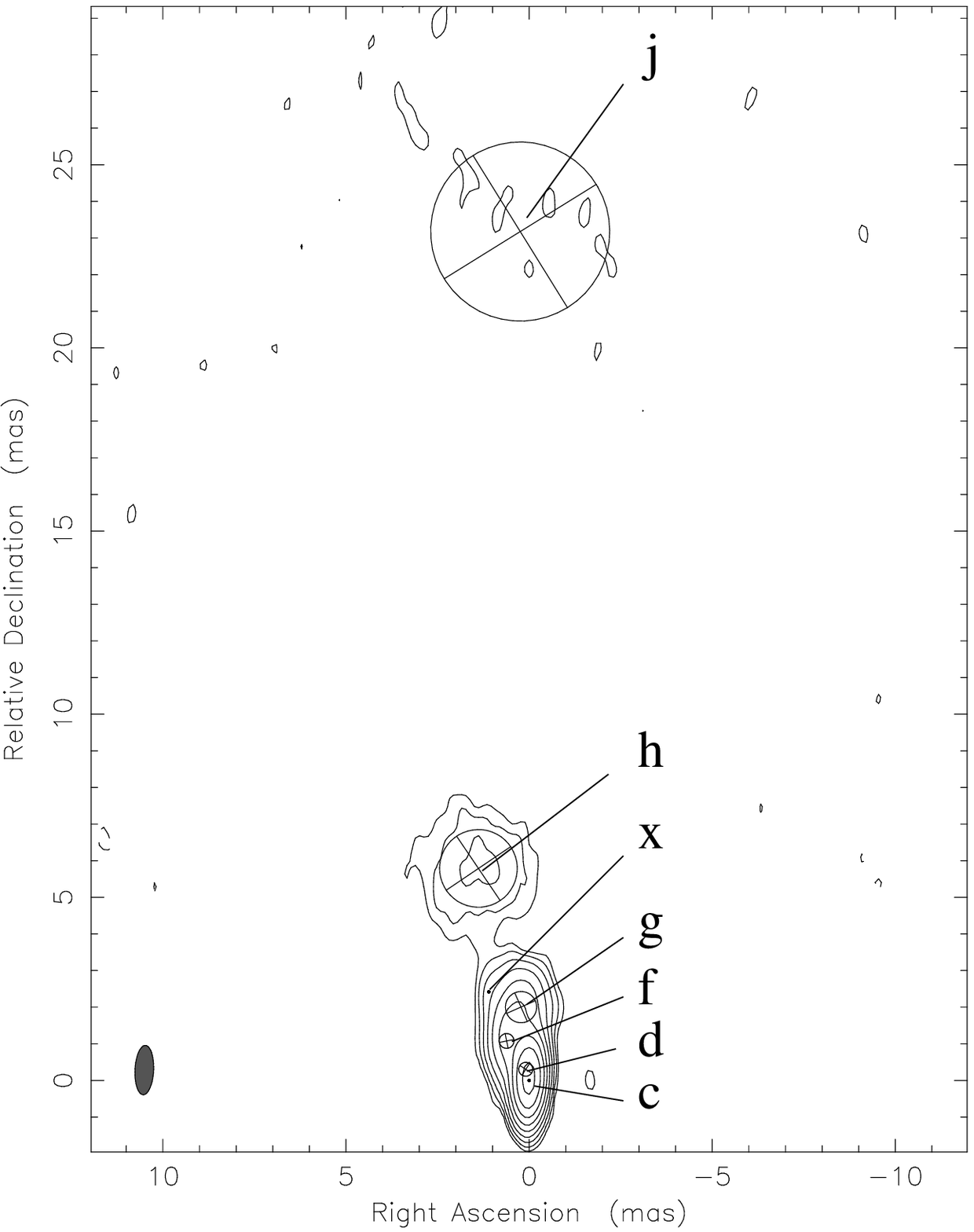}}\\
\hspace{0pt}%
\subfloat[][15\,GHz, 2002.773]{%
\label{fig:u-4}%
\includegraphics[width=0.30\textwidth,clip]{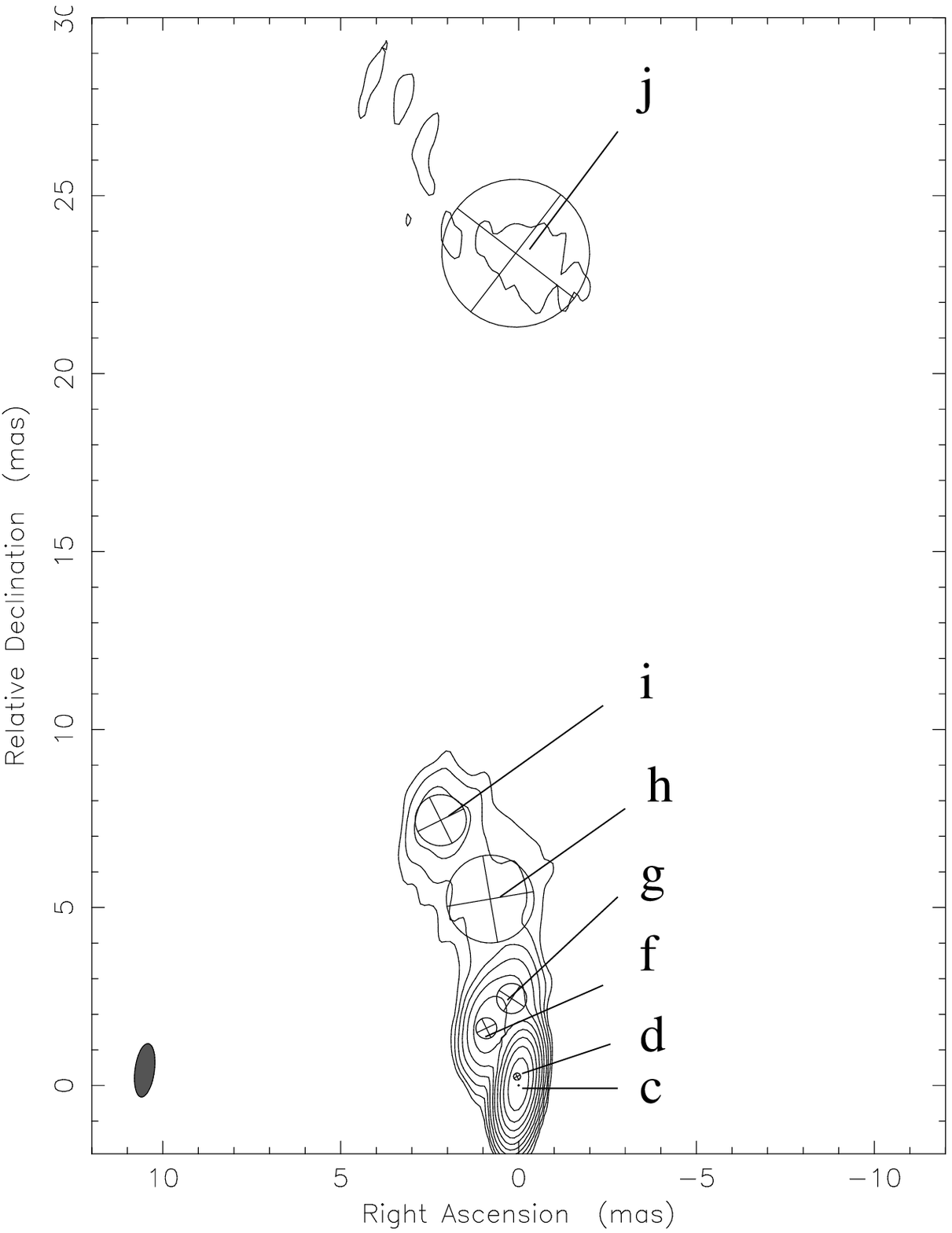}}
\hspace{3pt}%
\subfloat[][15\,GHz, 2003.258]{%
\label{fig:u-5}%
\includegraphics[width=0.30\textwidth,clip]{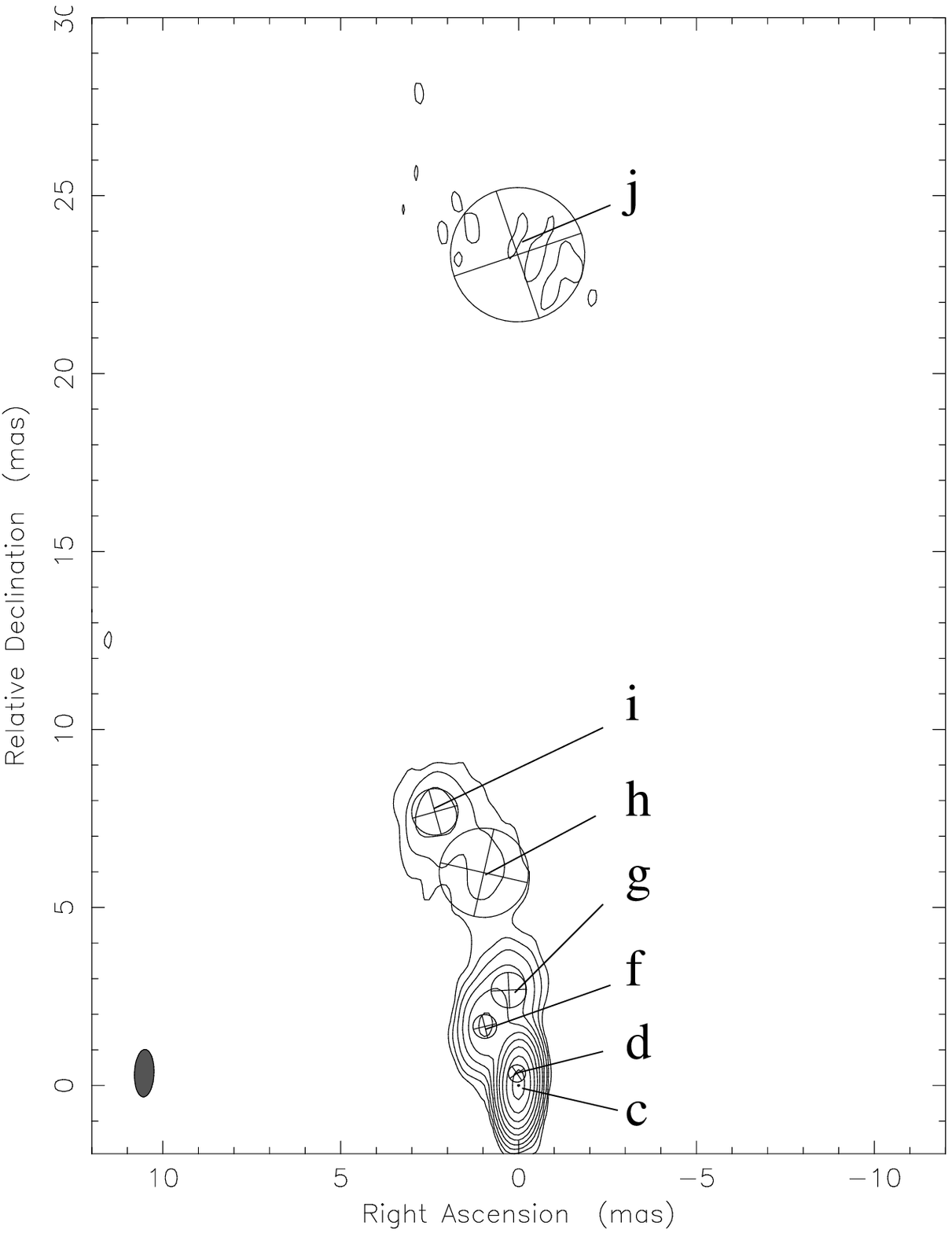}}
\hspace{3pt}%
\subfloat[][15\,GHz, 2003.337]{%
\label{fig:u-6}%
\includegraphics[width=0.30\textwidth,clip]{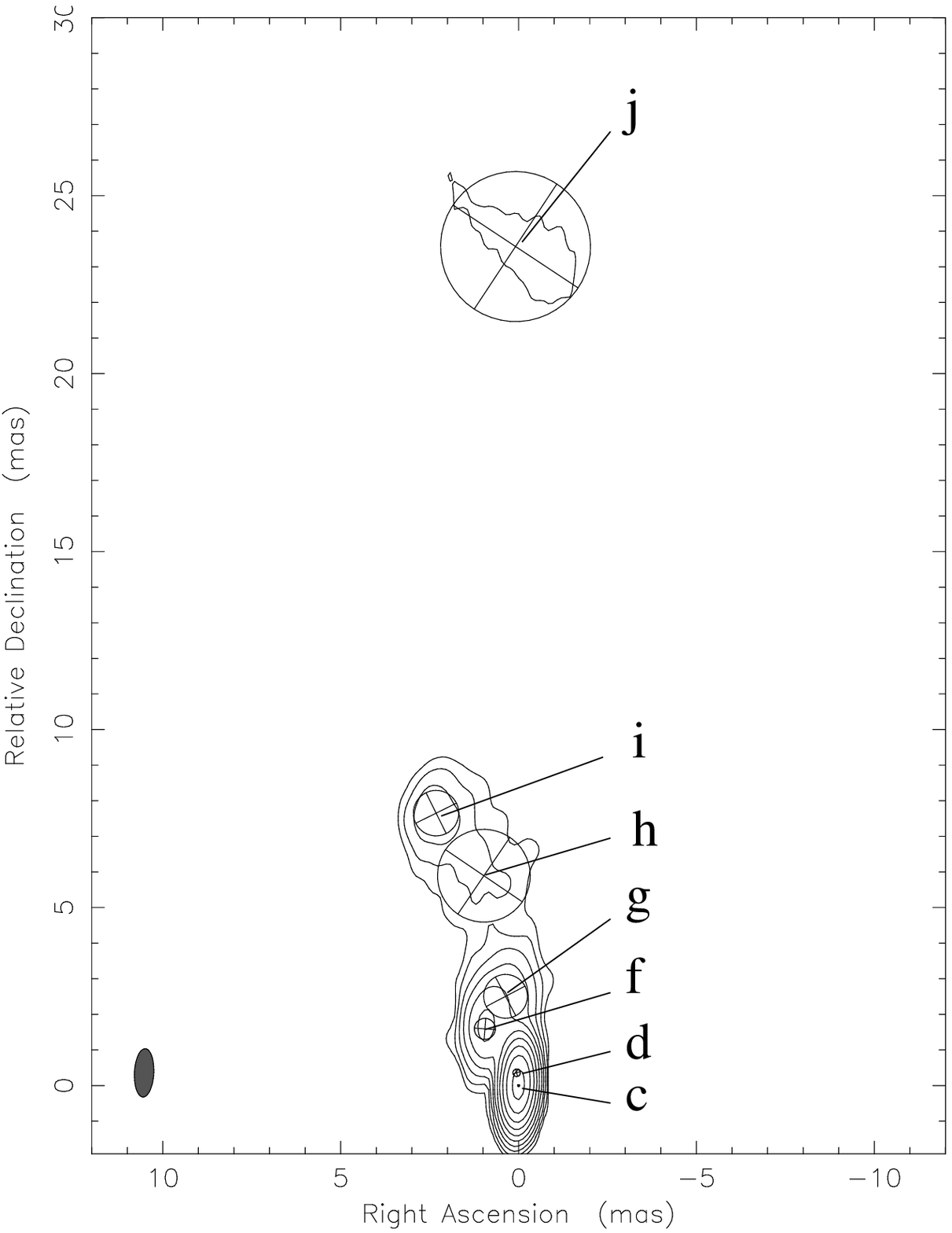}}\\
\subfloat[][15\,GHz, 2004.115]{%
\label{fig:u-7}%
\includegraphics[width=0.30\textwidth,clip]{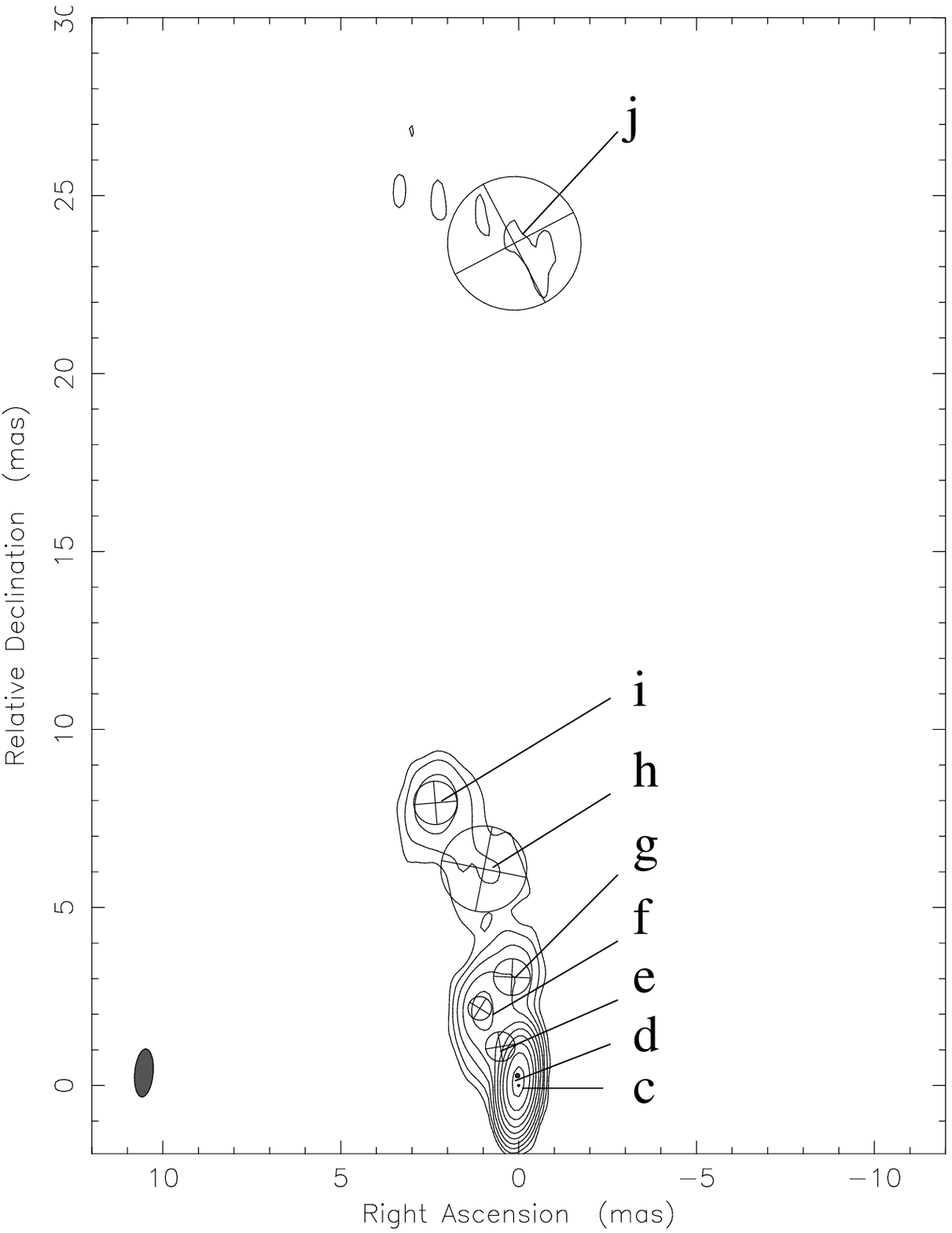}}
\hspace{3pt}%
\subfloat[][15\,GHz, 2005.225]{%
\label{fig:u-8}%
\includegraphics[width=0.30\textwidth,clip]{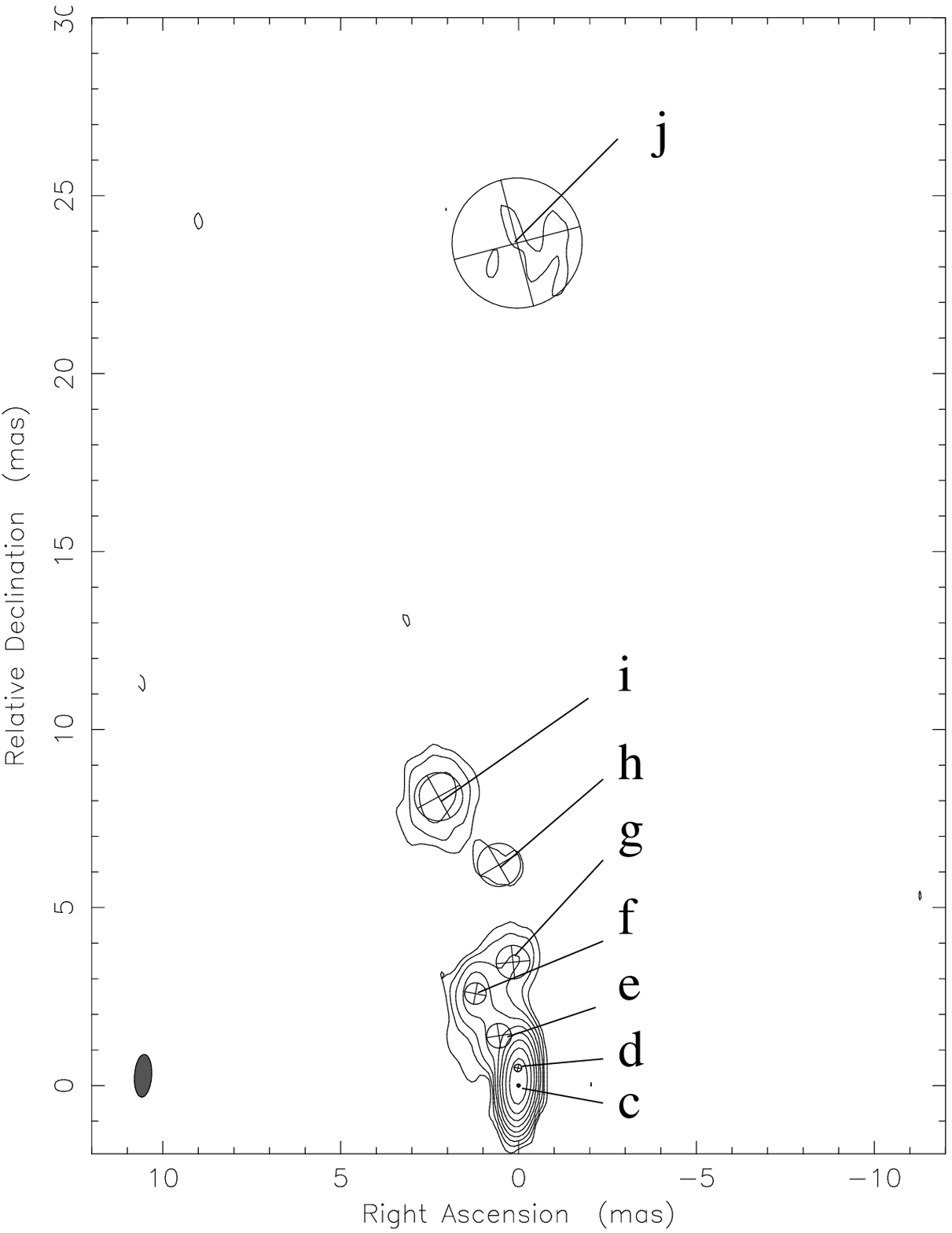}}
\hspace{3pt}%
\subfloat[][15\,GHz, 2005.562]{%
\label{fig:u-9}%
\includegraphics[width=0.30\textwidth,clip]{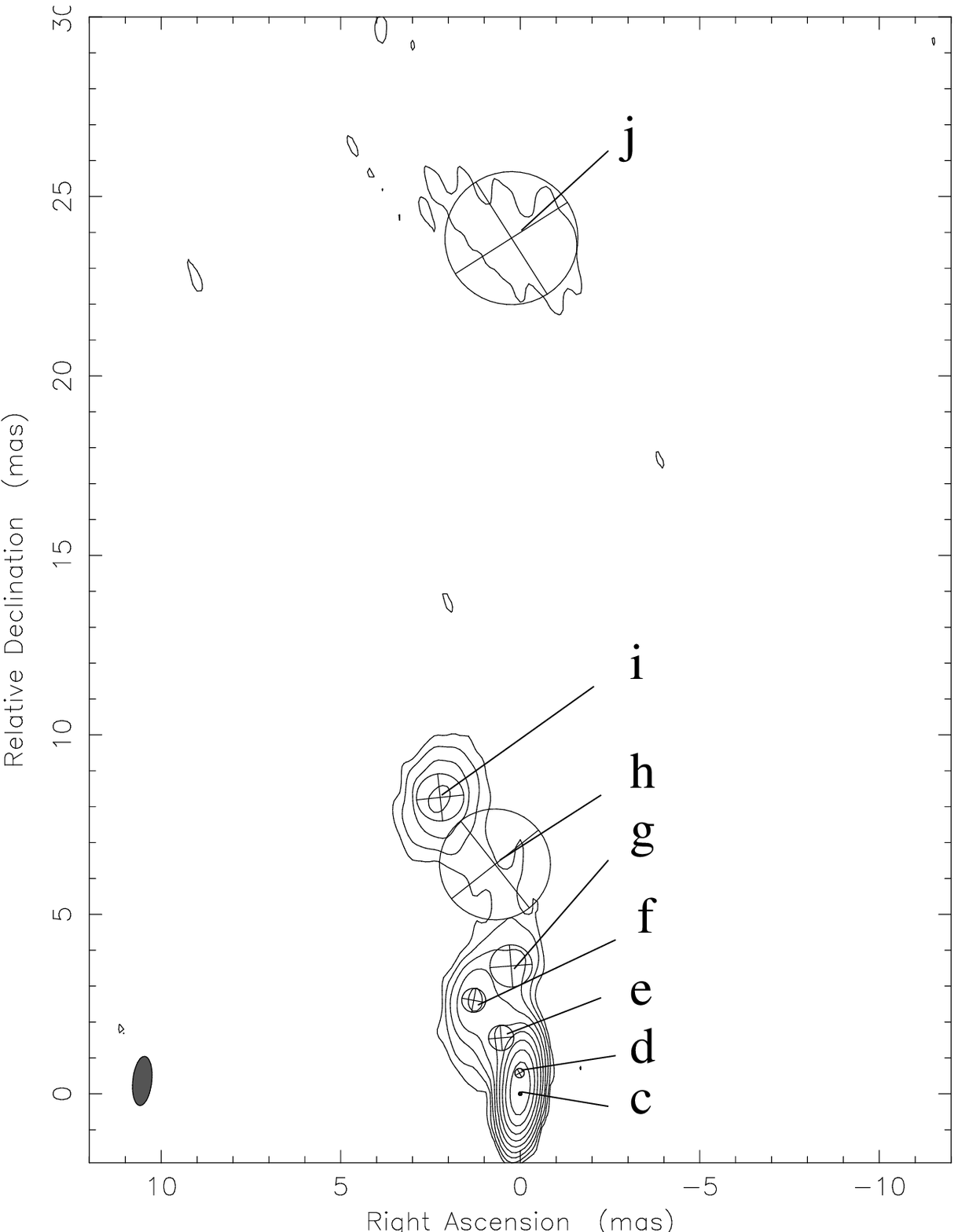}}
\caption{Clean maps of NRAO\,530 at 15\,GHz.}%
\label{fig:u_online}%
\end{figure*}
\begin{figure*}%
\centering
\ContinuedFloat
\hspace{0pt}%
\subfloat[][15\,GHz, 2006.515]{%
\label{fig:u-10}%
\includegraphics[width=0.30\textwidth,clip]{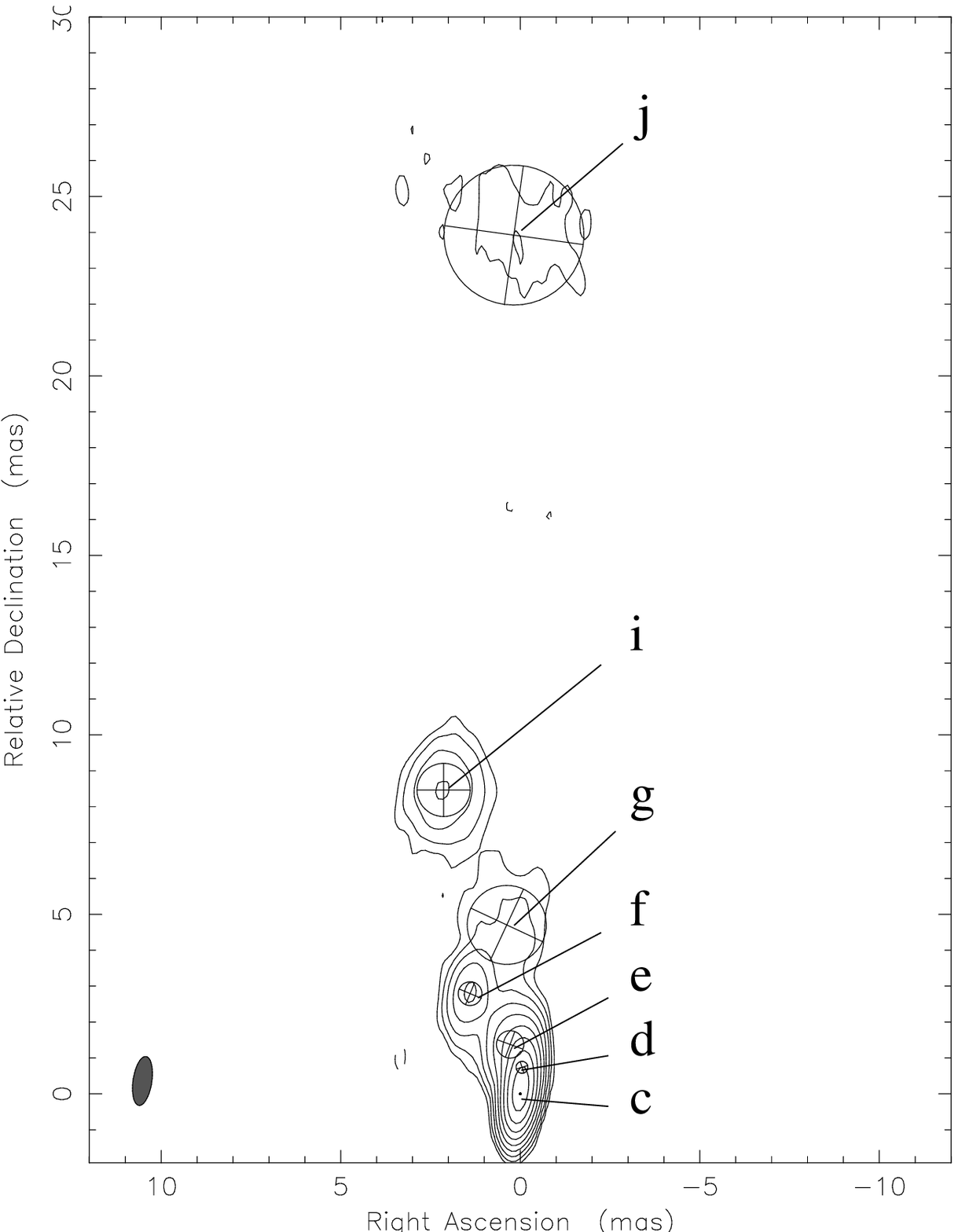}}
\hspace{3pt}
\subfloat[][15\,GHz, 2007.099]{%
\label{fig:u-11}%
\includegraphics[width=0.30\textwidth,clip]{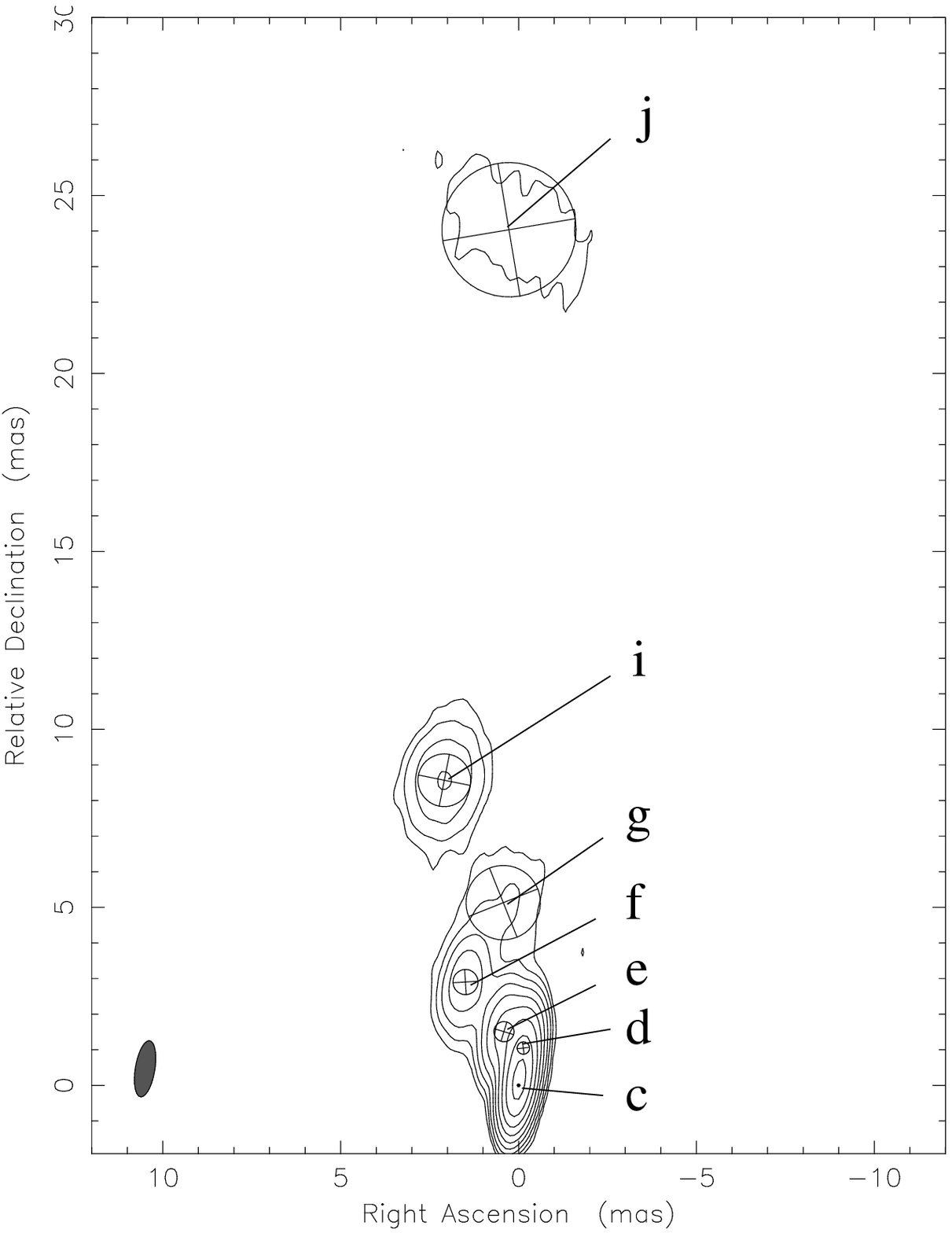}}
\hspace{3pt}%
\subfloat[][15\,GHz, 2007.441]{%
\label{fig:u-12}%
\includegraphics[width=0.30\textwidth,clip]{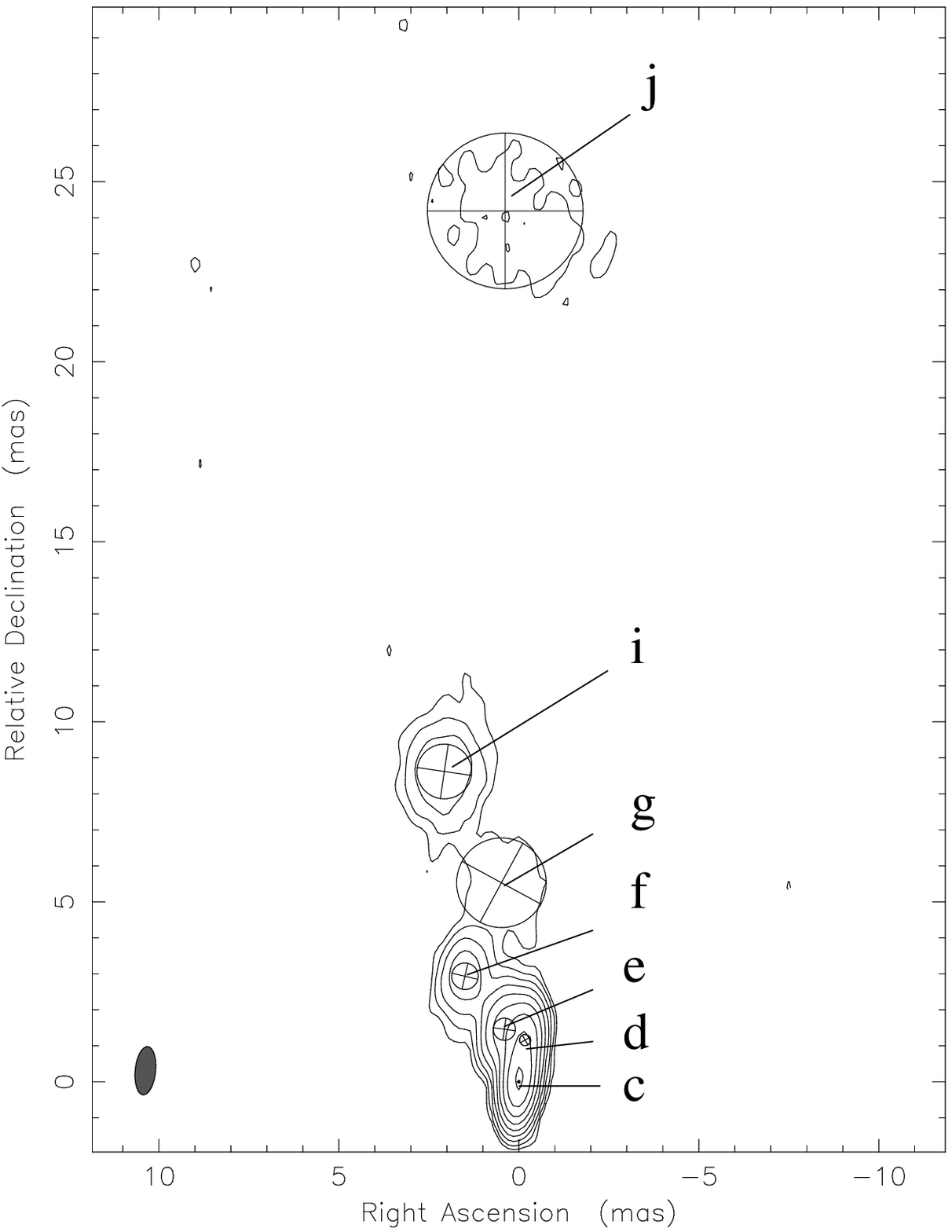}}\\
\hspace{0pt}%
\subfloat[][15\,GHz, 2008.333]{%
\label{fig:u-13}%
\includegraphics[width=0.30\textwidth,clip]{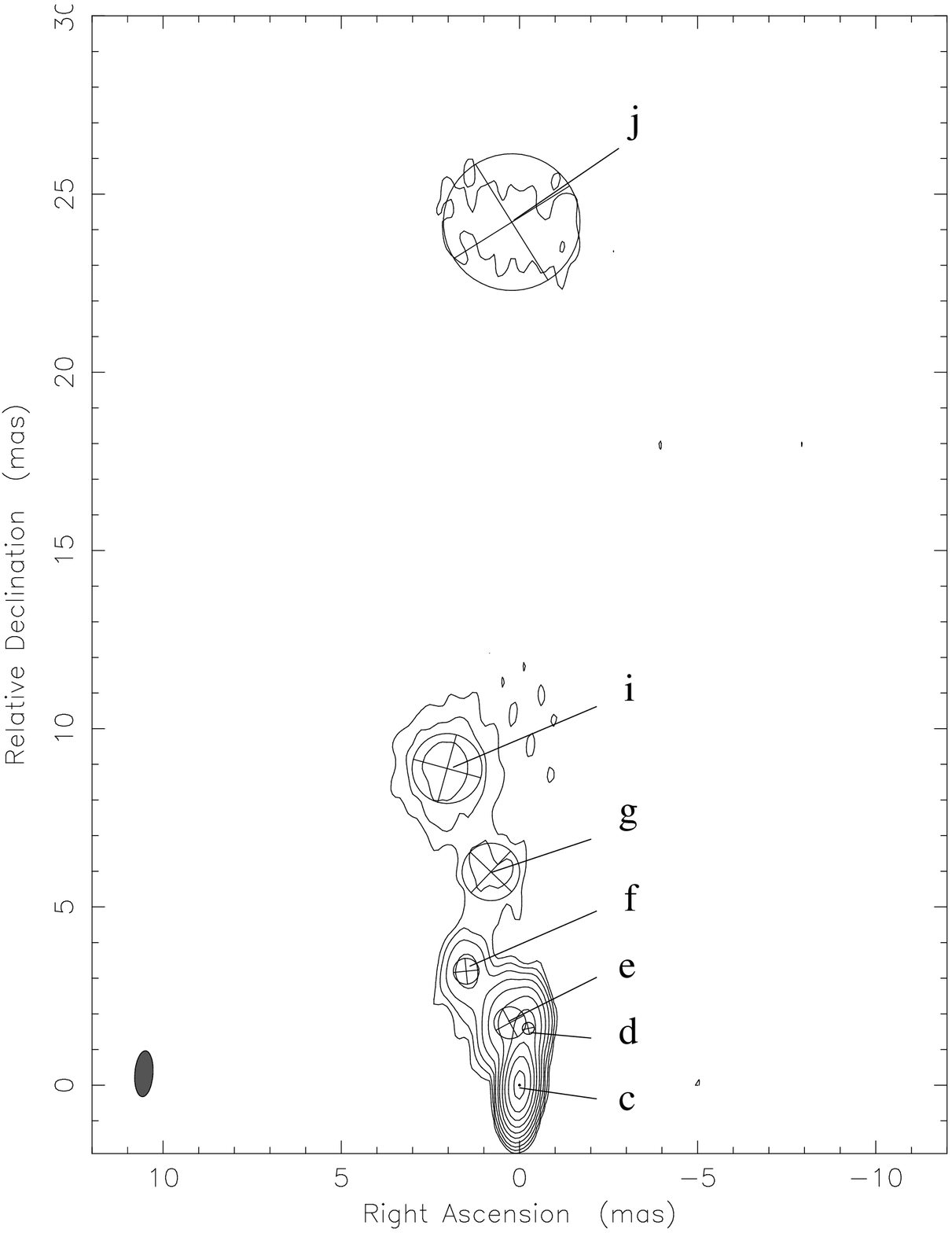}}
\hspace{3pt}%
\subfloat[][15\,GHz, 2008.544]{%
\label{fig:u-14}%
\includegraphics[width=0.30\textwidth,clip]{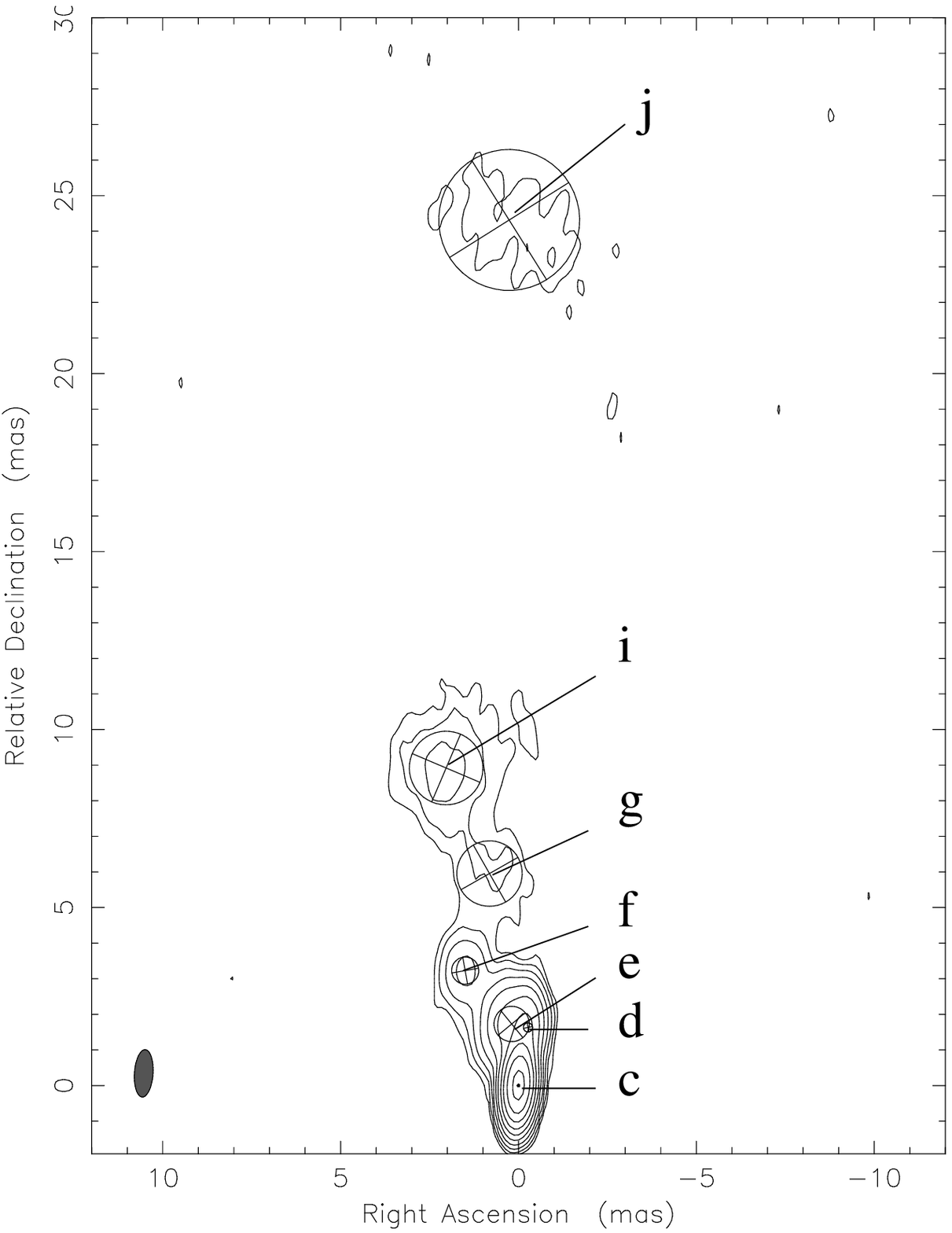}}
\hspace{3pt}%
\subfloat[][15\,GHz, 2008.757]{%
\label{fig:u-15}%
\includegraphics[width=0.30\textwidth,clip]{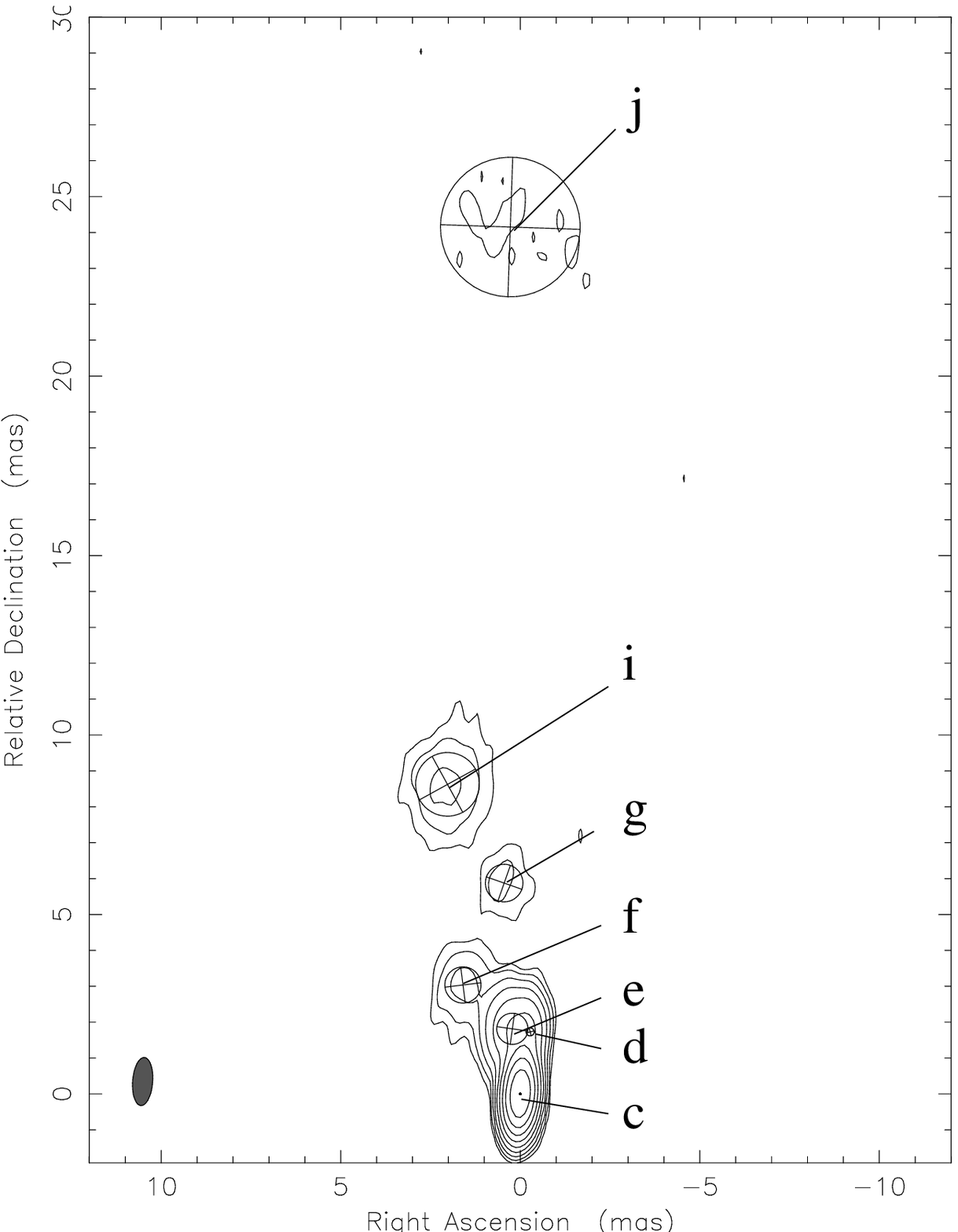}}\\
\hspace{3pt}%
\subfloat[][15\,GHz, 2009.153]{%
\label{fig:u-16}%
\includegraphics[width=0.30\textwidth,clip]{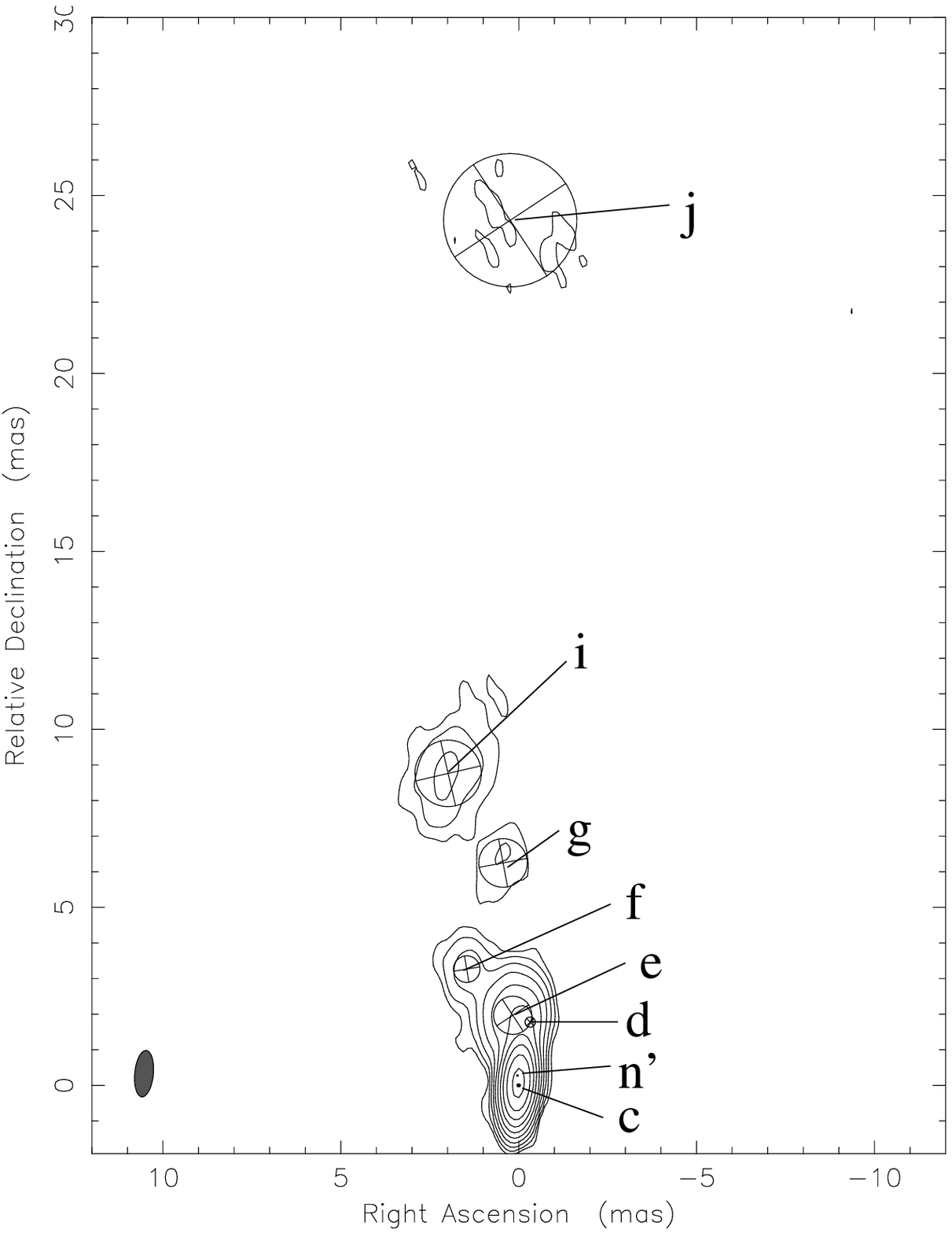}}
\hspace{3pt}%
\subfloat[][15\,GHz, 2009.482]{%
\label{fig:u-17}%
\includegraphics[width=0.30\textwidth,clip]{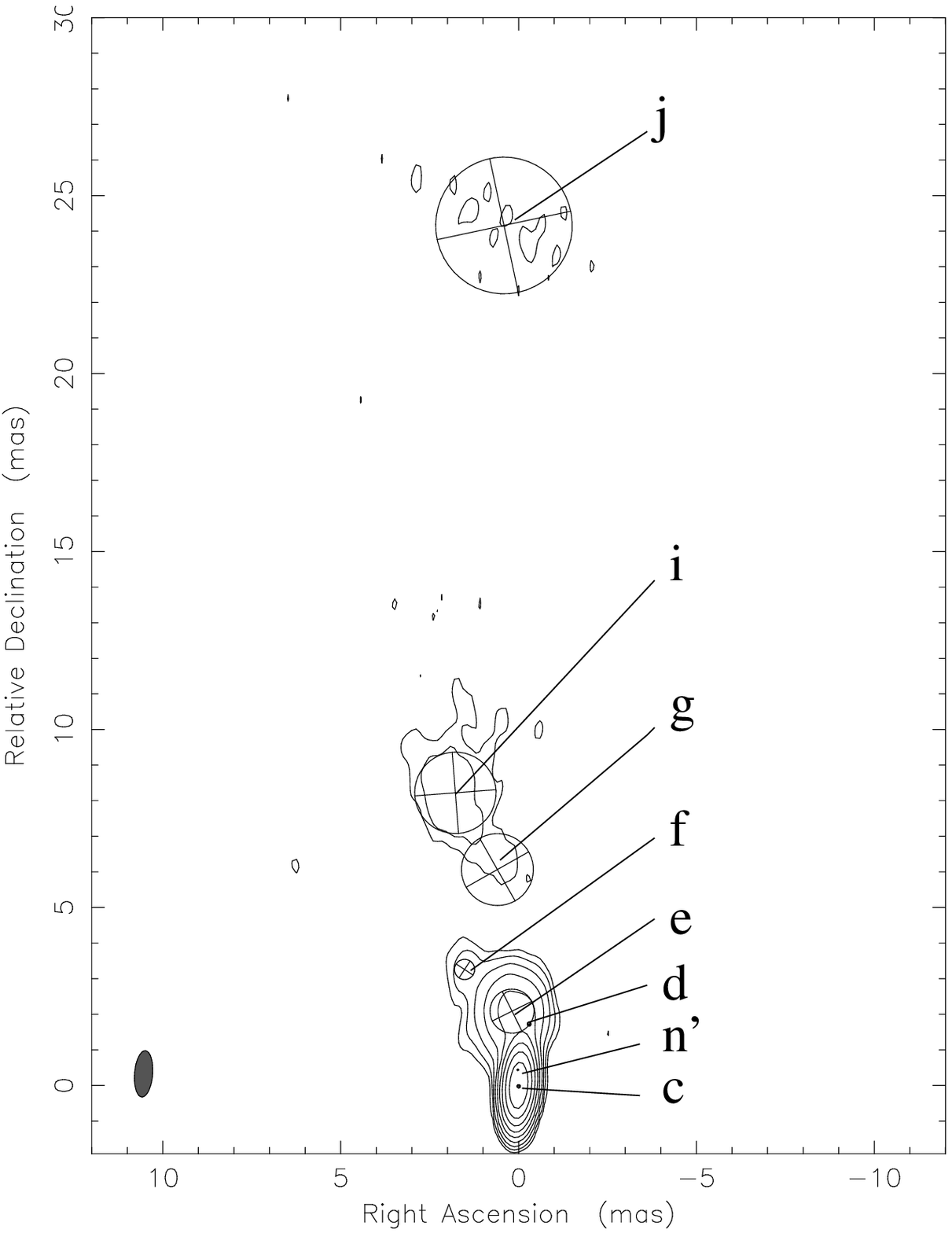}}
\hspace{0pt}%
\caption[\it{-continued}]{\it{-continued}.}%
\end{figure*}
\renewcommand{\thesubfigure}{\thefigure\,(\alph{subfigure})}
\captionsetup[subfigure]{labelformat=simple,labelsep=colon,
listofformat=subsimple}
\makeatletter
\renewcommand{\p@subfigure}{}
\makeatother
\clearpage
\onecolumn  
\begin{longtable}{lllllrr}
\caption[Description of VLBA images of NRAO\,530]{Description of VLBA images of NRAO\,530 shown in Fig.~\ref{fig:maps2007_online} and Fig.~\ref{fig:u_online}. The columns list the epoch of observation, the observing frequency, the peak flux density, the parameters of the restoring elliptical Gaussian beam: the full width at half maximum (FWHM) of the major and minor axes and the position angle of the major axis, and the contour levels of the image, expressed as a percentage of the peak intensity.}
\label{tab:para_n_online}\\
\hline
&&&\multicolumn{3}{c}{Restoring Beam}&\\
\cline{4-6}
Epoch&$\nu$&$S_{\rm peak}$&Major&Minor&P.A.&Contours\\
\hline
&GHz&Jy/beam&mas&mas&deg&\\
(1)&(2)&(3)&(4)&(5)&(6)&(7)\\
\endfirsthead
\multicolumn{7}{c}%
{{\bfseries \tablename\ \thetable{} -- continued}} \\
\hline
&&&\multicolumn{3}{c}{Restoring Beam}&\\
\cline{4-6}
Epoch&$\nu$&$S_{\rm peak}$&Major&Minor&P.A.&Contours\\
\hline
&GHz&Jy/beam&mas&mas&deg&\\
(1)&(2)&(3)&(4)&(5)&(6)&(7)\\
\hline
\endhead
\hline
\endfoot
\hline
\endlastfoot
\hline
2007.370&22&1.32&1.19&0.344&$-12.8$&$-0.3$, 0.3, 0.6, ..., 76.8\\
          &43&1.33&0.537&0.193&$-7.7$&$-0.3$, 0.3, 0.6, ..., 76.8\\
          &86&0.74&0.2&0.2&0&$-1$, 1, 2, ..., 64\\
\hline
2007.373&22&1.31&1.29&0.344&$-15.0$&$-0.3$, 0.3, 0.6, ..., 76.8\\
          &43&1.33&0.628&0.19&$-12.9$&$-0.3$, 0.3, 0.6, ..., 76.8\\
          &86&1.20&0.2&0.2&0&$-0.5$, 0.5, 1, ..., 64\\
\hline
2007.375&22&1.29&1.14&0.367&$-9.2$&$-0.3$, 0.3, 0.6, ..., 76.8\\
          &43&1.41&0.794&0.252&$-11.2$&-0.3, 0.3, 0.6, ..., 76.8\\
          &86&1.06&0.2&0.2&0&$-1$, 1, 2, ..., 64\\
\hline
2007.378&22&1.38&1.24&0.367&$-10.0$&$-0.3$, 0.3, 0.6, ..., 76.8\\
          &43&1.44&0.72&0.314&$-12.0$&$-0.3$, 0.3, 0.6, ..., 76.8\\
          &86&1.25&0.2&0.2&0&$-0.5$, 0.5, 1, ..., 64\\
\hline
2007.381&22&1.32&1.15&0.374&$-10.1$&$-0.3$, 0.3, 0.6, ..., 76.8\\
          &43&1.44&0.83&0.319&$-10.0$&$-0.3$, 0.3, 0.6, ..., 76.8\\
          &86&1.04&0.2&0.2&0&$-0.5$, 0.5, 1, ..., 64\\
\hline
2007.384&22&1.27&1.12&0.42&$-6.88$&$-0.3$, 0.3, 0.6, ..., 76.8\\
          &43&1.37&0.571&0.186&$-8.6$&0.3, 0.6, 1.2, ..., 76.8\\
          &86&1.13&0.2&0.2&0&$-0.5$, 0.5, 1, ..., 64\\
\hline
2007.386&22&1.31&1.09&0.403&$-5.95$&$-0.3$, 0.3, 0.6, ..., 76.8\\
          &43&1.37&0.604&0.198&$-8.03$&0.3, 0.6, 1.2, ..., 76.8\\
          &86&1.01&0.2&0.2&0&$-0.5$, 0.5, 1, ..., 64\\
\hline
2007.389&22&1.37&1.26&0.424&$-8.68$&$-0.3$, 0.3, 0.6, ..., 76.8\\
          &43&1.41&0.628&0.199&$-9.78$&0.3, 0.6, 1.2, ..., 76.8\\
          &86&1.04&0.2&0.2&0&$-0.5$, 0.5, 1, ..., 64\\
\hline
2007.392&22&1.34&1.20&0.43&$-6.14$&$-0.3$, 0.3, 0.6, ..., 76.8\\
          &43&1.37&0.553&0.197&$-7.02$&0.3, 0.6, 1.2, ..., 76.8\\
          &86&1.19&0.2&0.2&0&$-0.5$, 0.5, 1, ..., 64\\
\hline
2007.395&22&1.33&1.37&0.405&$-11.8$&$-0.3$, 0.3, 0.6, ..., 76.8\\
          &43&1.45&0.663&0.183&$-12.8$&0.3, 0.6, 1.2, ..., 76.8\\
          &86&1.13&0.2&0.2&0&$-0.5$, 0.5, 1, ..., 64\\
\hline
1999.712&15&2.07&1.35&0.495&$-6.8$&0.25, 0.5, 1, ..., 64\\
2000.030&15&1.83&1.42&0.505&$-9.0$&$-0.25$, 0.25 ,0.5, ..., 64\\
2000.333&15&1.99&1.35&0.509&$-2.6$&$-0.3$, 0.3 ,0.6, ..., 76.8\\
2002.773&15&4.73&1.51&0.547&$-7.7$&0.1, 0.2 ,0.4, ..., 51.2\\
2003.258&15&4.15&1.34&0.55&$-2.2$&0.15, 0.15 ,0.3, ..., 76.8\\
2003.337&15&4.23&1.36&0.544&$-2.8$&0.15, 0.3 ,0.6, ..., 76.8\\
2004.115&15&3.36&1.35&0.519&$-5.2$&0.15, 0.3 ,0.6, ..., 76.8\\
2005.225&15&2.81&1.2&0.483&$-3.9$&$-0.2$, 0.2 ,0.4, ..., 51.2\\
2005.562&15&2.29&1.37&0.529&$-6.0$&0.2, 0.4 ,0.8, ..., 51.2\\
2006.515&15&1.72&1.38&0.524&$-8.3$&$-0.25$, 0.25, 0.5, ..., 64\\
2007.099&15&1.54&1.59&0.551&$-9.3$&$-0.3$, 0.3, 0.6, ..., 76.8\\
2007.441&15&1.20&1.34&0.568&$-5.7$&$-0.35$, 0.35, 0.7, ..., 89.6\\
2008.333&15&2.76&1.29&0.509&$-4.2$&$-0.15$, 0.15, 0.3, ..., 76.8\\
2008.544&15&2.91&1.33&0.524&$-3.7$&$-0.15$, 0.15, 0.3, ..., 76.8\\
2008.757&15&3.19&1.34&0.559&$-4.2$&$-0.2$, 0.2 ,0.4, ..., 51.2\\
2009.153&15&3.63&1.31&0.524&$-5.4$&0.15, 0.3, 0.6, ..., 76.8\\
2009.482&15&3.19&1.30&0.51&$-4.5$&0.25, 0.5, 1, ..., 64\\
\hline
\end{longtable}
\clearpage
\onecolumn
\begin{longtable}{cclcrccc}
\caption[Model-fitting results for NRAO\,530]{Model-fitting results for NRAO\,530.}
\label{tab:nrao530_model_online} \\
\hline
\multicolumn{1}{c}{Epoch} & \multicolumn{1}{c}{Id.} & \multicolumn{1}{c}{Flux}& \multicolumn{1}{c}{Core Separation}& \multicolumn{1}{c}{P.A.}& \multicolumn{1}{c}{Size}\\
\multicolumn{1}{c}{}&\multicolumn{1}{c}{}& \multicolumn{1}{c}{[mJy]}&\multicolumn{1}{c}{[mas]}&\multicolumn{1}{c}{[degree]}&\multicolumn{1}{c}{[mas]}\\
\hline
\endfirsthead
\multicolumn{7}{c}%
{{\bfseries \tablename\ \thetable{} -- continued}} \\
\hline
\multicolumn{1}{c}{Epoch} & \multicolumn{1}{c}{Id.} & \multicolumn{1}{c}{Flux}& \multicolumn{1}{c}{Core Separation}& \multicolumn{1}{c}{P.A.}& \multicolumn{1}{c}{Size}\\
\multicolumn{1}{c}{}&\multicolumn{1}{c}{}& \multicolumn{1}{c}{[mJy]}&\multicolumn{1}{c}{[mas]}&\multicolumn{1}{c}{[degree]}&\multicolumn{1}{c}{[mas]}\\
\hline
\endhead
\hline
\endfoot
\hline 
\endlastfoot
&&\multicolumn{2}{|c|}{(I) $\nu$ = 22\,GHz}&&\\
\hline
2007.370&c&1332.9$\pm$23.6& 0.00$\pm$0.00&  0.0$\pm$0.0&0.08$\pm$0.01\\
          &d& 385.9$\pm$40.5& 1.10$\pm$0.06& $-9.5\pm$3.1&0.28$\pm$0.02\\
          &e& 377.8$\pm$72.1& 1.52$\pm$0.08&  9.6$\pm$2.9&0.87$\pm$0.15\\
          &f&  99.1$\pm$26.5& 3.41$\pm$0.10& 26.1$\pm$1.6&0.81$\pm$0.20\\
          &g&  29.5$\pm$11.2& 4.52$\pm$0.22&  7.0$\pm$2.8&0.32$\pm$0.09\\
          &i& 138.3$\pm$44.2& 8.75$\pm$0.30& 13.4$\pm$2.0&1.99$\pm$0.61\\
\hline
2007.373&c&1317.6$\pm$60.3& 0.00$\pm$0.00&  0.0$\pm$0.0&0.10$\pm$0.01\\
          &d& 390.3$\pm$45.4& 1.16$\pm$0.06& $-9.6\pm$3.0&0.28$\pm$0.03\\
          &e& 358.2$\pm$74.2& 1.48$\pm$0.08& 10.8$\pm$3.0&0.83$\pm$0.16\\
          &f&  76.1$\pm$19.8& 3.28$\pm$0.06& 28.1$\pm$1.1&0.56$\pm$0.12\\
          &g&  52.6$\pm$23.6& 4.56$\pm$0.31&  9.4$\pm$3.8&1.49$\pm$0.61\\
          &i& 116.7$\pm$38.6& 8.94$\pm$0.26& 13.6$\pm$1.6&1.64$\pm$0.52\\
\hline
2007.375&c&1289.7$\pm$43.2& 0.00$\pm$0.00&  0.0$\pm$0.0&0.09$\pm$0.01\\
           &d& 435.9$\pm$54.6& 1.15$\pm$0.06& $-9.6\pm$3.0&0.32$\pm$0.03\\
           &e& 298.9$\pm$58.2& 1.54$\pm$0.07& 13.2$\pm$2.5&0.78$\pm$0.14\\
           &f&  98.8$\pm$24.5& 3.27$\pm$0.10& 26.9$\pm$1.8&0.88$\pm$0.20\\
           &g&  26.3$\pm$16.2& 4.82$\pm$0.20&  6.4$\pm$2.4&0.41$\pm$0.19\\
           &i& 123.0$\pm$42.1& 8.88$\pm$0.28& 13.0$\pm$1.8&1.74$\pm$0.57\\
\hline
2007.378&c&1387.8$\pm$30.8& 0.00$\pm$0.00&  0.0$\pm$0.0&0.14$\pm$0.01\\
          &d& 413.1$\pm$36.2& 1.17$\pm$0.06&$-10.7\pm$2.9&0.25$\pm$0.02\\
          &e& 345.6$\pm$63.1& 1.54$\pm$0.06& 10.6$\pm$2.2&0.74$\pm$0.12\\
          &f&  97.8$\pm$20.9& 3.13$\pm$0.06& 29.6$\pm$1.1&0.48$\pm$0.08\\
          &g&  14.2$\pm$13.2& 5.19$\pm$0.16&  7.3$\pm$1.7&0.54$\pm$0.31\\
          &i&  83.3$\pm$29.1& 8.87$\pm$0.21& 15.1$\pm$1.3&1.29$\pm$0.42\\*
\hline
2007.381&c&1313.5$\pm$27.7& 0.00$\pm$0.00&  0.0$\pm$0.0&0.09$\pm$0.01\\
          &d& 392.4$\pm$25.0& 1.18$\pm$0.06& $-9.9\pm$2.9&0.27$\pm$0.01\\
          &e& 352.3$\pm$62.8& 1.42$\pm$0.07& 10.8$\pm$2.7&0.82$\pm$0.13\\
          &f&  89.0$\pm$16.9& 3.31$\pm$0.06& 27.3$\pm$1.0&0.70$\pm$0.12\\
          &g&  58.3$\pm$27.6& 4.31$\pm$0.39&  8.4$\pm$5.2&1.78$\pm$0.79\\
          &i& 121.1$\pm$35.8& 8.89$\pm$0.20& 13.4$\pm$1.6&1.73$\pm$0.49\\
\hline
2007.384&c&1249.7$\pm$28.4& 0.00$\pm$0.00&  0.0$\pm$0.0&0.09$\pm$0.01\\
          &d& 388.1$\pm$22.8& 1.16$\pm$0.06&$-10.7\pm$3.0&0.30$\pm$0.01\\
          &e& 364.1$\pm$52.5& 1.43$\pm$0.06& 11.2$\pm$2.4&0.85$\pm$0.10\\
          &f&  94.1$\pm$23.0& 3.33$\pm$0.08& 26.6$\pm$1.3&0.73$\pm$0.15\\
          &g&  22.0$\pm$16.7& 4.72$\pm$0.20&  6.5$\pm$2.4&0.35$\pm$0.20\\
          &i& 100.5$\pm$30.1& 8.95$\pm$0.20& 14.3$\pm$1.3&1.43$\pm$0.39\\
\hline
2007.386&c&1298.3$\pm$42.6& 0.00$\pm$0.00&  0.0$\pm$0.0&0.08$\pm$0.01\\
          &d& 409.7$\pm$47.1& 1.19$\pm$0.06& $-9.4\pm$2.9&0.32$\pm$0.03\\
          &e& 302.6$\pm$57.5& 1.49$\pm$0.07& 11.6$\pm$2.7&0.83$\pm$0.14\\
          &f&  89.7$\pm$23.6& 3.35$\pm$0.10& 26.6$\pm$1.7&0.82$\pm$0.19\\
          &g&  34.3$\pm$12.7& 4.58$\pm$0.26&  7.0$\pm$3.2&0.42$\pm$0.12\\
          &i& 112.0$\pm$36.9& 8.91$\pm$0.26& 13.6$\pm$1.7&1.66$\pm$0.52\\
\hline
2007.389&c&1336.8$\pm$41.5& 0.00$\pm$0.00&  0.0$\pm$0.0&0.06$\pm$0.01\\
          &d& 403.2$\pm$20.2& 1.22$\pm$0.06& $-9.8\pm$2.8&0.24$\pm$0.01\\
          &e& 338.1$\pm$54.8& 1.45$\pm$0.06& 11.8$\pm$2.4&0.74$\pm$0.11\\
          &f&  82.6$\pm$11.3& 3.22$\pm$0.06& 27.9$\pm$1.1&0.55$\pm$0.06\\
          &g&  80.3$\pm$30.5& 4.03$\pm$0.36& 10.9$\pm$5.1&2.02$\pm$0.72\\
          &i& 118.5$\pm$29.1& 9.04$\pm$0.18& 13.0$\pm$1.1&1.56$\pm$0.36\\
\hline
2007.392&c&1344.6$\pm$34.4& 0.00$\pm$0.00&  0.0$\pm$0.0&0.09$\pm$0.01\\
          &d& 399.8$\pm$31.5& 1.22$\pm$0.06&$-10.4\pm$2.8&0.26$\pm$0.02\\
          &e& 328.5$\pm$40.2& 1.48$\pm$0.06& 11.4$\pm$2.3&0.74$\pm$0.08\\
          &f&  79.9$\pm$15.2& 3.24$\pm$0.06& 27.9$\pm$1.1&0.56$\pm$0.09\\
          &g&  79.1$\pm$24.5& 3.86$\pm$0.24& 10.5$\pm$3.6&1.73$\pm$0.49\\
          &i& 140.2$\pm$30.5& 8.88$\pm$0.19& 12.9$\pm$1.2&1.83$\pm$0.37\\
\hline
2007.395&c&1294.6$\pm$37.0& 0.00$\pm$0.00&  0.0$\pm$0.0&0.08$\pm$0.01\\
          &d& 364.7$\pm$41.3& 1.15$\pm$0.06&$-10.0\pm$3.0&0.25$\pm$0.02\\
          &e& 416.2$\pm$75.9& 1.48$\pm$0.08&  7.7$\pm$3.2&0.98$\pm$0.17\\
          &f&  89.5$\pm$21.9& 3.35$\pm$0.07& 28.7$\pm$1.1&0.63$\pm$0.13\\
          &g&  39.9$\pm$19.2& 4.37$\pm$0.17&  8.1$\pm$2.2&0.83$\pm$0.34\\
          &i& 118.9$\pm$34.3& 8.92$\pm$0.22& 12.8$\pm$0.4&1.63$\pm$0.45\\
\hline
&&\multicolumn{2}{|c|}{(I) $\nu$ = 43\,GHz}&&\\
\hline
2007.370&c&1258.0$\pm$60.7& 0.00$\pm$0.00&  0.0$\pm$0.0&0.03$\pm$0.01\\
          &n& 272.9$\pm$22.4& 0.38$\pm$0.03& $-3.2\pm$4.5&0.08$\pm$0.01\\
          &d& 221.2$\pm$27.7& 1.31$\pm$0.03&$-10.4\pm$1.3&0.21$\pm$0.02\\
          &e& 250.3$\pm$40.2& 1.56$\pm$0.05&  7.5$\pm$2.0&0.70$\pm$0.11\\
          &f&  56.1$\pm$20.2& 3.44$\pm$0.12& 24.4$\pm$2.0&0.68$\pm$0.24\\
\hline
2007.373&c&1205.8$\pm$52.9& 0.00$\pm$0.00&  0.0$\pm$0.0&0.04$\pm$0.01\\
          &n& 295.5$\pm$17.4& 0.35$\pm$0.02& $-3.4\pm$4.9&0.10$\pm$0.01\\
          &d& 217.7$\pm$42.0& 1.32$\pm$0.03&$-10.2\pm$1.3&0.22$\pm$0.04\\
          &e& 245.8$\pm$49.2& 1.57$\pm$0.07&  8.6$\pm$2.5&0.72$\pm$0.14\\
          &f&  48.9$\pm$16.8& 3.50$\pm$0.11& 24.9$\pm$1.8&0.70$\pm$0.23\\
\hline
2007.375&c&1303.1$\pm$47.9& 0.00$\pm$0.00&  0.0$\pm$0.0&0.01$\pm$0.01\\
          &n& 233.1$\pm$19.2& 0.42$\pm$0.03& $-5.1\pm$4.1&0.12$\pm$0.01\\
          &d& 235.9$\pm$32.8& 1.34$\pm$0.03& $-9.4\pm$1.3&0.23$\pm$0.02\\
          &e& 195.5$\pm$35.4& 1.67$\pm$0.06& 11.1$\pm$1.9&0.67$\pm$0.11\\
          &f&  56.8$\pm$14.1& 3.47$\pm$0.07& 25.6$\pm$1.1&0.60$\pm$0.14\\
\hline
2007.378&c&1303.1$\pm$29.8& 0.00$\pm$0.00&  0.0$\pm$0.0&0.02$\pm$0.01\\
          &n& 261.9$\pm$22.0& 0.39$\pm$0.03& $-2.9\pm$4.4&0.11$\pm$0.01\\
          &d& 276.4$\pm$40.4& 1.36$\pm$0.03& $-8.3\pm$1.3&0.27$\pm$0.03\\
          &e& 181.0$\pm$41.3& 1.70$\pm$0.07& 12.5$\pm$2.3&0.66$\pm$0.14\\
          &f&  55.6$\pm$20.8& 3.45$\pm$0.13& 25.2$\pm$2.1&0.73$\pm$0.25\\
\hline
2007.381&c&1269.8$\pm$33.2& 0.00$\pm$0.00&  0.0$\pm$0.0&0.04$\pm$0.01\\
          &n& 302.8$\pm$19.1& 0.36$\pm$0.03& $-4.0\pm$4.8&0.12$\pm$0.01\\
          &d& 261.9$\pm$36.6& 1.36$\pm$0.03& $-8.5\pm$1.3&0.25$\pm$0.02\\
          &e& 181.1$\pm$35.6& 1.67$\pm$0.06& 11.9$\pm$2.0&0.66$\pm$0.11\\
          &f&  57.0$\pm$13.2& 3.50$\pm$0.07& 25.0$\pm$1.2&0.67$\pm$0.14\\
\hline
2007.384&c&1274.3$\pm$18.2& 0.00$\pm$0.00&  0.0$\pm$0.0&0.02$\pm$0.01\\
          &n& 234.4$\pm$22.0& 0.37$\pm$0.03& $-3.0\pm$4.6&0.05$\pm$0.01\\
          &d& 206.9$\pm$25.7& 1.33$\pm$0.03&$-10.5\pm$1.3&0.20$\pm$0.02\\
          &e& 253.5$\pm$57.9& 1.56$\pm$0.08&  8.0$\pm$2.9&0.73$\pm$0.15\\
          &f&  51.9$\pm$17.7& 3.33$\pm$0.10& 25.0$\pm$1.8&0.65$\pm$0.20\\
\hline
2007.386&c&1287.2$\pm$51.5& 0.00$\pm$0.00&  0.0$\pm$0.0&0.01$\pm$0.01\\
          &n& 251.1$\pm$16.2& 0.36$\pm$0.03& $-3.4\pm$4.8&0.06$\pm$0.01\\
          &d& 239.4$\pm$32.1& 1.34$\pm$0.03&$-10.0\pm$1.3&0.22$\pm$0.02\\
          &e& 226.2$\pm$55.1& 1.64$\pm$0.08&  9.9$\pm$2.9&0.71$\pm$0.16\\
          &f&  52.3$\pm$15.7& 3.34$\pm$0.10& 25.5$\pm$1.7&0.70$\pm$0.20\\
\hline
2007.389&c&1265.3$\pm$11.6& 0.00$\pm$0.00&  0.0$\pm$0.0&0.02$\pm$0.01\\
          &n& 241.7$\pm$21.8& 0.37$\pm$0.03& $-3.2\pm$4.6&0.06$\pm$0.01\\
          &d& 220.1$\pm$30.2& 1.34$\pm$0.03&$-10.1\pm$1.3&0.21$\pm$0.02\\
          &e& 246.8$\pm$58.8& 1.57$\pm$0.09&  9.0$\pm$3.1&0.75$\pm$0.17\\
          &f&  52.8$\pm$14.9& 3.18$\pm$0.08& 24.6$\pm$1.5&0.63$\pm$0.17\\
\hline
2007.392&c&1308.5$\pm$18.9& 0.00$\pm$0.00&  0.0$\pm$0.0&0.02$\pm$0.01\\
          &n& 214.4$\pm$29.7& 0.39$\pm$0.03& $-2.7\pm$4.4&0.02$\pm$0.01\\
          &d& 212.8$\pm$27.6& 1.33$\pm$0.03&$-10.2\pm$1.3&0.20$\pm$0.02\\
          &e& 243.7$\pm$38.4& 1.61$\pm$0.06&  8.7$\pm$2.0&0.76$\pm$0.11\\
          &f&  53.3$\pm$17.0& 3.42$\pm$0.11& 25.6$\pm$1.9&0.76$\pm$0.23\\
\hline
2007.395&c&1337.4$\pm$19.0& 0.00$\pm$0.00&  0.0$\pm$0.0&0.03$\pm$0.01\\
          &n& 210.5$\pm$13.4& 0.37$\pm$0.03& $-2.3\pm$4.6&0.02$\pm$0.01\\
          &d& 239.1$\pm$24.6& 1.34$\pm$0.03&$-10.5\pm$1.3&0.20$\pm$0.02\\
          &e& 245.5$\pm$32.9& 1.54$\pm$0.05& 10.2$\pm$2.0&0.83$\pm$0.11\\
          &f&  50.5$\pm$14.5& 3.43$\pm$0.09& 24.6$\pm$1.5&0.67$\pm$0.18\\
\hline
&&\multicolumn{2}{|c|}{(I) $\nu$ =  86\,GHz}&&\\
\hline
2007.370&c&1214.4$\pm$186.3&0.00$\pm$0.00&0.0  $\pm$0.0&0.16$\pm$0.02\\
\hline
2007.373&c&1187.5$\pm$106.8&0.00$\pm$0.00&0.0  $\pm$0.0&0.06$\pm$0.01\\
          &n&  73.1$\pm$37.0 & 0.45$\pm$0.02& $-9.6\pm$2.5&0.08$\pm$0.01\\
          &d&  107.4$\pm$21.4& 1.55$\pm$0.05& $-9.7\pm$1.8&0.08$\pm$0.02\\
\hline
2007.375&c&1077.2$\pm$94.0& 0.00$\pm$0.00&  0.0$\pm$0.0&0.03$\pm$0.01\\
          &n&  87.2$\pm$39.3& 0.37$\pm$0.02& $-6.7\pm$3.1&0.06$\pm$0.01\\
          &d& 116.6$\pm$49.0& 1.39$\pm$0.05& $-4.9\pm$2.1&0.08$\pm$0.02\\
\hline
2007.378&c&1194.6$\pm$70.3& 0.00$\pm$0.00&  0.0$\pm$0.0&0.05$\pm$0.01\\
          &n&  96.9$\pm$37.8& 0.35$\pm$0.04& $-4.5\pm$6.0&0.12$\pm$0.07\\
          &d& 109.8$\pm$56.2& 1.55$\pm$0.07& $-0.5\pm$2.6&0.30$\pm$0.14\\
\hline
2007.381&c&1088.2$\pm$39.0& 0.00$\pm$0.00&  0.0$\pm$0.0&0.03$\pm$0.01\\
          &n&  59.4$\pm$11.4& 0.37$\pm$0.02& $-9.3\pm$3.1&0.07$\pm$0.01\\
          &d& 114.6$\pm$52.4& 1.32$\pm$0.10&$-11.7\pm$4.1&0.44$\pm$0.19\\
\hline
2007.384&c&1170.0$\pm$54.9& 0.00$\pm$0.00&  0.0$\pm$0.0&0.04$\pm$0.01\\
          &n&  72.2$\pm$39.2& 0.35$\pm$0.02& $-6.8\pm$3.3&0.03$\pm$0.01\\
          &d&  53.9$\pm$27.1& 1.47$\pm$0.05& $-5.4\pm$1.9&0.06$\pm$0.02\\
\hline
2007.386&c&1067.4$\pm$88.2& 0.00$\pm$0.00&  0.0$\pm$0.0&0.05$\pm$0.01\\
          &n&  71.0$\pm$36.8& 0.31$\pm$0.02& $-4.3\pm$3.9&0.09$\pm$0.04\\
          &d&  53.4$\pm$24.5& 1.43$\pm$0.05& $-5.6\pm$2.0&0.15$\pm$0.05\\
\hline
2007.389&c&1091.5$\pm$81.2& 0.00$\pm$0.00&  0.0$\pm$0.0&0.04$\pm$0.01\\
          &n& 125.9$\pm$49.7& 0.43$\pm$0.06&$-12.2\pm$8.4&0.35$\pm$0.13\\
          &d&  86.3$\pm$30.4& 1.33$\pm$0.04&$-12.3\pm$1.6&0.24$\pm$0.08\\
\hline
2007.392&c&1192.8$\pm$75.8& 0.00$\pm$0.00&  0.0$\pm$0.0&0.01$\pm$0.01\\
          &n&  55.2$\pm$29.4& 0.35$\pm$0.02& $-7.3\pm$3.3&0.04$\pm$0.01\\
          &d& 142.5$\pm$40.7& 1.39$\pm$0.05& $-7.8\pm$2.1&0.21$\pm$0.05\\
\hline
2007.395&c&1131.2$\pm$60.9& 0.00$\pm$0.00&  0.0$\pm$0.0&0.01$\pm$0.01\\
          &n&  75.7$\pm$29.3& 0.36$\pm$0.02& $-0.9\pm$3.2&0.03$\pm$0.01\\
          &d& 116.1$\pm$47.9& 1.57$\pm$0.11&  4.8$\pm$3.9&0.59$\pm$0.21\\
\hline
&&\multicolumn{2}{|c|}{(II) $\nu$ = 15\,GHz}&&\\
\hline
1999.712&c&2013.6$\pm$48.7& 0.00$\pm$0.00&  0.0$\pm$0.0&0.09$\pm$0.01\\
       &d& 297.6$\pm$13.9& 0.47$\pm$0.06& 36.8$\pm$7.3&0.25$\pm$0.01\\
       &f& 145.5$\pm$12.0& 1.10$\pm$0.06& 29.1$\pm$3.1&0.09$\pm$0.01\\
       &g& 483.2$\pm$55.8& 1.55$\pm$0.06&  6.3$\pm$2.2&0.93$\pm$0.10\\
       &x&  36.5$\pm$9.7& 3.00$\pm$0.06& 16.0$\pm$1.1&0.43$\pm$0.09\\
       &h& 217.6$\pm$43.2& 5.82$\pm$0.20& 13.4$\pm$1.9&2.03$\pm$0.39\\
       &j & 108.9$\pm$46.3&23.06$\pm$0.93&  1.0$\pm$2.3&4.42$\pm$1.86\\
\hline
2000.030&c&1679.3$\pm$31.3& 0.00$\pm$0.00&  0.0$\pm$0.0&0.06$\pm$0.01\\
       &d& 303.5$\pm$25.5& 0.37$\pm$0.06& 18.6$\pm$9.2&0.45$\pm$0.03\\
       &f& 228.7$\pm$27.8& 1.12$\pm$0.06& 31.0$\pm$3.1&0.35$\pm$0.03\\
       &g&320.0$\pm$33.2& 1.88$\pm$0.06&  6.2$\pm$1.8&0.78$\pm$0.07\\
       &x&  22.9$\pm$6.8& 2.64$\pm$0.06& 26.4$\pm$1.3&0.11$\pm$0.02\\
       &h& 197.1$\pm$43.1& 5.95$\pm$0.24& 12.8$\pm$2.3&2.30$\pm$0.48\\
       &j &  98.2$\pm$43.5&23.79$\pm$1.31&  0.6$\pm$3.1&5.94$\pm$2.61\\
\hline
2000.333&c&1691.1$\pm$51.2& 0.00$\pm$0.00&  0.0$\pm$0.0&0.07$\pm$0.01\\
       &d& 489.5$\pm$25.3& 0.32$\pm$0.06& 17.0$\pm$10.6&0.38$\pm$0.02\\
       &f& 275.1$\pm$27.9& 1.24$\pm$0.06& 29.7$\pm$2.8&0.40$\pm$0.03\\
       &g& 391.6$\pm$42.3& 2.01$\pm$0.06&  6.3$\pm$1.7&0.85$\pm$0.08\\
       &x&  19.3$\pm$8.3& 2.66$\pm$0.06& 24.5$\pm$1.3&0.06$\pm$0.02\\
       &h& 219.6$\pm$40.8& 5.95$\pm$0.19& 13.4$\pm$1.8&2.12$\pm$0.38\\
       &j & 119.6$\pm$43.5&23.18$\pm$0.87&  0.6$\pm$2.2&4.89$\pm$1.75\\
\hline
2002.773&c&4150.2$\pm$81.5& 0.00$\pm$0.00&  0.0$\pm$0.0&0.03$\pm$0.01\\
       &d& 693.7$\pm$59.5& 0.25$\pm$0.06&  9.7$\pm$13.5&0.20$\pm$0.01\\
       &f& 390.5$\pm$41.1& 1.85$\pm$0.06& 29.4$\pm$1.9&0.59$\pm$0.05\\
       &g& 209.3$\pm$22.5& 2.45$\pm$0.06&  4.6$\pm$1.4&0.85$\pm$0.08\\
       &h& 129.8$\pm$23.0& 5.30$\pm$0.21&  8.7$\pm$2.3&2.47$\pm$0.42\\
       &i& 118.6$\pm$27.6& 7.77$\pm$0.15& 16.4$\pm$1.1&1.44$\pm$0.31\\
       &j& 128.1$\pm$33.3&23.38$\pm$0.53&  0.2$\pm$1.3&4.15$\pm$1.07\\
\hline
2003.258&c &3870.0$\pm$88.9& 0.00$\pm$0.00&  0.0$\pm$0.0&0.04$\pm$0.01\\
       &d & 494.0$\pm$26.3& 0.34$\pm$0.06&  7.8$\pm$10.0&0.48$\pm$0.02\\
       &f& 374.0$\pm$36.6& 1.91$\pm$0.06& 29.8$\pm$1.8&0.67$\pm$0.05\\
       &g& 198.5$\pm$31.8& 2.69$\pm$0.07&  5.9$\pm$1.5&0.99$\pm$0.14\\
       &h& 138.2$\pm$28.6& 6.06$\pm$0.25&  9.3$\pm$2.4&2.51$\pm$0.50\\
       &i & 118.2$\pm$25.6& 8.04$\pm$0.13& 17.0$\pm$0.9&1.31$\pm$0.26\\
       &j & 136.4$\pm$42.6&23.34$\pm$0.58&  0.1$\pm$1.4&3.78$\pm$1.16\\
\hline
2003.337&c&3907.3$\pm$79.1& 0.00$\pm$0.00&  0.0$\pm$0.0&0.05$\pm$0.01\\
       &d& 456.4$\pm$31.4& 0.36$\pm$0.06&  8.8$\pm$9.5&0.21$\pm$0.01\\
       &f& 348.7$\pm$38.3& 1.85$\pm$0.06& 30.8$\pm$1.9&0.59$\pm$0.05\\
       &g& 295.0$\pm$45.0& 2.53$\pm$0.09&  8.3$\pm$1.9&1.23$\pm$0.17\\
       &h& 146.0$\pm$25.9& 5.97$\pm$0.22&  9.4$\pm$2.2&2.61$\pm$0.45\\
       &i& 131.9$\pm$25.9& 8.00$\pm$0.11& 16.9$\pm$0.8&1.28$\pm$0.23\\
       &j& 176.1$\pm$51.8&23.57$\pm$0.61&  0.2$\pm$1.5&4.22$\pm$1.23\\
\hline
2004.115&c &2190.6$\pm$39.0& 0.00$\pm$0.00&  0.0$\pm$0.0&0.05$\pm$0.01\\
       &d &1315.0$\pm$31.0& 0.27$\pm$0.06&  6.9$\pm$12.5&0.11$\pm$0.01\\
       &e & 145.1$\pm$17.2& 1.21$\pm$0.06& 25.3$\pm$2.8&0.83$\pm$0.09\\
       &f & 167.5$\pm$16.1& 2.42$\pm$0.06& 26.8$\pm$1.4&0.66$\pm$0.05\\
       &g& 107.9$\pm$19.9& 3.05$\pm$0.09&  3.6$\pm$1.6&1.03$\pm$0.17\\
       &h&  96.5$\pm$19.1& 6.16$\pm$0.23&  9.1$\pm$2.1&2.42$\pm$0.46\\
       &i & 111.4$\pm$18.2& 8.28$\pm$0.09& 16.4$\pm$0.6&1.22$\pm$0.18\\
       &j & 108.1$\pm$31.0&23.66$\pm$0.53&  0.3$\pm$1.3&3.75$\pm$1.06\\
\hline
2005.225&c &2290.5$\pm$74.5& 0.00$\pm$0.00&  0.0$\pm$0.0&0.09$\pm$0.01\\
       &d & 874.2$\pm$48.5& 0.49$\pm$0.06&  2.0$\pm$7.0&0.21$\pm$0.01\\
       &e & 141.0$\pm$21.6& 1.50$\pm$0.06& 21.5$\pm$2.3&0.69$\pm$0.09\\
       &f & 138.9$\pm$19.0& 2.85$\pm$0.06& 25.2$\pm$1.2&0.60$\pm$0.07\\
       &g&  69.3$\pm$16.8& 3.47$\pm$0.11&  2.6$\pm$1.7&0.94$\pm$0.21\\
       &h&  31.7$\pm$12.9& 6.22$\pm$0.23&  5.1$\pm$2.1&1.21$\pm$0.46\\
       &i & 143.6$\pm$29.6& 8.42$\pm$0.13& 15.5$\pm$0.9&1.36$\pm$0.27\\
       &j & 133.8$\pm$46.1&23.67$\pm$0.63&  0.1$\pm$1.5&3.66$\pm$1.25\\
\hline
2005.562&c &1845.7$\pm$53.1& 0.00$\pm$0.00&  0.0$\pm$0.0&0.09$\pm$0.01\\
       &d & 795.9$\pm$40.0& 0.58$\pm$0.06&  1.9$\pm$5.9&0.26$\pm$0.01\\
       &e & 139.4$\pm$23.1& 1.65$\pm$0.06& 18.8$\pm$2.1&0.70$\pm$0.10\\
       &f& 139.9$\pm$17.6& 2.91$\pm$0.06&  26.5$\pm$1.2&0.67$\pm$0.07\\
       &g&  70.3$\pm$16.6& 3.57$\pm$0.13&  4.1$\pm$2.0&1.18$\pm$0.26\\
       &h&  69.0$\pm$26.7& 6.43$\pm$0.59&  6.3$\pm$5.3&3.09$\pm$1.19\\
       &i & 148.5$\pm$23.7& 8.55$\pm$0.10& 15.1$\pm$0.7&1.32$\pm$0.20\\
       &j & 131.1$\pm$40.4&23.84$\pm$0.56&  0.6$\pm$1.4&3.71$\pm$1.13\\
\hline
2006.515&c &1499.6$\pm$30.5& 0.00$\pm$0.00&  0.0$\pm$0.0&0.05$\pm$0.01\\
       &d & 502.0$\pm$22.9& 0.74$\pm$0.06& $-3.8\pm$4.6&0.33$\pm$0.01\\
       &e & 265.8$\pm$24.4& 1.41$\pm$0.06& 11.6$\pm$2.4&0.76$\pm$0.06\\
       &f & 132.2$\pm$15.7& 3.12$\pm$0.06& 26.6$\pm$1.1&0.66$\pm$0.07\\
       &g &  89.5$\pm$23.9& 4.72$\pm$0.28&  4.6$\pm$3.4&2.20$\pm$0.57\\
       &i & 157.8$\pm$23.5& 8.73$\pm$0.10& 14.2$\pm$0.7&1.48$\pm$0.21\\
       &j & 132.6$\pm$40.1&23.92$\pm$0.58&  0.4$\pm$1.4&3.90$\pm$1.17\\
\hline
2007.099&c&1378.7$\pm$51.5& 0.00$\pm$0.00&  0.0$\pm$0.0&0.07$\pm$0.01\\
       &d& 553.0$\pm$30.8& 1.06$\pm$0.06& $-7.3\pm$3.2&0.35$\pm$0.01\\
       &e& 282.7$\pm$25.8& 1.56$\pm$0.06& 15.3$\pm$2.2&0.56$\pm$0.04\\
       &f& 124.7$\pm$17.3& 3.26$\pm$0.06& 27.2$\pm$1.1&0.70$\pm$0.08\\
       &g&  66.8$\pm$14.5& 5.15$\pm$0.22&  4.8$\pm$2.4&2.09$\pm$0.43\\
       &i& 152.5$\pm$21.5& 8.82$\pm$0.10& 13.7$\pm$0.6&1.48$\pm$0.19\\
       &j& 124.4$\pm$35.3&24.04$\pm$0.53&  0.6$\pm$1.3&3.77$\pm$1.06\\
\hline
2007.441&c &1138.4$\pm$60.1& 0.00$\pm$0.00&  0.0$\pm$0.0&0.07$\pm$0.01\\
       &d & 545.9$\pm$20.6& 1.17$\pm$0.06& $-8.6\pm$2.9&0.31$\pm$0.01\\
       &e & 310.9$\pm$47.4& 1.52$\pm$0.06& 15.4$\pm$2.3&0.62$\pm$0.09\\
       &f & 109.6$\pm$19.5& 3.29$\pm$0.06& 27.0$\pm$1.0&0.74$\pm$0.11\\
       &g &  67.1$\pm$19.4& 5.55$\pm$0.35&  5.0$\pm$3.6&2.49$\pm$0.70\\
       &i & 139.5$\pm$22.7& 8.87$\pm$0.12& 13.5$\pm$0.8&1.52$\pm$0.23\\
       &j & 121.9$\pm$37.0&24.19$\pm$0.65&  0.9$\pm$1.5&4.33$\pm$1.30\\
\hline
2008.333&c &2765.9$\pm$57.8& 0.00$\pm$0.00&  0.0$\pm$0.0&0.04$\pm$0.01\\
       &d & 357.8$\pm$30.3& 1.61$\pm$0.06& $-8.8\pm$2.1&0.33$\pm$0.02\\
       &e & 419.9$\pm$35.6& 1.77$\pm$0.06&  8.7$\pm$1.9&0.90$\pm$0.07\\
       &f &  89.8$\pm$15.0& 3.53$\pm$0.06& 25.1$\pm$1.0&0.72$\pm$0.11\\
       &g &  52.4$\pm$12.6& 6.04$\pm$0.18&  7.6$\pm$1.7&1.61$\pm$0.36\\
       &i & 157.9$\pm$27.5& 9.11$\pm$0.16& 12.9$\pm$1.0&1.96$\pm$0.33\\
       &j & 136.8$\pm$32.0&24.21$\pm$0.44&  0.5$\pm$1.0&3.84$\pm$0.89\\
\hline
2008.544&c &2928.7$\pm$62.2& 0.00$\pm$0.00&  0.0$\pm$0.0&0.07$\pm$0.01\\
       &d & 225.0$\pm$24.5& 1.65$\pm$0.06& $-9.3\pm$2.1&0.26$\pm$0.02\\
       &e & 516.2$\pm$34.8& 1.74$\pm$0.06&  6.3$\pm$2.0&0.99$\pm$0.06\\
       &f &  87.5$\pm$12.8& 3.56$\pm$0.06& 24.9$\pm$1.0&0.76$\pm$0.10\\
       &g &  60.6$\pm$15.7& 6.01$\pm$0.22&  7.8$\pm$2.1&1.84$\pm$0.45\\
       &i & 161.2$\pm$30.0& 9.15$\pm$0.18& 12.9$\pm$1.2&2.07$\pm$0.37\\
       &j & 143.0$\pm$34.9&24.32$\pm$0.48&  0.6$\pm$1.1&3.96$\pm$0.95\\
\hline
2008.757&c &3197.9$\pm$39.9& 0.00$\pm$0.00&  0.0$\pm$0.0&0.05$\pm$0.01\\
       &d & 155.7$\pm$12.8& 1.75$\pm$0.06& $-9.1\pm$2.0&0.22$\pm$0.01\\
       &e & 456.9$\pm$33.9& 1.83$\pm$0.06&  6.9$\pm$1.9&0.87$\pm$0.06\\
       &f &  83.0$\pm$15.9& 3.43$\pm$0.09& 27.7$\pm$1.4&1.01$\pm$0.17\\
       &g &  42.4$\pm$13.6& 5.89$\pm$0.15&  4.4$\pm$1.4&1.05$\pm$0.29\\
       &i & 160.1$\pm$29.7& 8.86$\pm$0.16& 13.2$\pm$1.0&1.78$\pm$0.31\\
       &j & 140.1$\pm$41.5&24.15$\pm$0.57&  0.7$\pm$1.3&3.89$\pm$1.14\\
\hline
2009.153&c&2929.0$\pm$73.9& 0.00$\pm$0.00&  0.0$\pm$0.0&0.09$\pm$0.01\\
         &n$^\prime$& 842.8$\pm$42.8& 0.28$\pm$0.06&  8.1$\pm$12.1&0.03$\pm$0.01\\
         &d&  89.2$\pm$19.2& 1.80$\pm$0.06&$-10.3\pm$1.9&0.29$\pm$0.05\\
         &e& 489.0$\pm$57.5& 1.97$\pm$0.06&  4.9$\pm$1.7&1.07$\pm$0.12\\
         &f&  65.3$\pm$25.2& 3.58$\pm$0.13& 24.1$\pm$2.0&0.75$\pm$0.26\\
         &g&  43.3$\pm$16.8& 6.26$\pm$0.25&  4.0$\pm$2.3&1.36$\pm$0.51\\
         &i& 154.1$\pm$48.9& 8.99$\pm$0.29& 12.8$\pm$1.9&1.87$\pm$0.58\\
         &j& 126.4$\pm$53.5&24.31$\pm$0.79&  0.6$\pm$1.9&3.75$\pm$1.57\\
\hline
2009.482&c  &3125.8$\pm$86.9& 0.00$\pm$0.00&  0.0$\pm$0.0&0.09$\pm$0.01\\
       &n$^\prime$& 166.3$\pm$32.2& 0.46$\pm$0.06&  2.9$\pm$7.4&0.06$\pm$0.01\\
       &d  &  41.9$\pm$9.3& 1.78$\pm$0.06& $-9.4\pm$1.9&0.11$\pm$0.02\\
       &e  & 478.4$\pm$48.5& 2.11$\pm$0.06&  5.0$\pm$1.6&1.24$\pm$0.12\\
       &f  &  39.4$\pm$9.5& 3.65$\pm$0.07& 24.7$\pm$1.1&0.67$\pm$0.14\\
       &g  &  38.4$\pm$11.5& 6.47$\pm$0.12&  6.1$\pm$1.0&0.89$\pm$0.23\\
       &i  & 166.0$\pm$38.6& 8.83$\pm$0.26& 12.9$\pm$1.7&2.32$\pm$0.52\\
       &j  & 134.3$\pm$40.5&24.27$\pm$0.57&  1.0$\pm$1.3&3.84$\pm$1.14\\
\end{longtable}
\twocolumn 
\bsp
\label{lastpage}
\end{document}